\documentclass[aps,prb,reprint,twocolumn,floatfix,superscriptaddress,amsmath,amssymb,amsfonts]{revtex4-2}

\usepackage{url}
\usepackage{bm}
\usepackage{graphicx}
\usepackage{amsmath}
\usepackage{amstext}
\usepackage{amssymb}
\usepackage{amsfonts}
\usepackage{amsbsy}
\usepackage{verbatim}
\usepackage{color}
\usepackage[colorlinks=true, urlcolor=blue, linkcolor=blue, citecolor=blue, pdftex]{hyperref}
\usepackage{floatrow}
\usepackage{float}
\usepackage{tikz}
\usepackage{gensymb}
\usepackage{textcomp}
\usepackage{enumitem}
\usepackage[version=3]{mhchem}
\usepackage{afterpage}
\usepackage{pbox}
\usepackage{dsfont}
\usepackage{siunitx}
\usepackage{stackengine,wasysym,scalerel}
\usepackage{latexsym}
\usepackage{cancel}
\usepackage{comment}
\usepackage{array, multirow, bigdelim, makecell, booktabs}
\usepackage[misc]{ifsym}
\usepackage[capitalise]{cleveref}

\begin{document}
\title{Quantum spin liquids on the diamond lattice}
\author{Aishwarya Chauhan}
\thanks{These authors contributed equally.}
\affiliation{Department of Physics and Quantum Centre of Excellence for Diamond and Emergent Materials (QuCenDiEM), Indian Institute of Technology Madras, Chennai 600036, India}
\author{Atanu Maity}
\thanks{These authors contributed equally.}
\affiliation{Department of Physics and Quantum Centre of Excellence for Diamond and Emergent Materials (QuCenDiEM), Indian Institute of Technology Madras, Chennai 600036, India}
\author{Chunxiao Liu}
\thanks{These authors contributed equally.}
\affiliation{Department of Physics, University of California, Berkeley, California 94720, USA}
\author{Jonas Sonnenschein}
\affiliation{Theory of Quantum Matter Unit, Okinawa Institute of Science and Technology, 1919-1 Tancha, Onna-son, Okinawa 904-0495, Japan}
\author{Francesco Ferrari}
\affiliation{Institut f\"ur Theoretische Physik, Goethe Universit\"at Frankfurt, Max-von-Laue-Stra{\ss}e 1, 60438 Frankfurt am Main, Germany}
\affiliation{Department of Physics and Quantum Centre of Excellence for Diamond and Emergent Materials (QuCenDiEM), Indian Institute of Technology Madras, Chennai 600036, India}
\author{Yasir Iqbal}
\email{yiqbal@physics.iitm.ac.in}
\affiliation{Department of Physics and Quantum Centre of Excellence for Diamond and Emergent Materials (QuCenDiEM), Indian Institute of Technology Madras, Chennai 600036, India}

\begin{abstract} 
We perform a projective symmetry group classification of spin $S=1/2$ symmetric quantum spin liquids with different gauge groups on the diamond lattice. Employing the Abrikosov fermion representation, we obtain $8$ $SU(2)$, $62$ $U(1)$ and $80$ $\mathds{Z}_{2}$ algebraic PSGs. Constraining these solutions to mean-field parton \textit{Ans\"atze} with short-range amplitudes, the classification reduces to only $2$ $SU(2)$, $7$ $U(1)$ and $8$ $\mathds{Z}_{2}$ distinctly realizable phases. We obtain both the singlet and triplet fields for all \textit{Ans\"atze}, discuss the spinon dispersions, and present the dynamical spin structure factors within a self-consistent treatment of the Heisenberg Hamiltonian with up to third-nearest neighbor couplings. Interestingly, we find that a zero-flux $SU(2)$ state and some descendent $U(1)$ and $\mathds{Z}_{2}$ states host robust gapless nodal loops in their dispersion spectrum, owing their stability at the mean-field level to the projective implementation of rotoinversion and screw symmetries. A nontrivial connection is drawn between one of our $U(1)$ spinon Hamiltonians (belonging to the nonprojective class) and the Fu-Kane-Mele model for a three-dimensional topological insulator on the diamond lattice. We show that Gutzwiller projection of the 0- and $\pi$-flux $SU(2)$ spin liquids generates long-range N\'eel order.
\end{abstract}

\date{\today}

\maketitle

\section{Introduction}
Frustrated magnetic models and materials based on three-dimensional lattices are increasingly coming into limelight as fertile playgrounds in search of novel phases of matter. Indeed, the arrival of quantum materials based on the highly (geometrically) frustrated pyrochlore~\cite{Clark-2014,Plumb-2019,Sibille-2020,Smith-2022}, hyper-hyperkagome~\cite{Chillal-2020}, and trillium lattices~\cite{Ivica-2021} have evoked theoretical studies attempting to decipher the microscopic nature of the quantum paramagnetic ground states observed in experiment~\cite{Iqbal-2017,Zhang-2019,Chern-2021,fancelli2023classical,Bhardwaj-2022}. Numerical investigations of three-dimensional quantum Heisenberg antiferromagnets have identified quantum paramagnetic ground states on the geometrically frustrated pyrochlore~\cite{Iqbal-2019,Imre-2021,Astrakhantsev-2021,Schafer-2022} and face centered cubic~\cite{Kiese-2022} lattices. In three dimensions, the most exotic possibility is the realization of a quantum spin liquid (QSL) -- a magnetically disordered phase at zero temperature stabilized when the spins in a quantum magnet evade long-range magnetic order due to frustration and amplified quantum fluctuations~\cite{balents2010,Savary-2016,Kivelson-2023}.

In bipartite lattices where geometric frustration is absent, such exotic ground states can still occur when longer range antiferromagnetic interactions are included. These longer range interactions compete with the nearest-neighbor interactions, hence inducing frustration. Studies on simple cubic~\cite{Iqbal-2016frg,Farnell-2016} and body centered cubic~\cite{Ghosh-2019,Farnell-2016} lattices have found similar quantum disordered ground states in the presence of frustrating longer-range couplings. Given these results, it is natural to turn to another common 3D bipartite lattice -- the prototypical lattice for many familiar crystals -- the \emph{diamond lattice}, and ask about the nature of the antiferromagnetic ground states there. On the diamond lattice geometry, a pseudo fermion functional renormalization group analysis of the Heisenberg model with nearest neighbor $J_{1}$ and second nearest neighbor $J_{2}$ couplings reported a quantum paramagnetic ground state over a wide parameter regime for spin $S=1/2$~\cite{Buessen-2018}. The spin structure factor in this paramagnetic phase displays approximate spin spiral surfaces as remnants from the parent classical spin liquid of the $S\to\infty$ model~\cite{Bergman-2007}. Motivated by this finding, we take the first step towards addressing the following question: what is the precise microscopic nature of the $S=1/2$ quantum state on the diamond lattice which underpins the observed spiral surfaces? 

A powerful statement can already be made by solely examining the symmetry and lattice geometry: the  Lieb--Schultz--Mattis theorems (LSM) \cite{LIEB1961407,PhysRevLett.84.1535,PhysRevB.69.104431}, in its original form, state that if a lattice unit cell contains odd number of $S=1/2$ spins the ground state cannot be a trivial gapped paramagnet. While this seems to place the $S=1/2$ diamond magnet on the uninteresting side, as the diamond unit cell contains two (i.e., even number of) sublattices, the tale does not end here. A seminal work by Parameswaran {\it et al.} \cite{Sid-2013} claims that the diamond lattice is subject to a generalized LSM theorem: the ground state on this lattice is nontrivial provided the electron filling per unit cell is not in multiples of 4. The key insight here is to realize that the diamond lattice space group contains \emph{nonsymmorphic} symmetry elements such as a twofold screw, which produce filling constraints leading to a nontrivial ground state, even if the electronic filling is predicted to be trivial by the orignal LSM argument. Since then, many more lattice magnets have been predicted (in a systematic fashion) to have a nontrivial ground state \cite{doi:10.1073/pnas.1514665112,PhysRevLett.119.127202,PhysRevB.101.224437,10.21468/SciPostPhys.13.3.066}. Here, the LSM theorem provides a strong evidence for the putative spin liquid ground state, albeit other states that break some lattice symmetries are possible e.g., a lattice nematic liquid or a valence bond crystal, possibly described as the dimerization of a parent QSL state.

Quantum spin liquids, being paramagnetic in nature, cannot be distinguished by patterns of symmetry breaking and physical order parameters. They are host to excitations characterized by fractional quantum numbers (\textit{spinons}) and to emergent gauge fields which mediate their interactions~\cite{Wenbook}. Thus QSLs can broadly be categorized by the nature of the matter fields (gapped or gapless) and gauge fields [gapless $SU(2)$/$U(1)$ or gapped $\mathds{Z}_{2}$] excitations \cite{Zhou-2017}. Crucially, due to the emergence of a gauge structure, lattice symmetries are realized {\it projectively} in the Hilbert space of spinon excitations. This implies that a projective implementation of a lattice symmetry group (and time-reversal), together with a given gauge group, allows one to systematically classify different QSLs which possess the same physical symmetry, but belong to distinct projective representations. This classification framework goes under the name of projective symmetry groups (PSGs)~\cite{Wen-2002}. The PSG approach has been extensively applied on two- and three-dimensional lattices~\cite{Wang-2006,Lawler-2008,Lu-2011a,Lu-2011b,Yang-2012,Messio-2013,Yang-2016,Lu-2016a,Lu-2016b,Bieri-2016,Huang-2017,Huang-2018,Liu-2019,Jin-2020,Sonnenschein-2020,Sahoo-2020,Liu-2021,Chern-2021,Chern-2022,Benedikt-2022,Maity-2022,liu2023schwinger}, and has met with wide success in revealing the microscopic nature of the ground state of two-dimensional antiferromagnets on highly frustrated lattices~\cite{Iqbal-2011b,Iqbal-2013,Iqbal_2014,Iqbal-2011a,Iqbal-2012,Iqbal-2016,Iqbal-2018_bk,Iqbal-2021,Hu-2013,Kiese-2023,Ferrari-2023}

In this work, we employ the PSG framework for the Abrikosov fermion representation of $S=1/2$ spin operators~\cite{Abrikosov-1965} to provide a complete and systematic classification of symmetric QSLs on the diamond lattice with $SU(2)$, $U(1)$, and $\mathds{Z}_{2}$ low-energy gauge groups. For each of these gauge groups, we incorporate the full lattice space group symmetry as well as impose time-reversal symmetry. While our analysis principally focuses on singlet QSLs, in general, we also give the symmetry allowed triplet amplitudes thus allowing for a treatment of spin-orbit coupling in the original spin system. By following the PSG scheme to solve the gauge-symmetry consistency equations, we find that there are 8 $SU(2)$, 62 $U(1)$, and 80 $\mathds{Z}_{2}$ algebraic PSGs. Upon restricting to short-range (up to third nearest-neighbor) amplitudes in the symmetry-allowed spinon mean-field Hamiltonian, we find that the classification reduces to a limited number of distinct PSGs \textendash with only $2$ $SU(2)$, $7$ $U(1)$, and $8$ $\mathds{Z}_{2}$ distinct QSLs being realizable. Interestingly, we find that a zero-flux $SU(2)$ spinon Hamiltonian features a gapless nodal loop in the Brillouin zone. This nodal loop is shown to be robust at the mean-field level being protected by the projective rotoinversion and screw symmetries of the lattice. For the frustrated $J_{1}$-$J_{2}$-$J_{3}$ Heisenberg Hamiltonian, we self-consistently determine the spinon dispersions and dynamical spin structure factors.

The paper is organized as follows. In Sec.~\ref{sec:method}, we present the Heisenberg model, discuss the fermionic representation of spins, and explain the essence of the PSG framework. In Sec.~\ref{sec:lattice_time}, we discuss the symmetry properties of the lattice. In Sec.~\ref{sec:classification}, we apply the PSG method to the classification of QSLs on the diamond lattice and obtain the $SU(2)$, $U(1)$, and $\mathds{Z}_{2}$ algebraic PSGs. In Sec.~\ref{sec:flux_op}, we construct the $SU(2)$ flux operators which are crucial towards characterizing the different \textit{Ans\"atze}. In Sec.~\ref{sec:mean_field_amplitude}, we construct symmetric mean-field fermionic spinon Hamiltonians for $SU(2)$, $U(1)$, and $\mathds{Z}_{2}$ invariant gauge groups (IGGs), as well as perform a self-consistent mean-field analysis  to obtain the spinon band structures and dynamical spin structure factors. In Sec.~\ref{sec:topological_insulator}, we draw a nontrivial connection between one of our parton Hamiltonian and the Fu-Kane-Mele model for a three-dimensional topological insulator on the diamond lattice. In Sec.~\ref{sec:projection}, we discuss how Gutzwiller projection of the $0$- and $\pi$-flux $SU(2)$ \textit{Ans\"atze} gives rise to long-range N\'eel order. Finally, we summarize our results and present an outlook in Sec.~\ref{sec:discussion}.

\section{Model and method}
\label{sec:method}
In this section, we describe the fermionic representation of a spin Hamiltonian and the basic idea of projective symmetry groups. Our starting point is a generic Heisenberg model on a lattice 
\begin{equation}
\hat{\mathcal{H}}=\sum_{\langle i,j\rangle}J_{ij}\hat{\mathbf{S}}_{i} \cdot \hat{\mathbf{S}}_{j},
\label{eq:mod-ham}
\end{equation}
with $S=1/2$ spins on each site.

\subsection{Abrikosov fermion mean-field theory}
\label{sec:afmft}
To construct fermionic mean-field theories~\cite{Baskaran-1988} for QSLs from the spin Hamiltonian given by Eq.~\eqref{eq:mod-ham}, the first step is to express the spin operators in terms of spin-$1/2$ (charge neutral) fermionic spinon quasiparticles $\hat{f}_{i \sigma}$~\cite{Abrikosov-1965}
\begin{equation}
\label{eq:abrikosov}
\hat{S}^{\alpha}_{i}=\frac{1}{2}\sum_{\sigma\sigma^{\prime}}\hat{f}^\dagger_{i\sigma}\tau^{\alpha}_{\sigma\sigma^\prime}\hat{f}_{i\sigma^\prime},
\end{equation}
where  $\sigma=\{\uparrow,\downarrow$\}, $\alpha=1,2,3$, and $\tau^\alpha$ are the Pauli matrices. 
Formally, splitting a spin operator into two spinon operators captures  the fractional character of the spinon excitation in QSLs. While the local Hilbert space of the original spin model is two-dimensional, consisting of $\uparrow$ or $\downarrow$ states, the dimensionality of the fermionic Hilbert space is larger, as it includes the unphysical doubly occupied and empty states which carry spin $S=0$. A \textit{bonafide} wave function for spins can thus be obtained by projecting a fermionic state onto the Hilbert subspace with one fermion per site. This can be implemented by imposing the following constraints
\begin{align}\label{constraint}
\sum_{\alpha} \hat{f}_{i\alpha}^{\dagger}\hat{f}_{i\alpha} = 1 \; , \; \sum_{\alpha,\beta} \hat{f}_{i\alpha}\hat{f}_{i\beta}\epsilon_{\alpha\beta} = 0. \;
\end{align}
Employing the doublet representation $\hat{\psi}_i=(\hat{\phi}_i,\hat{\bar{\phi}}_i)$ with $\hat{\phi}_i=(\hat{f}_{i,\uparrow},\hat{f}_{i,\downarrow})^T$ and $\hat{\bar{\phi}}_i=(\hat{f}^\dagger_{i,\downarrow},-\hat{f}^\dagger_{i,\uparrow})^T$, Eq.~\eqref{eq:abrikosov} can be recast as~\cite{Affleck-1988}
\begin{equation}
\label{eq:abrikoshov_doublet}
\hat{S}^{\alpha}_{i}=\frac{1}{2}\text{Tr}[\hat{\psi}^\dagger_i\tau^\alpha\hat{\psi}_i].
\end{equation}

In this form, the symmetries of the fermionic representation are manifest. In particular, under the operation of right multiplication of $\hat{\psi}_{i}$ by $W_{i}$, i.e., $\hat{\psi}_i\rightarrow\hat{\psi}_i W_i$ with $W_i\in U(2)$, the spin operators in Eq.~\eqref{eq:abrikoshov_doublet} remain invariant. Furthermore, to preserve the fermionic character this symmetry must be restricted to $SU(2)$~\cite{Liu-2010}. Since this local site-dependent transformation acts internally on the spin operators, it corresponds to an emergent $SU(2)$ gauge symmetry of the fermionic representation. On the other hand, a left multiplication operation $\hat{\psi}_i\rightarrow G\hat{\psi}_{i}$ corresponds to a $SU(2)$ rotation of spin operators. This property enables us to define the manifestly $SU(2)$ spin rotational invariant (singlet) fields $\hat{U}_{ij}=\hat{\psi}^\dagger_i\hat{\psi}_j$, which live on the bonds of the lattice. These singlet link operators are composed of two terms, a singlet hopping field $\hat{\chi}_{ij}=\hat{f}^\dagger_{i,\uparrow}\hat{f}_{j,\uparrow}+\hat{f}^\dagger_{i,\downarrow}\hat{f}_{j,\downarrow}$ and a singlet pairing field $\hat{\Delta}_{ij}=\hat{f}_{i,\downarrow}\hat{f}_{j,\uparrow}-\hat{f}_{i,\uparrow}\hat{f}_{j,\downarrow}$, arranged in the following matrix structure:
\begin{eqnarray}
\hat{U}_{ij} =
\begin{bmatrix}
\hat{\chi}_{ij} & \hat{\Delta}^\dagger_{ij}  \\
\hat{\Delta}_{ij} & -\hat{\chi}^\dagger_{ij}
\end{bmatrix}.
\label{eq:ansatz}
\end{eqnarray}

Inserting Eq.~\eqref{eq:abrikoshov_doublet} into Eq.~\eqref{eq:mod-ham} results in a quartic Hamiltonian in the fermionic operators, which can be turned into an analytically solvable model through a quadratic mean-field decomposition~\cite{Wen-2002}. In order to obtain a $SU(2)$ rotationally invariant mean-field model, suitable for QSLs of isotropic systems, one restricts the mean-field approach to the quadratic singlet fields $\hat{\chi}_{ij},\hat{\Delta}_{ij}$. The resulting mean-field setting involves the link parameters $u_{ij}=\langle\hat{U}_{ij}\rangle$ (i.e., a matrix containing the terms $\chi_{ij}=\langle\hat{\chi}_{ij}\rangle$, $\Delta_{ij}=\langle\hat{\Delta}_{ij}\rangle$), and requires 
 the constraints of Eq.~\eqref{constraint} to be fulfilled on average, namely ${\sum_\alpha\langle \hat{f}_{i\alpha}^{\dagger}\hat{f}_{i\alpha}\rangle = 1} \; , \; {\sum_{\alpha,\beta}\langle \hat{f}_{i\alpha}\hat{f}_{i\beta}\rangle\epsilon_{\alpha\beta} = 0}$. In terms of the doublet representation, the latter mean-field expressions for the one-fermion-per-site constraints are formulated in the compact form $\langle\hat{\psi}_{i}\tau^{\alpha}\hat{\psi}^{\dagger}_{i}\rangle=0   $ ($\alpha\in\{1,2,3\}, \; \forall~i$). These terms can be incorporated in the Hamiltonian by means of three Lagrange multipliers $a_\mu$. Finally, the mean-field Hamiltonian takes the form,
\begin{align}\label{eq:mf_ham}
    \hat{H}_{MF} = & \sum_{\langle ij \rangle} \frac{3}{8} J_{ij} \left[\frac{1}{2}\text{Tr}(u_{ij}^{\dagger} u_{ij}) - \text{Tr}(\hat{\psi}_{i} u_{ij} \hat{\psi}^{\dagger}_{j} + \text{H.c.}) \right] \; \notag \\
    & + \sum_{i} \sum_\alpha a_{\alpha} \text{Tr}[\hat{\psi}_{i} \tau^{\alpha} \hat{\psi}^{\dagger}_{i} ]\;.
\end{align}

Here, $(u_{ij},a_\alpha)$ represents an \textit{Ansatz} for QSLs in the zeroth order mean-field setting. The link mean-field $u_{ij}$ can be re-expressed in terms of complex ``h'' and ``p'' parameters, 
\begin{equation}\label{eq:link_singlet}
    u_{ij}=\dot{\iota}\text{Im}h_{ij} \tau^0 + \text{Re}h_{ij} \tau^3+\text{Re}p_{ij} \tau^1+\text{Im}p_{ij} \tau^2,
\end{equation}
where $\tau^0$ is $2\times2$ identity matrix. Notice that the aforementioned mean-field Hamiltonian is invariant under the local $SU(2)$ gauge transformation 
\begin{equation}
    \hat{\psi}_i\rightarrow \hat{\psi}_iW_i, \; u_{ij}\rightarrow W^\dagger_iu_{ij}W_j,\;a_\alpha\tau^\alpha \rightarrow a_\alpha W^\dagger_i\tau^\alpha W_i.
    \label{eq:gauge_symmetry}
\end{equation}
 Therefore, in the mean-field construction, the internal symmetry group transforms into a local gauge symmetry.  

\subsection{Projective Symmetry group}
In this section, we describe the  PSG framework to classify the QSL mean-field \textit{Ans\"atze}. As discussed earlier, the fermionic mean-field Hamiltonian given by Eq.~\eqref{eq:mf_ham} has a local SU(2) gauge redundancy. This means that any \textit{Ansatz} $u_{ij}$ and its local $SU(2)$ gauge transformed \textit{Ansatz} $W_iu_{ij}W^\dagger_j$ represent the same physical QSL phase. They can be simply viewed as different ways of labeling the same QSL~\cite{Wen-2002}. This gauge redundancy can be exploited to distinguish physically different QSL phases. Indeed, two different \textit{Ans\"atze} label two different QSL phases \textit{iff} they are not related to each other via a local $SU(2)$ gauge transformation. This concept is the backbone of the PSG approach.

Let us consider an element $\mathcal{O}$ of the space group of a certain lattice and let us apply it to a given \textit{Ansatz}, such that $\mathcal{O}(u_{ij})=u_{\mathcal{O}(i)\mathcal{O}(j)}$. If $u_{ij}\neq u_{\mathcal{O}(i)\mathcal{O}(j)}$ one may conclude that the \textit{Ansatz} breaks the lattice symmetry $\mathcal{O}$. However,  this is not necessarily true, since the local $SU(2)$ gauge redundancy allows any symmetry to act not only linearly but also projectively into the local $SU(2)$ gauge space. In other words, even if $u_{ij}\neq u_{\mathcal{O}(i)\mathcal{O}(j)}$, the symmetry can be restored if we are able to associate a suitable local gauge transformation $G_{\mathcal{O}}(i)\in SU(2)$ to the lattice symmetry $\mathcal{O}$ such that 
\begin{align}\label{eq:sym_con_gauge}
G_{\mathcal{O}}^\dagger(\mathcal{O}(i)) u_{\mathcal{O}(i)\mathcal{O}(j)} G_{\mathcal{O}}(\mathcal{O}(j))= u_{ij}.
\end{align}
The combined operation of the symmetry element $\mathcal{O}$ and a local gauge transformation $G_{\mathcal{O}}(i)\in SU(2)$ constitutes a symmetry group known as the PSG. Different gauge inequivalent PSGs label different QSL \textit{Ans\"atze} and thus all different \textit{Ans\"atze} can be distinguished using PSGs. Therefore the PSG is a mathematical tool to characterize different QSLs, analogously to what regular symmetries do for distinct phases of matter within the Landau paradigm. 

The projective construction also requires the definition of the identity operation $\mathbb{I}$ in the projective space. This can be achieved by defining a pure local gauge group $\mathcal{G}\in SU(2)$ such that 
\begin{equation}    \mathcal{G}_{i}^{\dagger}u_{ij}\mathcal{G}_{j}=u_{ij}.
    \label{eq:IGG}
\end{equation}
 This is known as the invariant gauge group (IGG). Comparing with the definition of PSG given by Eq.~\eqref{eq:sym_con_gauge}, one can conclude that $\mathcal{G}$ is the PSG extension corresponding to $\mathcal{O}=\mathbb{I}$. Thus, in projective space, an identity can be defined up to an element of $\mathcal{G}$. For mean-field \textit{Ans\"atze} containing only imaginary hopping terms, i.e., $u_{ij}=\dot{\iota}\text{Im}h_{ij} \tau^0 $, the IGG contains all possible global $SU(2)$ transformations, i.e., $\mathcal{G}=\{e^{\dot{\iota}\phi\hat{n}.\hat{\tau}}\}$. If we add real hopping terms to the \textit{Ans\"atze}, i.e., $u_{ij}=\dot{\iota}\text{Im}h_{ij} \tau^0 + \text{Re}h_{ij} \tau^3$, the associated IGG is broken down to $U(1)$, i.e., $\mathcal{G}=\{e^{\dot{\iota}\phi\tau^3}\}$. Similarly, adding pairing terms, i.e., $u_{ij}=\dot{\iota}\text{Im}h_{ij} \tau^0 + \text{Re}h_{ij} \tau^3+\text{Re}p_{ij} \tau^1+\text{Im}p_{ij} \tau^2$, leads to a further lowering of the IGG to $\mathds{Z}_2$, i.e., $\mathcal{G}=\{\pm1\}$. The QSL mean-field \textit{Ans\"atze} are generally labeled by their IGGs: if one says $SU(2)$, $U(1)$, or $\mathds{Z}_2$ QSL mean-field \textit{Ansatz}, this means that the IGG of the \textit{Ansatz} is $SU(2)$, $U(1)$, or $\mathds{Z}_2$, respectively.

\section{Lattice and time-reversal symmetries}
\label{sec:lattice_time}

\begin{figure}
	\includegraphics[width=0.7\linewidth]{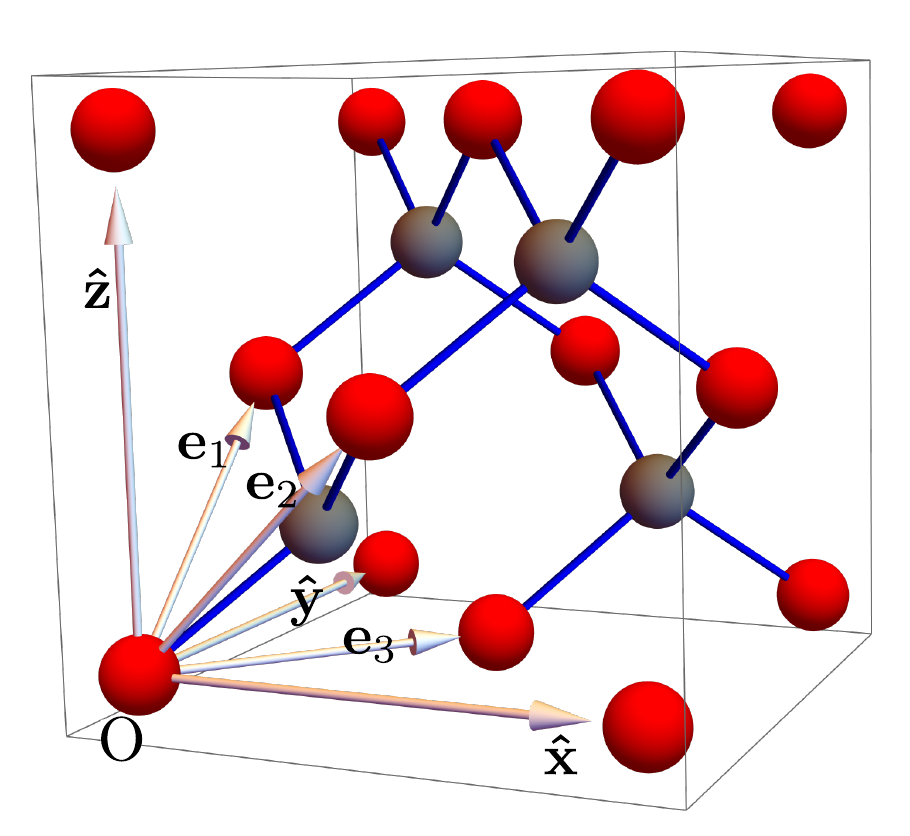}
	\caption{An illustration of the diamond lattice with red and grey spheres denoting the two sublattices ($\mu=0,1$, respectively). The Cartesian coordinate system has its basis vectors $\bf{\hat x}$, $\bf{\hat y}$, and $\bf{\hat z}$ aligned with the sides of the cubic conventional cell of the diamond lattice. The origin is placed on a $\mu=0$ site. $\mathbf{e}_{1}$, $\mathbf{e}_{2}$, and $\mathbf{e}_{3}$ are the fcc Bravais lattice vectors.}
	\label{fig:fig1}
\end{figure}

The diamond lattice is a bipartite lattice composed of two fcc sublattices which we label by $\mu = 0,1$. The fcc Bravais lattice vectors $\mathbf{e}_{1}$, $\mathbf{e}_{2}$, and $\mathbf{e}_{3}$ are defined as
\begin{align}\label{eq:fcc_vector}
	 &\mathbf{e}_{1}=\frac{a}{2}(\mathbf{\hat y}+\mathbf{\hat z}) \;, \\
	&\mathbf{e}_{2}=\frac{a}{2}(\mathbf{\hat z}+\mathbf{\hat x}) \;, \\
     &\mathbf{e}_{3}=\frac{a}{2}(\mathbf{\hat x}+\mathbf{\hat y}) \;, 
 \end{align}
where $a$ is the cubic lattice constant, $\hat{\mathbf{x}}$, $\hat{\mathbf{y}}$, and $\hat{\mathbf{z}}$ are the basis vectors of the Cartesian coordinates aligned with the cubic system of the diamond lattice, with its origin placed at a $\mu=0$ site (see Fig.~\ref{fig:fig1}). The sublattice-dependent site coordinates are defined as
\begin{align}\label{eq:GCC}
	(r_1,r_2,r_3,\mu) & \equiv \mathbf{r}_\mu \equiv  r_1\mathbf{e}_{1} +r_2\mathbf{e}_{2}+r_3\mathbf{e}_{3} +\bm{\epsilon}_\mu\;
\end{align}
with
\begin{align}
    &\bm{\epsilon}_{0}=(0,0,0) \; , \\
    &\bm{\epsilon}_{1}=\frac{a}{4} (\hat{\mathbf{x}}+\hat{\mathbf{y}}+\hat{\mathbf{z}}).
\end{align}

The space group of the diamond lattice is $Fd\bar{3}m$ (No. 227) (see Appendix~\ref{app:symmetry_properties}), which is generated by the following five symmetry elements:
\begin{equation}
  \label{eq:trans}
	\left.\begin{aligned}
		&T_{1}:(r_1,r_2,r_3,\mu)\rightarrow (r_1+1,r_2,r_3,\mu),\\
		&T_{2}:(r_1,r_2,r_3,\mu)\rightarrow (r_1,r_2+1,r_3,\mu),\\
		&T_{3}:(r_1,r_2,r_3,\mu)\rightarrow (r_1,r_2,r_3+1,\mu),\\
		&S_{4}:(r_1,r_2,r_3,\mu)\rightarrow (r_1+r_2+r_3+\mu,-r_3,-r_1,\bar{\mu}),\\
		&\bar{C}_{3}:(r_1,r_2,r_3,\mu)\rightarrow (-r_3,-r_1,-r_2,\bar{\mu}),\\
	\end{aligned}\right.
\end{equation}
where $\bar{\mu}=1-\mu$. The operations $T_{1}$, $T_{2}$, and $T_{3}$ denote the translations along the lattice vectors $\mathbf{e}_{1}$, $\mathbf{e}_{2}$, and $\mathbf{e}_{3}$, respectively; $S_4$ is a nonsymmorphic screw symmetry which is the composition of a $C_4$ rotation about the $z$ axis anchored at $(0,\frac{1}{4},0)$, followed by a translation $(0, 0, \frac{1}{4})$; $\bar{C}_{3}$ is a sixfold rotoinversion composed of a $C_3$ rotation around the $[1, 1, 1]$ axis and an inversion with respect to $(1/8,1/8,1/8)$. In Appendix~\ref{app:symmetry_properties}, we show that the entire space group can be generated from $S_4$ and $\bar{C}_3$ only, which thus constitute the minimal set of generators. 

 In addition to $\{T_1,T_2,T_3,S_4,\bar{C}_3\}$, fully symmetric QSL mean-field \textit{Ans\"atze} require the inclusion of time-reversal symmetry $\mathcal{T}$, which can be defined such that its action results in a sign change for the \textit{Ansatz}, i.e., $\mathcal{T}(u_{ij},a_\mu)=-(u_{ij},a_\mu)$~\cite{Wen-2002,Bieri-2016}. This implies that the projective construction given by Eq.~\eqref{eq:sym_con_gauge} becomes $G_{\mathcal{T}}^\dagger(i) u_{ij} G_{\mathcal{T}}(j)=-u_{ij}$ for $\mathcal{O}=\mathcal{T}$. Since it does not affect site positions, $\mathcal{T}$ commutes with all space group symmetries (and with gauge transformations~\cite{Bieri-2016}). The symmetry group of the diamond lattice is thus completely characterized by the following group relations:
\begin{subequations}
\label{eq:id_relation}
\begin{align}
  T_iT_{i+1}T^{-1}_iT^{-1}_{i+1}&=\mathbb{I}\ , \; \\
  T_i\bar{C}^{-1}_3T_{i+1}\bar{C}_3&=\mathbb{I}\ , \; \\
(\bar{C_3})^6 & = \mathbb{I} \ , \; \\
T^{-1}_1S_4T_2S^{-1}_4&=\mathbb{I} \ , \; \\
T^{-1}_2S_4T_2T^{-1}_3S^{-1}_4&=\mathbb{I}\ , \; \\
  S_4T^{-1}_1T_2S^{-1}_4T^{-1}_3&=\mathbb{I}\ , \; \\
  S_4T_1S^{-1}_4T^{-1}_1T_3&=\mathbb{I}\ , \; \\
  S_4T^{-1}_3S^{-1}_4T^{-1}_2T_1&=\mathbb{I}\ , \; \\
  S^4_4T^{-1}_1T^{-1}_2T_3&=\mathbb{I}\ , \; \\
(S_4 \bar{C_3})^2 & = \mathbb{I} \ , \; \\
(\bar{C_3}^2 S_4^2)^3 & = \mathbb{I} \ , \; \\
(\bar{C_3}^3 S_4)^4 & = \mathbb{I} \ , \; \\
(\bar{C_3} S_4^2)^6 & = \mathbb{I} \ , \; \\
\mathcal{T}^2 & = \mathbb{I} \ , \; \\
\mathcal{T} \mathcal{O} \mathcal{T}^{-1} \mathcal{O}^{-1} = \mathbb{I} \ , \; \mathcal{O} \in &\{T_1, T_2, T_3, S_4, \bar{C_3}\} \
.  
\end{align}
\end{subequations}
Here, the notation $i+1$ is intended as the modular arithmetic operation $\textrm{mod}(i-1,3)+1$, which permutes the three translation directions.

\section{PSG classification for diamond lattice}
\label{sec:classification}

\begin{table*}[t]
	\caption{The eight gauge inequivalent $SU(2)$ PSGs.}
	\begin{ruledtabular}
		\begin{tabular}{ccccccccccc}
			$\eta_1$&$\eta_{ST_3}$&$\eta_{\bar{C}_3T_1}$&$\eta_{\bar{C}_3T_2}$& $g_\mathcal{T}(0)$& $g_\mathcal{T}(1)$ &$g_{S_4}(0)$ & $g_{S_4}(1)$ &$g_{\bar{C}_3}(0)$ &$g_{\bar{C}_3}(1)$ & \# of PSGs\\
			\hline
			$\pm1$&$\eta_1$&$+1$&$+1$& $g_\mathcal{T}$& $-g_\mathcal{T}$ &$g_{S_4}$ & $\eta_{S_4}g_{S_4}$ &$g_{\bar{C_3}}$ &$\eta_{\bar{C}_3}g_{\bar{C_3}}$ & 8\\
		\end{tabular}
	\end{ruledtabular}
	\label{table:su2_psg}
\end{table*}

\begin{table*}
	\caption{The 62 $U(1)$ PSGs. $m_1,m_2,m_3$ can take values $0\;{\rm or}\;1$ and $n_c=0,1,2,3,4,5$. For $w_\mathcal{T}=0$, $G_\mathcal{T}(r,\mu)=(-1)^\mu\tau^0$ and for $w_\mathcal{T}=1$, $G_\mathcal{T}(r,\mu)=\dot{\iota}\tau^1$.}
	\begin{ruledtabular}
		\begin{tabular}{cccccccc}
			$w_{S_4}$&$w_{\bar{C}_3}$&$w_{\mathcal{T}}$&$\chi_1$&$\{\chi_{ST_3},\chi_{\bar{C}_3T_1},\chi_{\bar{C}_3T_2}\}$&$g_{3}(\theta_\mu)$&$g_{3}(\rho_\mu)$ & \# of PSGs\\
			\hline
			0&0&0&$\chi_1=0,\pi$&$\{-\chi_1,0,0\}$&$\{\tau^0,g_3(\theta_1)\}$ & $\{\tau^0,(-1)^{m_3} g_3(\theta_1)\}$&	4\\
			0&1&0&$\chi_1=0,\pi$&$\{-\chi_1,0,0\}$&$\{\tau^0,g_3(\theta_1)\}$ & $\{\tau^0, g_3(n_c\pi/3)\}$&	12\\
			1&0&0&$\chi_1=0,\pi,\frac{\pi}{2}$&$\{\chi_1,0,-2\chi_1\}$&$\{\tau^0,(-1)^{m_1}g_3(\chi_1)\}$ & $\{\tau^0, g_3(\rho_c)\}$&	6\\
			1&1&0&$\chi_1=0,\pi$&$\{\chi_1,0,0\}$&$\{\tau^0,(-1)^{m_2}\tau^0\}$ & $\{\tau^0, (-1)^{m_3}\tau^0\}$&	8\\
			0&0&1&$\chi_1=0,\pi$&$\{-\chi_1,0,0\}$&$\{\tau^0,(-1)^{m_1}\tau^0\}$ & $\{\tau^0,(-1)^{m_2} \tau^0\}$&	8\\
			0&1&1&$\chi_1=0,\pi$&$\{-\chi_1,0,0\}$&$\{\tau^0,(-1)^{m_1}\tau^0\}$ & $\{\tau^0,(-1)^{m_2} \tau^0\}$&	8\\
			1&0&1&$\chi_1=0,\pi$&$\{\chi_1,0,0\}$&$\{\tau^0,(-1)^{m_1}\tau^0\}$ & $\{\tau^0,(-1)^{m_2} \tau^0\}$&	8\\
			1&1&1&$\chi_1=0,\pi$&$\{\chi_1,0,0\}$&$\{\tau^0,(-1)^{m_1}\tau^0\}$ & $\{\tau^0, (-1)^{m_2}\tau^0\}$&	8\\
		\end{tabular}
	\end{ruledtabular}
	\label{table:u1_psg}
\end{table*}

	\begin{table*}
	\caption{ The 80 gauge inequivalent $\mathds{Z}_2$ PSGs.}
	\begin{ruledtabular}
		\begin{tabular}{ccccccccccc}
	$\eta_1$&$\eta_{ST_3}$&$\eta_{\bar{C}_3T_1}$&$\eta_{\bar{C}_3T_2}$& $g_\mathcal{T}(0)$& $g_\mathcal{T}(1)$ &$g_{S_4}(0)$ & $g_{S_4}(1)$ &$g_{\bar{C}_3}(0)$ &$g_{\bar{C}_3}(1)$ &  \# of PSGs\\
			\hline
			$\pm1$&$\eta_1$&$+1$&$+1$& $\tau^0$& $\eta_\mathcal{T}\tau^0$ &$\tau^0$ & $\eta_{S_4}\tau^0$ &$\tau^0$ &$\eta_{\bar{C}_3}\tau^0$ & 16\\
			$\pm1$&$\eta_1$&$+1$&$+1$& $\tau^0$& $\eta_\mathcal{T}\tau^0$ &$\tau^0$ & $\eta_{S_4}\tau^0$ &$\dot{\iota}\tau^2$ &$\eta_{\bar{C}_3}\dot{\iota}\tau^2$ & 16\\
			$\pm1$&$\eta_1$&$+1$&$+1$& $\dot{\iota}\tau^2$& $\dot{\iota}\eta_\mathcal{T}\tau^2$ &$\tau^0$ & $\eta_{S_4}\tau^0$ &$\tau^0$ &$\eta_{\bar{C}_3}\tau^0$ & 16\\
			$\pm1$&$\eta_1$&$+1$&$+1$& $\dot{\iota}\tau^2$& $\dot{\iota}\eta_\mathcal{T}\tau^2$ &$\tau^0$ & $\eta_{S_4}\tau^0$ &$\dot{\iota}\tau^2$ &$\eta_{\bar{C}_3}\dot{\iota}\tau^2$ & 16\\
			$\pm1$&$\eta_1$&$+1$&$+1$& $\dot{\iota}\tau^2$& $\dot{\iota}\eta_\mathcal{T}\tau^2$ &$\tau^0$ & $\eta_{S_4}\tau^0$ &$\dot{\iota}\tau^3$ &$\eta_{\bar{C}_3}\dot{\iota}\tau^3$ & 16\\
		\end{tabular}
	\end{ruledtabular}
	\label{table:z2_psg}
\end{table*}

Given the symmetry group $\mathcal{O}\in\{T_1,T_2,T_3,S_4,\bar{C}_3,\mathcal{T}\}$, the allowed PSG solutions for the diamond lattice are obtained using the algebraic relations given in Eqs.~\eqref{eq:id_relation} by associating a gauge transformation $G_\mathcal{O}$ with each of the symmetry operations $\mathcal{O}$. The important point is that these relations are required to be fulfilled only up to the IGG, and not the identity. The resulting generalized algebraic relations incorporating the corresponding gauge transformations are given in Appendix~\ref{sec:genric_gauge_con}.

Since the diamond lattice is bipartite, one can define unfrustrated \textit{Ans\"atze} with IGG~$\in SU(2)$, corresponding to the following PSG solutions [see Appendix~\ref{app:su2_psg_derivation}].
\begin{subequations}
	\label{eq:su2_gauge_sol}
	\begin{align}
		&G_{T_1}(r,\mu)=\tau^0,\;G_{T_2}(r,\mu)=\eta^{r_1}_1\tau^0,\;G_{T_3}(r,\mu)=\eta^{r_1+r_2}_1\tau^0,\\
		&G_{\mathcal{T}}(r,\mu)=g_\mathcal{T}(\mu),\\
		&G_{S_4}(r,\mu)=\eta^{r_3}_{ST_3}\eta^{r_2\left(\frac{r_2-1}{2}+r_3-\bar{\mu}\right)-r_3\left(\frac{r_3-1}{2}+r_1\right)}_1g_{S_4}(\mu),\\
		&G_{\bar{C}_3}(r,\mu)=\eta^{r_2}_{\bar{C}_3T_1}\eta^{r_3}_{\bar{C}_3T_2}\eta^{r_1(r_2+r_3)}_1g_{\bar{C}_3}(\mu)\, .
	\end{align}
\end{subequations}
Here, the $\eta$-parameters are integers which take values $\pm1$ and $(g_\mathcal{T}(\mu), g_{S_4}(\mu), g_{\bar{C}_3}(\mu))\in SU(2)$. As a consequence of the different possible values of some $\eta$ parameters, we find a total of eight $SU(2)$ PSGs which are summarized in Table~\ref{table:su2_psg}.

Upon fixing the IGG~$\in U(1)$, and working in the canonical gauge~\cite{Wen-2002}, both real and imaginary hopping terms are allowed and the corresponding PSG associated with a symmetry operation $\mathcal{O}$ takes the generic form $G_\mathcal{O}(r,\mu)=g_3(\phi_\mathcal{O}(r,\mu))(\dot{\iota}\tau^1)^{w_\mathcal{O}}$. Here, we use the notation $g_3(\xi)=e^{\dot{\iota}\xi\tau^3}$, where $\xi$ is a $U(1)$ phase; $w_\mathcal{O}$ can take values $0$ and $1$.  We obtain the following solutions (see Appendix~\ref{app:u1_psg_derivation}).
\begin{equation}
\label{eq:u1_trans_gauge}
\left.\begin{aligned}
&G_{T_1}(r,\mu)=\tau^0,\\
&G_{T_2}(r,\mu)=g_3(-r_1\chi_1),\\
&G_{T_3}(r,\mu)=g_3((r_1-r_2)\chi_1),\\
\end{aligned}\right.
\end{equation}
 \begin{equation}
\label{eq:u1_s4_gauge}
\left.\begin{aligned}
   G_{S_4}(r,\mu)&=g_3\left(\theta_{\mu}+r_3\chi_{ST_3}-(-1)^{w_{S_4}}\zeta_{r,\mu}\chi_1\right)(\dot{\iota}\tau^1)^{w_{S_4}}\\ {\rm with}~
   \zeta_{r,\mu}&=r_2\left(\frac{r_2-1}{2}-r_3-\bar{\mu}\right)+r_3\left(r_1+\frac{r_3-1}{2}\right),\\
    \end{aligned}\right.
\end{equation}

\begin{equation}
\label{eq:u1_cb3_gauge}
\left.\begin{aligned}
   G_{\bar{C}_3}(r,\mu)&=g_3\left(r_1(r_3-r_2)\chi_1+\rho_{\mu}\right.\\
   &\left.-(-1)^{w_{\bar{C}_3}}(r_2\chi_{\bar{C}_3T_1}+r_3\chi_{\bar{C}_3T_2})\right)(\dot{\iota}\tau^1)^{w_{\bar{C}_3}},\\
       \end{aligned}\right.
\end{equation}

\begin{equation}
\label{eq:u1_time_gauge}
\left.\begin{aligned}
&G_\mathcal{T}(r,\mu)=(-1)^{\mu}\tau^0,\dot{\iota}\tau^1 \, .\\
\end{aligned}\right.   
\end{equation}
We find a total of 62 $U(1)$ PSGs which are listed in Table~\ref{table:u1_psg}. It is worth noting the presence of solutions corresponding to $w_\mathcal{T}=0$ is a unique feature of the diamond lattice and is connected to it bipartite nature. Such solutions do not exist on non bipartite geometries such as the pyrochlore~\cite{Liu-2021} and hyperkagome~\cite{Huang-2017} lattices. From a technical perspective, such a characteristic is accompanied by the appearance of a sublattice-dependant projective solution of time-reversal symmetry. Consequently, this permits the existence of nonvanishing time-reversal symmetric \textit{Ans\"atze}. Another unique feature of the diamond lattice is the appearance of an \textit{Ansatz} with $\chi_1=\pi/2$ (referred to as Class-C) which respects both the lattice space group as well as time-reversal symmetry. On the contrary, on the pyrochlore lattice~\cite{Liu-2021}, this class of solutions respects the lattice space group symmetry but breaks time-reversal symmetry, thus giving rise to chiral spin liquid \textit{Ans\"atze}.

Upon further lowering the IGG to $\mathds{Z}_2$, we obtain the following PSG solutions (see Appendix~\ref{app:z2_psg_derivation})
\begin{subequations}
	\label{eq:z2_gauge_sol}
    \begin{align}
		&G_{T_1}(r,\mu)=\tau^0,\;G_{T_2}(r,\mu)=\eta^{r_1}_1\tau^0,\;G_{T_3}(r,\mu)=\eta^{r_1+r_2}_1\tau^0,\\
		&G_{\mathcal{T}}(r,\mu)=g_\mathcal{T}(\mu),\\
		&G_{S_4}(r,\mu)=\eta^{r_3}_{ST_3}\eta^{r_2\left(\frac{r_2-1}{2}+r_3-\bar{\mu}\right)-r_3\left(\frac{r_3-1}{2}+r_1\right)}_1g_{S_4}(\mu),\\
		&G_{\bar{C}_3}(r,\mu)=\eta^{r_2}_{\bar{C}_3T_1}\eta^{r_3}_{\bar{C}_3T_2}\eta^{r_1(r_2+r_3)}_1g_{\bar{C}_3}(\mu)\,.
	\end{align}
\end{subequations}
Here, $(g_\mathcal{T}(\mu), g_{S_4}(\mu), g_{\bar{C}_3}(\mu))\in SU(2)$ and the $\eta$ parameters take values $\pm1$. We find a total of 80 $\mathds{Z}_2$ PSGs which are listed in Table~\ref{table:z2_psg}. Similar to the $U(1)$ PSGs, for the $\mathds{Z}_{2}$ case also we find the existence of sublattice-dependent time-reversal projective gauge solutions which are absent for the pyrochlore lattice~\cite{Liu-2021}. 

Here, we would like to stress that, one novelty of the $U(1)$ PSG on the diamond lattice is the existence of those $\chi_1=\pi/2$ classes (classes labeled by C), whose parton unit cell is quadrupled in two out of the three translation directions. To our knowledge, this ``quadruply enlarging'' of unit cell have only been observed in the diamond and pyrochlore lattice. The physical origin for it remains unclear and is an interesting question for future studies. Interestingly, Wen in his seminal paper~\cite{Wen-2002} already pointed out that, for the square lattice PSG, an infinite number of chiral symmetry classes may exist corresponding to $\chi_1=\frac{m\pi}{n}$ (U1$^m_n$ class) with $m,n\in\mathds{Z}$. On the other hand, we point out that there have been lattices, when, assuming full lattice symmetry, forbids any unit cell enlargement in the U(1) PSG: For example, in hyperhoneycomb lattice, the presence of glide reflection symmetry forbids such classes.

\section{$SU(2)$-flux operators}
\label{sec:flux_op}
Given the $SU(2)$ gauge redundancy, the mean-field \textit{Ans\"atze} may not be expressed in their canonical IGG form. In such cases, the IGG can be traced out by introducing $SU(2)$ flux operators~\cite{Bieri-2016,Wen-2002} $P_{\mathcal{C}_j}$ which are defined around the loops ``$\mathcal{C}$'' (comprised of sites $j,k,l,\cdots,j$) with respect to a base site ``$j$''. 
\begin{equation}
  P_{\mathcal{C}_j}=P(j,k,l,\cdots,j)=u_{jk}u_{kl}\cdots u_{mj}.  
\end{equation}
The $\mathcal{G}\in$~IGG for an \textit{Ansatz} in a generic gauge form can be traced by the following condition~\cite{Bieri-2016}:
\begin{equation}
    [\mathcal{G},P_{\mathcal{C}_j}]=0.
\end{equation}
In general, $P_{C_j}$ for a ``$q$''-sided loop takes the form,
\begin{equation}
    P_{\mathcal{C}_j}(\phi_{\mathcal{C}})\propto     g_je^{\dot{\iota}\phi_{\mathcal{C}}\tau^3}(\tau^3)^qg^\dagger_j\;\text{with}\; g_j\in SU(2). 
\end{equation}
Here, the phase ``$\phi_\mathcal{C}$'' can be interpreted as a flux threading the loop $\mathcal{C}$. Complementary to the PSG and IGG, the \textit{Ans\"atze} can also be characterized based on these fluxes. In the following, we define the relevant loop operators on the diamond lattice, based on which we characterize and distinguish the \textit{Ans\"atze}.

 \begin{figure}[b]
	\includegraphics[width=1.0\linewidth]{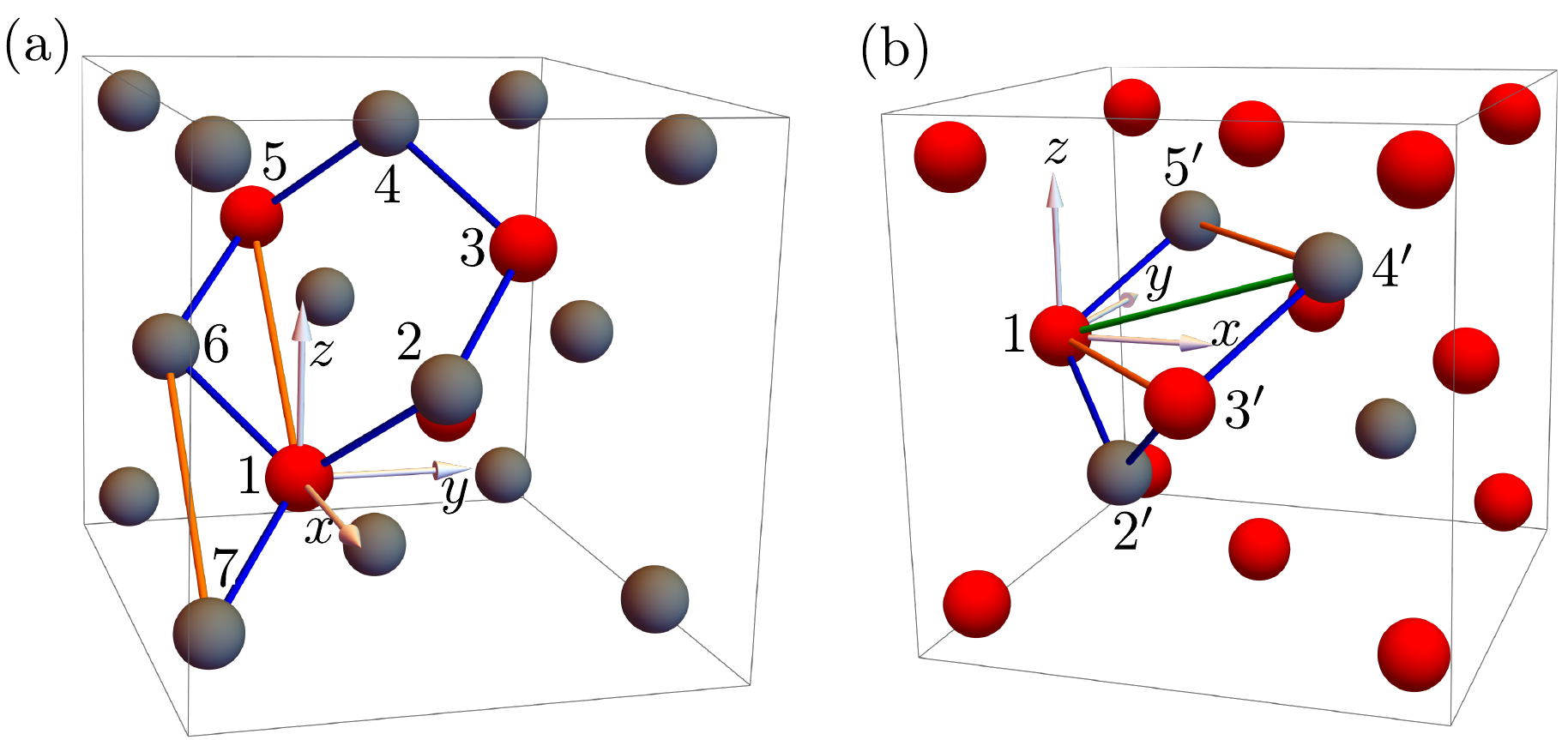}
	\caption{Loops formed when considering up to (a) second nearest neighbor bonds and (b) third nearest neighbor bonds. The $\mu=0$ and~$1$ sublattice sites are marked with red and gray spheres, respectively. The 1NN, 2NN, and 3NN bonds are drawn in blue, orange and green, respectively.}
	\label{fig:fig2}
\end{figure}

\begin{table*}
\begin{ruledtabular}
\begin{tabular}{ccccc}
\multirow{3}{*}{Label} & 1NN & 2NN & 3NN & Onsite\\
& \{$u_{OA},u_{OB},u_{OC},u_{OD}$\} & \{$u^0_{O1},u^0_{O2},u^0_{O3},u^0_{O4},u^0_{O5},u^0_{O6},$ & \{$u_{Oa},u_{Ob},u_{Oc},u_{Od},u_{Oe},u_{Of},$ & $\{a_\alpha(0),a_\alpha(1)\}$  \\
& & $u^1_{O1},u^1_{O2},u^1_{O3},u^1_{O4},u^1_{O5},u^1_{O6}\}$ & $u_{Og},u_{Oh},u_{Oi},u_{Oj},u_{Ok},u_{Ol}\}$ & \\
\hline
$SA3$& $\{u_1=h_1\tau^3\}$ & $\{0\}$ & $\{u_3=h_3\tau^3\}$ & $\{0\}$ \\
$SB1$& $\{u_1=h_1\tau^3\}$ & $\{0\}$ & $\{0\}$ & $\{0\}$ \\
$UA01$& $\{u_1=h_1\tau^3\}$ & $\{u_2=h_2\tau^3\}$ & $\{u_3=h_3\tau^3\}$ & $\{a_3\}$ \\
$UA31$& $\{u_1=h_1\tau^3\}$ & $\{u_2=h_2\tau^3,u_2,u_2,u_2,u_2,u_2,$ & $\{u_3=h_3\tau^3\}$ & $\{a_3,-a_3\}$ \\
& & $u'_2=-u_2,u'_2,u'_2,u'_2,u'_2,u'_2\}$ & & \\
$UA30$& $\{u_1=h_1\tau^3\}$ & $\{u_2=0\}$ & $\{u_3=\dot{\iota} h_3\tau^0+h'_3\tau^3\}$ & $\{0\}$ \\
$UB01$& $\{u_1=h_1\tau^3\}$ & $\{u_2=h_2\tau^3,-u_2,u_2,u_2,u_2,-u_2,$ & $\{0\}$ & $\{a_3\}$ \\
& & $u'_2=u_2,u'_2,u'_2,-u'_2,u'_2,u'_2\}$& & \\
$UB31$& $\{u_1=h_1\tau^3\}$ & $\{u_2=h_2\tau^3,-u_2,u_2,u_2,u_2,-u_2,$ & $\{0\}$ & $\{a_3,-a_3\}$ \\
& & $u'_2=-u_2,u'_2,u'_2,-u'_2,u'_2,u'_2\}$ &  & \\
$UB20$& $\{u_1=h_1\tau^3\}$ & $\{u_2=0\}$ & $\{u_3=\dot{\iota} h_3\tau^0,u_3,-u_3,u_3,u_3,-u_3\}$ & $\{0\}$ \\
& &  & $-u_3,-u_3,-u^\dagger_3,-u_3,-u_3,-u_3\}$ & \\
$UC20$& $\{u_1=h_1\tau^3\}$ & $\{0\}$ & $\{0\}$ & $\{0\}$ \\

$ZA000$& $\{u_1=h_1\tau^3\}$ & $\{u_2=h_2\tau^3+p_2\tau^1\}$ & $\{u_3=h_3\tau^3+p_3\tau^1\}$ & $\{a_{1,3}\}$ \\
$ZA310$& $\{u_1=h_1\tau^3\}$ & $\{u_2=h_2\tau^3+p_2\tau^1,u_2,u_2,u_2,u_2,u_2,$ & $\{u_3=h_3\tau^3\}$ & $\{a_3\}$ \\
& & $u'_2=\tau^3u_2\tau^3,u'_2,u'_2,u'_2,u'_2,u'_2\}$ & & \\
$ZA010$& $\{u_1=h_1\tau^3\}$ & $\{u_2=h_2\tau^3+p_2\tau^1,u_2,u_2,u_2,u_2,u_2,$ & $\{u_3=h_3\tau^3\}$ & $\{a_3,-a_3\}$ \\
&&$u'_2=\tau^1u_2\tau^1,u'_2,u'_2,u'_2,u'_2,u'_2\}$ & & \\
$ZB200$& $\{u_1=h_1\tau^3\}$ & $\{u_2=h_2\tau^3+p_2\tau^1,-u_2,u_2,u_2,u_2,-u_2,$ & $\{0\}$ & $\{a_{1,3}\}$ \\
&&$u_2,u_2,u_2,-u_2,u_2,u_2\}$ & & \\
$ZB110$& $\{u_1=h_1\tau^3\}$ & $\{u_2=h_2\tau^3+p_2\tau^1,-u_2,u_2,u_2,u_2,-u_2,$ & $\{0\}$ & $\{a_3\}$ \\
&&$u'_2=\tau^3u_2\tau^3,u'_2,u'_2,-u'_2,u'_2,u'_2\}$ & & \\
$ZB210$& $\{u_1=h_1\tau^3\}$ & $\{u_2=h_2\tau^3+p_2\tau^1,-u_2,u_2,u_2,u_2,-u_2,$ & $\{0\}$ & $\{a_3,-a_3\}$ \\
&&$u'_2=\tau^1u_2\tau^1,u'_2,u'_2,-u'_2,u'_2,u'_2\}$ & & \\
$ZB303$& $\{u_1=h_1\tau^3\}$ & $\{u_2=p_2\tau^1,-u_2,u_2,u_2,u_2,-u_2,$ & $\{u_3=\dot{\iota} h_3\tau^0,u_3,-u_3,u_3,u_3,-u_3\}$ & $\{a_1\}$ \\
& &$u'_2=u_2,u'_2,u'_2,-u'_2,u'_2,u'_2\}$ & $-u_3,-u_3,-u^\dagger_3,-u_3,-u_3,-u_3\}$ & \\
$ZB203$& $\{u_1=h_1\tau^3\}$ & $\{u_2=p_2\tau^1,-u_2,u_2,u_2,u_2,-u_2,$ & $\{u_3=\dot{\iota} h_3\tau^0,u_3,-u_3,u_3,u_3,-u_3\}$ & $\{a_1,-a_1\}$ \\
& &  $u'_2=\tau^3u_2\tau^3,u'_2,u'_2,-u'_2,u'_2,u'_2\}$ & $-u_3,-u_3,-u^\dagger_3,-u_3,-u_3,-u_3\}$ & \\
\end{tabular}
\end{ruledtabular}
\caption{All possible \textit{Ans\"atze} on the diamond lattice when restricted upto third nearest neighbours. There are in total 2 $SU(2)$, 7 $U(1)$ and 8 $\mathds{Z}_2$ mean-field \textit{Ans\"atze}. The notation ($u_{OA},u_{OB},u_{OC},u_{OD}$), ($u^\mu_{Oi}$ with $\mu=0,1$ and $i=1,2,..,6$) and ($u_{O\alpha}$ with $\alpha=a,b,c,...,l$) is defined in the Appendix~\ref{app:ref_bond}. The labelling scheme is given at the beginning of Sec.~\ref{sec:mean_field_amplitude}.}
\label{table:ansatz}
\end{table*}

\begin{table*}
\begin{ruledtabular}
\begin{tabular}{cccccccc}
Label & $h_1$ & $h_2$ & $p_2$ & $\textbf{Re} h_3$ & $\textbf{Im} h_3$ & $p_3$ & $e_0$\\
\hline
$SA3$& 0.168509 & 0.0 & 0.0 & -0.013143 & 0.0 & 0.0 & -0.158353\\
$UA01$& 0.168509 & 1.64$\times 10^{-4}$ & 0.0 & -0.013143 & 0.0 & 0.0 &-0.158353\\
$UA31$& 0.168486 & 0.001336 & 0.0 & -0.013140 & 0.0 & 0.0 &-0.158358\\
$UA30$& 0.168509 & 0.0 & 0.0 & -0.013143 & 3.1$\times 10^{-5}$ & 0.0 & -0.158353\\
$ZA000$& 0.168486 & 8.$\times 10^{-6}$ & 0.001339 & -0.013140& 0.0 & 3.1$\times 10^{-5}$ & -0.158358\\
$ZA310$& 0.168509 & 1.28$\times 10^{-4}$  & 1.28$\times 10^{-4}$  & -0.013143 & 0.0 & 0.0 & -0.158353\\
$ZA010$& 0.168481 & 0.001037 & 0.001037 & -0.012589 & 0.0 & 0.0 & -0.158358\\
\hline
$SB1$& 0.159108 & 0.0 & 0.0 & 0.0 & 0.0 & 0.0 &-0.135016\\
$UB01$& 0.159108 & 0.0 & 0.0 & 0.0 & 0.0 & 0.0 & -0.135016\\
$UB31$& 0.148013 & 0.027952 & 0.0 & 0.0 & 0.0 & 0.0 & -0.139943\\
$UB20$& 0.159108 &0.0& 0.0 & 0.0 & 7.2$\times 10^{-5}$ & 0.0 & -0.135016\\
$ZB200$& 0.148017 & 5$\times 10^{-5}$ & 0.027948 & 0.0 & 0.0 & 0.0 & -0.139943\\
$ZB110$& 0.159108 & 0.0 & 0.0  & 0.0 & 0.0 & 0.0 & -0.135016\\
$ZB210$& 0.147977 & 0.022403 & 0.016817 & 0.0 & 0.0 & 0.0 & -0.139943\\
$ZB303$& 0.148017 & 0.0 & 0.027948 & 0.0& 3.2$\times 10^{-5}$ & 0.0 &-0.139943\\
$ZB203$& 0.159108 &0.0 & 0.0 & 0.0 & 7.4$\times 10^{-5}$ & 0.0 & -0.135016\\
\hline
$UC20$& 0.159905 &0.0 & 0.0 & 0.0 & 0.0 & 0.0 & -0.136372\\
\end{tabular}
\end{ruledtabular}
\caption{The self-consistent mean-field amplitudes and ground state energies calculated for the parameters $(J_2/J_1,J_3/J_1)=(0.54,0.40)$}
\label{table:self_consistent_values}
\end{table*}

Consider a base site ``$1$'' [e.g., $(0,0,0,0)$, see Fig~\ref{fig:fig2}(a)]. Starting from this base site and considering only the first nearest neighbor (1NN) bonds [drawn in blue] we find that there are only six possible hexagonal loops [shown is the loop formed by sites $(1,2,3,4,5,6)$] which are connected to each other by symmetries. We can thus consider any one of them to define the flux structure when considering only 1NN amplitudes. Similarly, the inclusion of second nearest neighbor bonds (2NN) [drawn in orange] gives rise to multiple triangular loops all composed of two 1NN bonds and one 2NN bond. In order to define the flux structure, we consider two possible distinct triangular loop\; the $(1,5,6)$ loop contains a 2NN bond connecting ``0'' sublattice sites [$1\leftrightarrow 5$], while the $(1,6,7)$ loop contains a 2NN bond connecting ``1'' sublattice sites [$6\leftrightarrow 7$]. The hexagonal loop has been defined as
\begin{equation}
  	\left.\begin{aligned}
  		P_h&=P(1,2,3,4,5,6,1)\\
  		&=\langle(0,0,0,0)(0,0,0,1)\rangle\langle(0,0,0,1)(1,0,0,0)\rangle\\
  		&\langle(1,0,0,0)(1,0,-1,1)\rangle\langle(1,0,-1,1)(1,0,-1,0)\rangle\\
  		&\langle(1,0,-1,0)(0,0,-1,1)\rangle\langle(0,0,-1,1)(0,0,0,0)\rangle,\\
  	\end{aligned}\right.
  \end{equation} 
where $\langle(r_1,r_2,r_3,\mu)(r'_1,r'_2,r'_3,\mu')\rangle$ denotes the \textit{Ans\"atze} $u_{ij}$ on the bond connecting sites $i\equiv(r_1,r_2,r_3,\mu)$ and $j\equiv(r'_1,r'_2,r'_3,\mu')$. The two triangular loops, schematically shown in Fig.~\ref{fig:fig2}(a), are defined as
  \begin{equation}
	\left.\begin{aligned}
		P_{t_0}&=P(1,5,6,1)=\langle(0,0,0,0)(1,0,-1,0)\rangle\\
		&\langle(1,0,-1,0)(0,0,-1,1)\rangle\langle(0,0,-1,1)(0,0,0,0)\rangle\\
	\end{aligned}\right.
\end{equation} 
and
  \begin{equation}
	\left.\begin{aligned}
		P_{t_1}&=P(1,6,7,1)=\langle(0,0,0,0)(0,0,-1,1)\rangle\\
		&\langle(0,0,-1,1)(-1,0,0,1)\rangle\langle(-1,0,0,1)(0,0,0,0)\rangle.\\
	\end{aligned}\right.
\end{equation} 

Employing translations, the hexagonal loop operator can be written in terms of $u_{OA}$, $u_{OB}$, $u_{OC}$, $u_{OD}$  and similarly the triangular loops can be expressed in terms of translations and $u^\mu_{Oi}$ [see Appendix~\ref{app:ref_bond} for the definition of bonds]:
\begin{equation}
  	\label{eq:loop_hex}
	\left.\begin{aligned}
		P_h&=u_{OA}u^\dagger_{OB}u_{OD}u^\dagger_{OA}g_3(-\chi_1)u_{OB}u^\dagger_{OD}\\
	\end{aligned}\right.
\end{equation} 
\begin{equation}
	\label{eq:loop_tri}
	\left.\begin{aligned}
		P_{t_0}&=g_3(\chi_1)(u^0_{O4})^\dagger g_3(-\chi_1)u_{OB}u^\dagger_{OD}\\
		P_{t_1}&=u_{OD}g_3(-\chi_1)u^1_{O4}u_{OB}.\\
	\end{aligned}\right.
\end{equation} 

Once we include third nearest neighbor (3NN) bonds, two distinct triangular loops whose sides are composed of 1NN, 2NN and 3NN bonds one each, can be defined [see Fig.~\ref{fig:fig2}(b)]. Starting with a given base point ``$1$'', e.g. $(0,0,0,0)$, one can distinguish these two triangular loops [$(1,3',4')$ and $(1,4',5')$] by the fact that they contain distinct 2NN bonds, which connect either two $\mu=0$ sublattice sites [$1\leftrightarrow 3'$ in $(1,3',4')$] or two $\mu=1$ sublattice sites [$4'\leftrightarrow 5'$ in $(1,4',5')$] . The corresponding flux operators are
  \begin{equation}
	\left.\begin{aligned}
		P_{3t_0}&=P(1,4',3',1)=\langle(0,0,0,0)(-1,1,0,1)\rangle\\
		&\langle(-1,1,0,1)(-1,1,0,0)\rangle\langle(-1,1,0,0)(0,0,0,0)\rangle,\\
		P_{3t_1}&=P(1,4',5',1)=\langle(0,0,0,0)(-1,1,0,1)\rangle\\
		&\langle(-1,1,0,1)(0,0,0,1)\rangle\langle(0,0,0,1)(0,0,0,0)\rangle,\\
	\end{aligned}\right.
\end{equation} 
which can be written as,
\begin{equation}
	\label{eq:loop_tri_third}
	\left.\begin{aligned}
		P_{3t_0}&=u_{Ob}u^\dagger_{OA}g_3(\chi_1)u^0_{O6},\\
		P_{3t_1}&=u_{Ob}g_3(\chi_1)u^1_{O6}u^\dagger_{OA}.\\
	\end{aligned}\right.
\end{equation} 

Furthermore, there is a four-site loop ($(1,4',3',2')$) composed of one 3NN bond and three 1NN bonds [see Fig.~\ref{fig:fig2}(b)] 
  \begin{equation}
	\left.\begin{aligned}
		P_{3r}&=P(1,4',3',2',1)\\
  &=\langle(0,0,0,0)(-1,1,0,1)\rangle\langle(-1,1,0,1)(-1,1,0,0)\rangle\\
		&\langle(-1,1,0,0)(-1,0,0,1)\langle(-1,0,0,1)(0,0,0,0)\rangle,\\
	\end{aligned}\right.
\end{equation} 
which can be expressed as
  \begin{equation}
  	\label{eq:loop_rec_third}
	\left.\begin{aligned}
		P_{3r}&=u_{Ob}u^\dagger_{OA}u_{OC}u^\dagger_{OB}.\\
	\end{aligned}\right.
\end{equation} 

All the aforementioned loop operators $(P_h,P_{3r})$-$(P_{t_0},P_{t_1})$-$(P_{3t_0},P_{3t_1})$ have been utilized to trace the IGGs of all the \textit{Ans\"atze} in arbitrary gauge form. However, for characterizing an \textit{Ansatz} by its flux structure, it suffices to use the following fluxes $(P_h,P_{3r})$-$(P_{t_0},P_{t_1})$.

\begin{figure*}
	\includegraphics[width=1.0\linewidth]{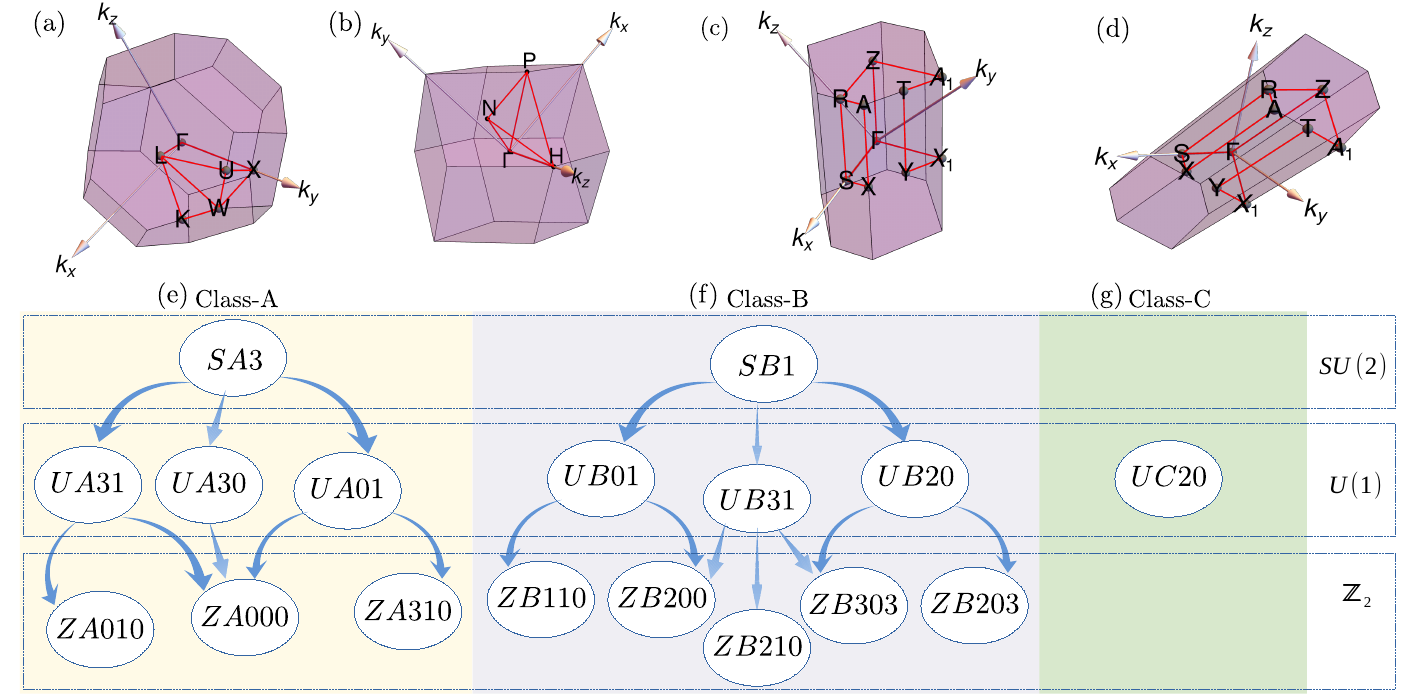}
	\caption{Illustration of (a) first Brillouin zone for class-A \textit{Ans\"atze}, (b) the corresponding extended Brillouin zone. Illustration of the reduced Brillouin zones required for class-B (c) and class-C (d) \textit{Ans\"atze} involving doubling and quadrupling of unit cells, respectively. The high-symmetry points marked have the following coordinates: (a), $\Gamma=(0,0,0)$, $X=(0,2\pi,0)$, $U=(\pi/2,2\pi,\pi/2)$, $W=(\pi,2\pi,0)$, $K=(3\pi/2,3\pi/2,0)$, and $L= (\pi,\pi,\pi)$; (b), $\Gamma=(0,0,0)$, $H=(0,0,4\pi)$, $N=(0,2\pi,2\pi)$ and $P=(2\pi,2\pi,2\pi)$. In (c), the coordinates are $\Gamma=(0,0,0)$, $X=(\pi,\pi/4,-\pi/4)$, $Z=(0,\pi,\pi)$, $Y=(\pi/2,\pi/2,-\pi/2)$, $S=(\pi,0,0)$, $A_{1}=(0,7\pi/4,\pi/4)$, $T=(\pi/2,3\pi/2,\pi/2)$, $A=(\pi,5\pi/4,3\pi/4)$, $R=(\pi,\pi,\pi)$, and $X_{1}=(0,3\pi/4,-3\pi/4)$. In (d), $\Gamma=(0,0,0)$, $X=(\pi/2,\pi/8,-\pi/8)$, $Z=(0,\pi,\pi)$, $Y=(\pi/4,\pi/4,-\pi/4)$, $S=(\pi/2,0,0)$, $A_{1}=(0,11\pi/8,5\pi/8)$, $T=(\pi/4,5\pi/4,3\pi/4)$, $A=(\pi/2,9\pi/8,7\pi/8)$, $R=(\pi/2,\pi,\pi)$, and $X_{1}=(0,3\pi/8,-3\pi/8)$. The family tree of spin liquids showing the interconnection between the parent and descendent states with different gauge groups for (e) class A, (f) class B, and (g) class C.}
	\label{fig:fig3}
\end{figure*}

\section{Short-range mean-field ans\"atze}
\label{sec:mean_field_amplitude}
We now proceed towards obtaining the mean-field \textit{Ans\"atze} of quantum spin liquids realizing the different PSGs employing the symmetry conditions in Appendix~\ref{app:sym_cond_mfa}. The different \textit{Ans\"atze} are characterized based on the gauge fluxes threading through elementary plaquettes. In our analysis, we restrict the mean-field amplitudes up to third nearest neighbor bonds which leads to a total of $17$ distinct \textit{Ans\"atze} (see Table~\ref{table:ansatz} and Appendix~\ref{app:ref_bond} for the definition of bonds). Furthermore, we perform self-consistent mean-field calculations for the $J_{1}$-$J_{2}$-$J_{3}$ Heisenberg Hamiltonian at a point in parameter space, in order to determine the spinon band structure and dynamical spin structure factors. We choose $(J_{2}/J_{1},J_{3}/J_{1})=(0.54,0.4)$, which is proximate to a classical triple point between different orders, since here strong quantum fluctuations for $S=1/2$ could potentially realize a nonmagnetic ground state. The self-consistently determined mean-field parameters and ground state energies are given in Table~\ref{table:self_consistent_values}.   

In the ensuing discussion, our choice of the labeling scheme for the \textit{Ans\"atze} is such that it reflects the information on the corresponding gauge inequivalent PSGs. We thus adopt the following labeling scheme for $SU(2)$ \textit{Ans\"atze}:
\begin{equation}
	\label{eq:su2_label}
	SA/B\eta_{S_4\bar{C}_3},
\end{equation}
where the $SU(2)$ classes $A$ and $B$ refer to the cases with $\eta_1=+1$ and $\eta_1=-1$, respectively. The case $\eta_1=-1$ corresponds to a scenario where the mean-field unit cell needed to accommodate the gauge flux structure is doubled along two fcc directions [$(r_2,r_3)$ in Fig.~\ref{fig:fig1}(a)] compared to the original geometrical unit cell. The label $\eta_{S_4\bar{C}_3}$ takes the values $0$, $1$, $2$ and $3$ for the combinations $(\eta_{S_4},\eta_{\bar{C}_3})=$ $(+1,+1)$, $(+1,-1)$, $(-1,+1)$ and $(-1,-1)$, respectively. 

The labeling scheme for $U(1)$ \textit{Ans\"atze} is
\begin{equation}
	\label{eq:u1_label}
	UA/B/Cw_{S_4\bar{C}_3}w_\mathcal{T},
\end{equation}
where the $U(1)$ classes $A$ and $B$ refer to the cases $\chi_1=0$ and $\chi_1=\pi$, respectively. Since, $\chi_1$ is related to $\eta_1$ as $\eta_1=g_3(\chi_1)$ for these two classes, the $A$ and $B$ labels for $U(1)$ \textit{Ans\"atze} have the same reference as for the $SU(2)$ case (this carries over also to the $\mathds{Z}_2$ \textit{Ans\"atze} discussed below). Furthermore, for the $U(1)$ PSGs, there appears an extra class $C$ referring to $\chi_1=\pi/2$ and corresponding to a quadrupling of the unit cell along two fcc directions [$(r_2,r_3)$ in Fig.~\ref{fig:fig1}(a)]. The label $w_{S_4\bar{C}_3}$ takes the values $0$, $1$, $2$ and $3$ for the combinations $(w_{S_4},w_{\bar{C}_3})=$ $(0,0)$, $(0,1)$, $(1,0)$, and $(1,1)$, respectively. 

Finally, the labelling scheme for the $\mathds{Z}_2$ states follows
\begin{equation}
	\label{eq:z2_label}
	ZA/B\eta_{S_4\bar{C}_3}\eta_\mathcal{T}g_{\bar{C}_3}(0),
\end{equation}
where akin to the $SU(2)$ case we shall use $0$, $1$, $2$, and $3$ to denote different combinations of parameters $\eta_{S_4},\eta_{\bar{C}_3}$. The $0$ and $1$ values denote positive and negative signs of $\eta_\mathcal{T}$. If $g_{\bar{C}_3}(0) \propto \tau^\alpha$, we shall write $\alpha$ in its place. We therefore divide the following discussion of the \textit{Ans\"atze} into three parts based on the classes A, B, and C.

\begin{figure*}
	\includegraphics[width=1.0\linewidth]{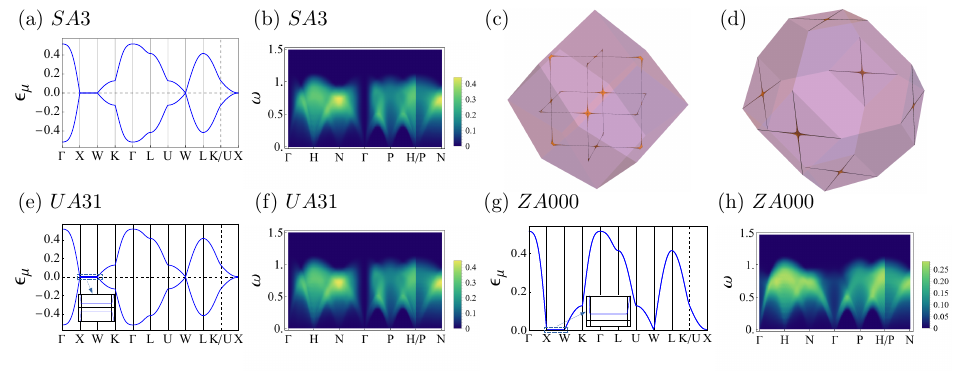}
	\caption{Spinon dispersion and dynamical spin structure factors plotted along the high-symmetry path for the states $SA3$ [(a), (b)], $UA31$ [(e), (f)] and $ZA000$ [(g), (h)]. Here, the dispersion spectrum is plotted in the first Brillouin zone [see Fig.~\ref{fig:fig3}(a)] while the dynamical structure factor is plotted in the extended Brillouin zone [see Fig.~\ref{fig:fig3}(b)]. Shown in (c) is the nodal loop in the extended Brillouin zone which appears as a nodal cross in the first Brillouin zone (d).}
	\label{fig:disp1}
\end{figure*}

\subsection{Class-A {\it{Ans\"atze}}}
\label{sec:class-a}
In this class, all the \textit{Ans\"atze} can be realized within the crystallographic two-site unit cell. The first Brillouin zone (FBZ) of the diamond lattice [see Fig.~\ref{fig:fig3}(a)], corresponding to the underlying fcc Bravais lattice, is a truncated octahedron. However, due to the presence of two sublattices in the diamond lattice,  the periodicity of the structure factors is dictated by an extended Brillouin zone (EBZ), a rhombic dodecahedron (the first Brillouin zone of the body-centered cubic lattice), which is depicted together with the high-symmetry points in Fig.~\ref{fig:fig3}(b).

\subsubsection{$SU(2)$}
\label{sec:class-a-su2}
We begin with the $SU(2)$ state given by the first row in Table~\ref{table:ansatz}, which belongs to Class-$A$. This \textit{Ansatz} corresponds to a uniform RVB state. The mean-field parameters vanish on 2NN bonds. Thus, $P_{t_0}(\phi_{t_0})$, $P_{t_1}(\phi_{t_1})$, $P_{3t_0}(\phi_{3t_0})$ and $P_{3t_1}(\phi_{3t_1})$ are not defined, and the $SA3$ state can be characterized by $(\phi_h,\phi_{3r})=(0,0)$ flux threading the loops associated with $(P_{h},P_{3_r})$. Hence, we refer to this uniform RVB state as simply $0$-flux \textit{Ansatz}. The spinon excitation spectrum is gapless and shown in Fig.~\ref{fig:disp1}(a) along the high symmetry path. It consists of two 2-fold degenerate bands given by
\begin{equation}
	\label{eq:su2_0_bands}	
 \epsilon_{\mathbf{k},\mu}=\pm|A_\mathbf{k}|
\end{equation}
where, $A_\mathbf{k}=h_1\sum_{i\in1NN} e^{\dot{\iota}\mathbf{k}\cdot\delta_i}+h_3\sum_{i\in3NN} e^{\dot{\iota}\mathbf{k}\cdot\delta_i}$, and $\delta_i$ denotes the position vectors of the 1NN sites. To fulfill the (mean-field) one-particle per-site constraint, the Fermi level, shown by a dashed line in Fig~\ref{fig:disp1}(a), is determined to be such that the lower-half energy eigenstates are filled (this is the case for all \textit{Ans\"atze} consisting of hopping only terms, i.e., the $SU(2)$ and $U(1)$ states, see Appendix~\ref{app:quadratic_ham}). In the $SU(2)$ $0$-flux state, the gapless $\mathbf{k}$ points are given by $|A_\mathbf{k}|=0$. This gives rise to fourfold degenerate nodal loops placed at the Fermi energy [shown in the EBZ in Fig.~\ref{fig:disp1}(c)] which in the FBZ appear as a nodal cross-like structure [see Fig.~\ref{fig:disp1}(d)]. We have established the robustness of such a nodal structure at the mean-field level upon inclusion of up to tenth nearest neighbor amplitudes. Thus, the presence of a fourfold degenerate nodal loop is not accidental, but rather protected by the projective symmetries of screw rotation and rotoinversion as demonstrated in Appendix~\ref{app:robust_nodal_loops}. 

It is worth noting that due to the explicit presence of real hopping terms, the \textit{Ansatz} is not in its canonical gauge form [Eq.~\eqref{eq:canonical_su2}]. However, by effecting a suitable gauge transformation it can be recast in the following canonical $SU(2)$ form:
\begin{equation}	\label{eq:sa3}
SA3\;:\;u_1=\dot{\iota} h_1\tau^0,\;u_2=0,\;u_3=\dot{\iota} h_3\tau^0,\;a_\alpha=0\, .
\end{equation}

\begin{figure*}
	\includegraphics[width=1.0\linewidth]{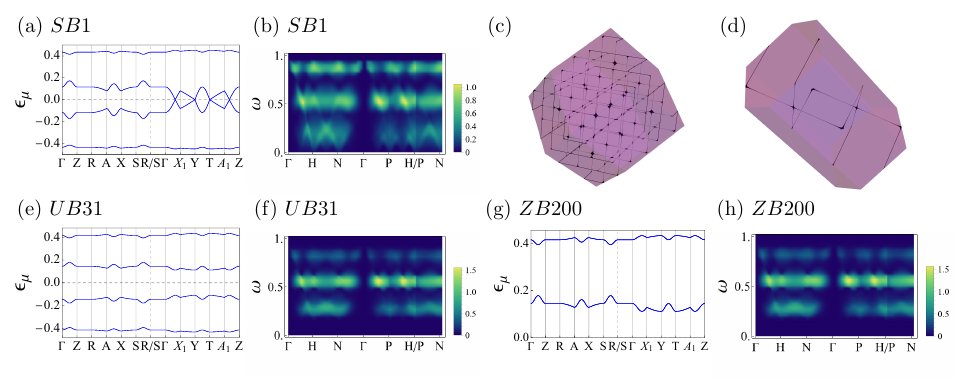}
	\caption{Spinon dispersion and dynamical spin structure factors plotted along the high-symmetry path for the states $SB1$ [(a) and (b)], $UB31$ [(e) and (f)], and $ZB200$ [(g) and (h)]. Here, the dispersion is plotted in the reduced Brillouin zone [Fig.~\ref{fig:fig3}(c), and the dynamical spin structure factor is plotted in the extended Brillouin zone [see Fig.~\ref{fig:fig3}(b)]. Shown in (c) are the nodal lines in the extended Brillouin zone and in (d) the nodal lines in the reduced Brillouin zone.}
	\label{fig:disp2}
\end{figure*}

\subsubsection{$U(1)$}
\label{sec:class-a-u1}

We now proceed towards identifying the symmetric $U(1)$ \textit{Ans\"atze} which lie in the neighborhood of the $0$-flux $SU(2)$ state. We find three such \textit{Ans\"atze} given by the rows $3$\textendash$5$ of Table~\ref{table:ansatz} and labeled $UA01$, $UA31$, and $UA30$, respectively. The presence of non vanishing 2NN mean-field amplitudes for the $UA01$ and $UA31$ Ans\"atze permits them to be characterized by the fluxes $(\phi_{h},\phi_{3r})$ and $(\phi_{t_0},\phi_{t_1})$.

For the $UA01$ \textit{Ansatz}, there is zero flux threading the loops corresponding to $(\phi_h,\phi_{3r})$, and also the fluxes threading the triangles are $(\phi_{t_0},\phi_{t_1})=(0,0)$. We thus refer to this state as $U(1)$ $(0,0)$-$(0,0)$-flux state. This state, in general, is comprised of gapless spinon excitation and the nodal manifold characteristic of the parent $SU(2)$ state is also present for this \textit{Ansatz}. Furthermore, the self-consistently determined mean-field parameters are such that this state effectively transforms into the parent zero-flux $SU(2)$ state and thus the spinon excitation spectrum is similar to the parent state.   
The $UA31$ state is characterized by the flux structure $(\phi_h,\phi_{3r})$-$(\phi_{t_0},\phi_{t_1})$=$(0,0)$-$(0,\pi)$, and thus we refer to this state as $U(1)$ $(0,0)$-$(0,\pi)$-flux state. The spinon excitation spectrum is shown in Fig.~\ref{fig:disp1}(e) where we observe that the nodal manifold of the parent $SU(2)$ state is no longer present and there appears a (tiny, but finite) gap in the excitation spectrum associated with the lowering of the IGG from $SU(2)$ to $U(1)$.

For $UA30$ \textit{Ansatz}, due to vanishing 2NN parameters, $P_{t_0}$ and $P_{t_1}$ do not exist. This state can be distinguished from the parent state by a nontrivial flux ($\phi_{3r}\neq0,\pi$) associated with $P_{3r}$. Thus we refer to this state as the $U(1)$ $(0,\phi)$-$(*,*)$ state. Here the notation ``$*$'' has been used to denote the non existence of the flux operators. Like $UA01$, this state, in general, features gapless excitation which retains the nodal manifold. Also, the self-consistently determined properties effectively transform into its parent $SU(2)$ state.

\subsubsection{$\mathds{Z}_2$}
\label{sec:class-a-z2}

We find that the $0$-flux $SU(2)$ state has three $\mathds{Z}_2$ descendants given by the rows $10$\textendash$12$ of Table~\ref{table:ansatz} which are labelled as $ZA000$, $ZA310$ and $ZA010$, respectively. However, these \textit{Ans\"atze} are not direct descendants of the $SU(2)$ state. Rather, the lowering of the IGG from $SU(2)$ to $\mathds{Z}_2$ takes place via $U(1)$ states [schematically shown in Fig.~\ref{fig:fig3}(e)]. However, the connections between these \textit{Ans\"atze} may not always be explicit, being masked by different gauge choices. For example, the connection of the $ZA000$ with the two $U(1)$ states given by $UA31$ and $UA30$ is not apparent from their structure given in Table~\ref{table:ansatz}. However, after a suitable gauge transformation, the $UA31$ and $UA30$ states can be rewritten as 
\begin{equation}\label{eq:ua31_pairing}
\left.\begin{aligned}
&UA31\;:\;u_1=h_1\tau^3,\;u_2=p_2\tau^1,\;u_3=h_3\tau^3,\;a_1\neq0.\\
&UA30\;:\;u_1=h_1\tau^3,\;u_2=0,\;u_3=h_3\tau^3+p_3\tau^1,\;a_\mu=0.\\
\end{aligned}\right.
\end{equation}
whereby it becomes manifest that the inclusion of second nearest neighbor amplitudes\textendash a hopping term $\propto\tau^{3}$ (for the $UA31$ \textit{Ansatz}), while a concurrently occurring pairing $\propto\tau^{1}$ and hopping term $\propto\tau^{3}$ (for the $UA30$ \textit{Ansatz)}\textendash are responsible for lowering the IGG from $U(1)$ to $\mathds{Z}_{2}$.

The spinon excitation spectrum of the $ZA000$ state is shown in Fig.~\ref{fig:disp1}(g), where we plot the positive energies of the Bogoliubov quasiparticles (see Appendix~\ref{app:quadratic_ham}), which implies that the Fermi level lies at zero energy. For the remainder of the paper, this plotting scheme is applied to all spinon dispersions of $\mathds{Z}_{2}$ states. The excitation spectrum of $ZA000$ displays a small gap. Notice from the self-consistently determined mean-field parameters [row 5 of Table~\ref{table:self_consistent_values}] that $h_2,h_3,p_3\approx0$. With this setting, the state essentially takes the form of its parent $U(1)$ state $UA31$ [see Eq.~\eqref{eq:ua31_pairing}], which has the lowest energy. A similar conclusion can be made for the state $ZA010$ which is the other descendent of $UA31$. On the other hand, the $ZA310$ \textit{Ansatz} descending from the $UA01$ state develops a vanishingly small (though finite) pairing terms (see row 6 of Table~\ref{table:self_consistent_values}). This leads to the appearance of a finite gap in the spectrum, which then appears similar to Fig.~\ref{fig:disp1}(g). Generally speaking, all the $\mathds{Z}_2$ states discussed above can potentially develop a spinon gap, whose magnitude is determined by the optimal pairing obtained for a given Hamiltonian. In particular, for the $ZA000$ and $ZA010$ \textit{Ans\"atze}, the gapped nature of the excitations within our mean-field calculations is expected, since they descend from the gapped parent $U(1)$ $UA31$ \textit{Ansatz}.

\subsection{Class-B {\it{Ans\"atze}}}
\label{sec:class-b}
For the \textit{Ans\"atze} belonging to this class,  a doubling of the unit cell along two fcc directions [$\mathbf{e}_2$ and $\mathbf{e}_3$] is needed in order to accommodate the gauge magnetic flux. Hence, the resulting unit cell is four times larger compared to the crystallographic unit cell. The corresponding reduced Brillouin zone is shown in Fig.~\ref{fig:fig3}(c) together with a labeling of high-symmetry points.

\subsubsection{$SU(2)$}
Similar to class-$A$, we first discuss the parent $SU(2)$ state in this class. This \textit{Ansatz} is labeled as $SB1$ and given by row 2 of Table~\ref{table:ansatz}. Since the mean-field parameters vanish on 2NN and 3NN bonds, we characterize $SB1$ by $\phi_h=\pi$ flux threading hexagonal loops, in contrast to the $\phi_h=0$ flux for the corresponding $SU(2)$ \textit{Ansatz} of class-$A$. We thus refer to this state as the $\pi$-flux \textit{Ansatz}, and it is characterized by gapless excitation as shown in Fig.~\ref{fig:disp2}(a). Similar to the zero flux state, there is a multi-nodal band structure present for this $\pi$-flux \textit{Ansatz}, shown in the extended and reduced Brillouin zone in Figs.~\ref{fig:disp2}(c) and \ref{fig:disp2}(d), respectively, for the case of first neighbor hoppings. However, in contrast to the zero flux state, here the nodal structure is not a robust characteristic, but its appearance is rather an artefact of considering only short-range interactions. Indeed, a finite gap opens if we add, e.g., fifth nearest neighbor amplitudes (see Appendix~\ref{app:robust_nodal_loops}).

\subsubsection{$U(1)$}
We obtain three $U(1)$ descendants of the $\pi$-flux $SU(2)$ state. In Table~\ref{table:ansatz}, these are given by rows 6-8 and labeled as $UB01$, $UB31$, and $UB20$, respectively. For the first two \textit{Ans\"atze}, the mean-field parameters on the 3NN bonds are not allowed by symmetry. We thus characterize their flux structure as $(\phi_h,\phi_{3r})$-$(\phi_{t0},\phi_{t1})=$ $(\pi,*)$-$(0,0)$ and $(\pi,*)$-$(0,\pi)$, respectively. The state $UB01$ transforms into the parent $SU(2)$ state under self-consistent treatment. On the contrary, for the $UB31$ state, a finite mean-field amplitude develops on the 2NN bonds. This results in a gapped spinon excitation spectrum [see Fig.~\ref{fig:disp2}(e)]. Notice that in this case, the addition of 2NN interaction lowers the ground state energy noticeably (see row 10 of Table~\ref{table:self_consistent_values}).

For $UB20$ Ansatz, the mean-field parameters vanish on 2NN bonds, i.e., $u_2=0$. It can thus be characterized via a flux $\pi$ and $\pi/2$, on the loops $P_{h}$ and $P_{3r}$, respectively. We thus refer to this state as the $U(1)$ $(\pi,\pi/2)$-$(*,*)$ flux state. Within a self-consistent treatment, the behavior of this state is similar to $UB01$.

\subsubsection{$\mathds{Z}_2$}
We now discuss the five $\mathds{Z}_2$ \textit{Ans\"atze} which are continuously connected to the aforementioned $U(1)$ states. The corresponding $\mathds{Z}_{2}$ PSGs are listed in rows 13-17 of Table~\ref{table:ansatz}. Among these, the $ZB200$ and $ZB110$ both descend from the parent $UB01$ state. Interestingly, $ZB200$ has another parent $U(1)$ state, namely, the $UB31$ \textit{Ansatz}. Other descendants of $UB31$ \textit{Ansatz} are $ZB210$ and $ZB303$. The remaining two $ZB303$ and $ZB203$ are continuously connected to the $UB20$ \textit{Ansatz}. Notice that $ZB303$ is the common descendent of both $UB31$ and $UB20$. This hierarchy and interconnections among \textit{Ans\"atze} is schematically illustrated in Fig.~\ref{fig:fig3}(f).

Similar to the class-A $\mathds{Z}_2$ \textit{Ans\"atze}, all these states, in general, exhibit gapped excitations. For the states $ZB200$ and $ZB210$, their parent $U(1)$ state $UB31$ also features gapped excitations. Notice that the lowering of the IGG in the other three $\mathds{Z}_2$ states is accompanied by an opening of a gap. Interestingly, the solutions within a saddle point approximation are such that $ZB110$ and $ZB203$ flow back to the $\pi$-flux $SU(2)$ state, similar to their parent $U(1)$ states $UB01$ and $UB20$, respectively. The other $\mathds{Z}_2$ states effectively go to their lowest energy $U(1)$ parent $UB31$. Their spectrum is illustrated in Fig.~\ref{fig:disp2}(g).

\begin{figure}
	\includegraphics[width=1.0\columnwidth]{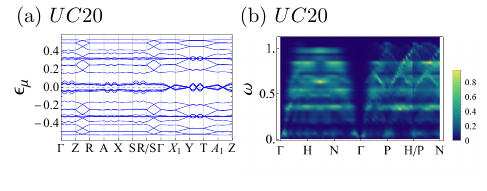}
	\caption{Spinon dispersion and dynamical spin structure factors plotted along the high-symmetry path for the state $UC20$ [(a) and (b)]. Here, the dispersion is plotted in the reduced Brillouin zone Fig.~\ref{fig:fig3}(d), and the dynamical spin structure factor is plotted in the extended Brillouin zone [see Fig.~\ref{fig:fig3}(b)].}
	\label{fig:disp3}
\end{figure}

\subsection{Class-C {\it{Ans\"atze}}}
\label{sec:class-c}
The gauge magnetic flux for an \textit{Ansatz} belonging to this class can be accommodated only if one quadruples the unit cell along two fcc directions [$\mathbf{e}_2$ and $\mathbf{e}_3$]. The resulting unit cell is thus sixteen times enlarged compared to the crystallographic unit cell. The corresponding reduced Brillouin zone is shown in Fig.~\ref{fig:fig3}(d) together with the high-symmetry points.

There is only one member belonging to this class given by row 9 of Table~\ref{table:ansatz} and labeled as $UC20$. Interestingly, this state appears directly with $U(1)$ IGG and has no parent $SU(2)$ state or descendent $\mathds{Z}_2$ \textit{Ans\"atze}. It is characterized by a $\phi_h=\pi/2$ flux threading the hexagonal plaquettes and hence, we refer to it as the $U(1)$ $\pi/2$-flux state. This \textit{Ansatz} does not allow for mean-field amplitudes on 2NN and 3NN bonds, and thus no other flux operator can be defined. Its spectrum is gapless at isolated $\mathbf{k}$-points shown in Fig.~\ref{fig:disp3}(a) with the associated dynamical spin structure factor shown in Fig.~\ref{fig:disp3}(b).

\vspace{0.2cm}
Within a self-consistent treatment, it is seen from Table~\ref{table:self_consistent_values}, that the class-A $U(1)$ \textit{Ansatz} labeled by $UA31$ represents the minimum energy QSL state.

We have seen in Sec.~\ref{sec:classification} that as a consequence of the appearance of sublattice-dependent projective realization of time-reversal symmetry corresponding to $w_\mathcal{T}=0$, there exist three symmetric $U(1)$ \textit{Ans\"atze} labeled by $UA30$, $UB20$ and $UC20$ (considering up to third nearest neighbor amplitudes). This is also the case for the two $SU(2)$ \textit{Ans\"atze} labeled by $SA3$ and $SB1$.

In general, a sublattice-dependent time-reversal PSG solution with $w_{\mathcal{T}}=0$ lead to \textit{Ans\"atze} which feature nonvanishing mean-field amplitudes only on bonds connecting different sublattices. This suggests the projective symmetry associated with time-reversal can be identified with a chiral symmetry (i.e., inter sublattice symmetry).  

This is a generic phenomenon for bipartite lattices, and has been observed in honeycomb~\cite{Lu-2011b,You-2012}, square~\cite{Wen-2002}, square-octagon~\cite{Maity-2022}, hyperhoneycomb~\cite{Huang-2018}, and simple cubic~\cite{Sonnenschein-2020} lattices, to name a few.
 
\section{Realization of spinonic topological insulator in diamond lattice}
\label{sec:topological_insulator}
In this section, we will address a non trivial aspect of the parton mean-field theory. We will show that one of the PSG-classified quadratic parton Hamiltonian yields the Fu-Kane-Mele model for a three-dimensional topological insulator on the diamond lattice~\cite{PhysRevLett.98.106803}. The Fu-Kane-Mele model is written as
\begin{equation}\label{FKM_model}
\hat{H} = t\sum_{\langle ij\rangle} \hat{c}^\dag_i \hat{c}_j
+ \dot{\iota}\lambda_{\text{SO}}\sum_{\langle\langle ij\rangle\rangle}
\hat{c}^\dag_i \mathbf{s}\cdot (\mathbf{d}_{ij}^1\times \mathbf{d}_{ij}^2) \hat{c}_j,
\end{equation}
where $\mathbf{d}_{ij}^1$ and $\mathbf{d}_{ij}^2$  are the two norm-one nearest neighbor
bond vectors traversed between a pair of second nearest neighbor sites $i$ and $j$.

To see this, we recall the quadratic decomposition of the generic fermionic quartic Hamiltonian. In Sec.~\ref{sec:afmft}, we consider a decomposition in terms of the spin rotation invariant or singlet field $\hat{U}_{ij}=\hat{\psi}^\dagger_i\hat{\psi}_j$. Another possible Hubbard–Stratonovich decomposition can be performed in terms of the fields $\hat{U}^{(\alpha)}_{ij}=\hat{\psi}^\dagger_i\tau^\alpha\hat{\psi}_j$ with $\alpha=\{x,y,z\}$ which are not invariant under spin rotation, thus, representing triplet fields. In two dimensions, it is well known that including these triplet fields can lead to a plethora of topologically non trivial spinon models and possibly spin nematic states \cite{Reuther-2014, Sonnenschein-2017, Shindou-2009,Dodds-2013}. In a mean-field setting, i.e., $u^{(\alpha)}_{ij}=\langle\hat{U}^{(\alpha)}\rangle$, the combined singlet-triplet quadratic parton Hamiltonian can be expressed as
\begin{align}\label{eq:mf_ham_singlet_triplet}
    \hat{H}^{s+\text{t}}_{MF} = & \sum_{\langle ij \rangle} \text{Tr}\left[\hat{\psi}_{i} u_{ij} \hat{\psi}^{\dagger}_{j}+\tau^\alpha\hat{\psi}_iu^{(\alpha)}_{ij}\hat{\psi}^\dagger_j + \text{H.c.} \right] \; \notag \\
    & + \sum_{i} a_{\gamma} \text{Tr}[\hat{\psi}_{i} \tau^{\gamma} \hat{\psi}^{\dagger}_{i} ].\;
\end{align}
Similar to Eq.~\eqref{eq:link_singlet}, $u^{(\alpha)}_{ij}$ can be expressed in terms of complex parameters $h^\alpha_{ij}$ and $p^\alpha_{ij}$:
\begin{align}\label{eq:triplet_fields}
    u^{(x)}_{ij} = & \mathrm{Re}h^x_{ij}\tau^0+\dot{\iota}(\mathrm{Im}h^x_{ij}\tau^3+\mathrm{Re}p^x_{ij}\tau^1+\mathrm{Im}p^x_{ij}\tau^2) \;, \notag \\
    u^{(y)}_{ij} = & \mathrm{Re}h^y_{ij}\tau^0+\dot{\iota}(\mathrm{Im}h^y_{ij}\tau^3+\mathrm{Re}p^y_{ij}\tau^1+\mathrm{Im}p^y_{ij}\tau^2) \;, \notag \\
    u^{(z)}_{ij} = & \mathrm{Re}h^z_{ij}\tau^0+\dot{\iota}(\mathrm{Im}h^z_{ij}\tau^3+\mathrm{Re}p^z_{ij}\tau^1+\mathrm{Im}p^z_{ij}\tau^2) \;.
\end{align}
The definition of these parameters for 1NN, 2NN and 3NN bonds along with the associated constraints within the $SU(2)$, $U(1)$, and $\mathds{Z}_2$ PSG classes is given in Appendix~\ref{app:singlet_trplet}. Note that the non projective $U(1)$ class (the first class that appears in Table \ref{tab:u1_triplet_singlet} in Appendix~\ref{app:singlet_trplet}) contains the Fu-Kane-Mele model~\eqref{FKM_model}, 
with the corresponding values for the onsite term $h=0$, nearest neighbor bond $\mathrm{Re}h_1 = t$, and second nearest neighbor bonds $\mathrm{Im}h^y_2 = -\mathrm{Im}h^z_2 = -\frac{2}{3}\lambda_{\text{SO}}$ and $\mathrm{Re}h_2=0$.

\section{Gutzwiller Projection of flux phases}\label{sec:projection}

\begin{figure}[t]
\includegraphics[width=\linewidth]{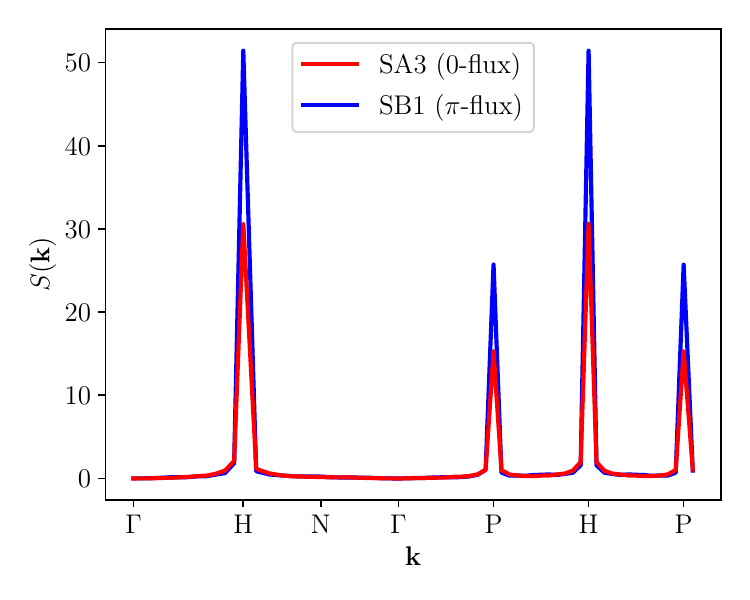}
\caption{Static spin structure factor $S(\mathbf{k})$ along a high-symmetry path in the extended Brillouin zone. The results for the Gutzwiller-projected $SU(2)$ flux states, $SA3$ and $SB1$, are shown. Note that, although $S(\mathbf{k})$ is consistently larger for $SA3$ than $SB1$ along this path, the integral over all momenta $\mathbf{k}$ in the extended Brillouin zone fulfills the sum rule for both states ($\sum_\mathbf{k} S(\mathbf{k})=4 S(S+1) L^3=3L^3$, where the factor $4$ comes from the ratio between the areas of the extended and first Brillouin zones).}
\label{fig:sq_path}
\end{figure}

\begin{figure}[t]
\includegraphics[width=\linewidth]{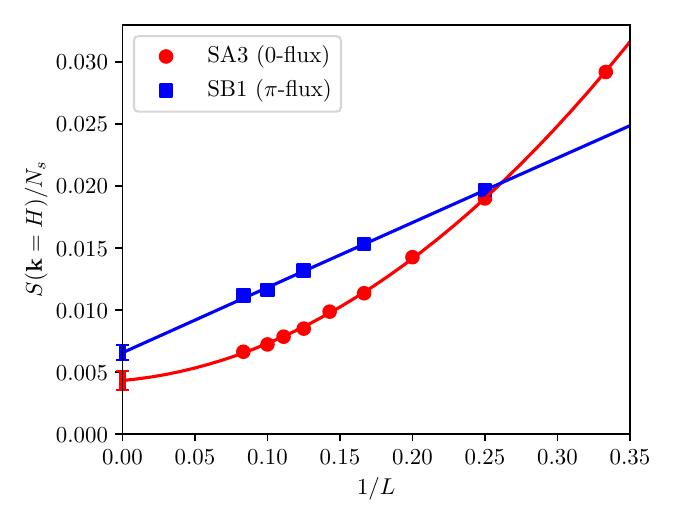}
\caption{Finite-size scaling of the static spin structure factor $S(\mathbf{k})$  at the momentum $\mathbf{k}=H=(0,0,4\pi)$. The results for the Gutzwiller-projected $SU(2)$ flux states, $SA3$ and $SB1$, are shown and indicate a finite N\'eel magnetic order in the thermodynamic limit.}
\label{fig:sq_scaling}
\end{figure}

As previously mentioned, starting from the fermionic ground state $|\Psi_{MF}\rangle$ of the mean-field Hamiltonian~\eqref{eq:ansatz}, a \textit{bonafide} wave function for spins $|\Psi\rangle$ can be obtained by enforcing the constraints of Eq.~\eqref{constraint} exactly (i.e., not on average). The spin state is defined as $|\Psi\rangle=\hat{\mathcal{P}}_G|\Psi_{MF}\rangle$, where the operator $\hat{\mathcal{P}}_G=\prod_i \hat{n}_i(2-\hat{n}_i)$ is the Gutzwiller projector (with $\hat{n}_i=\sum_\alpha \hat{f}_{i\alpha}^{\dagger}\hat{f}_{i\alpha}$), which is analytically intractable, in general, but can be numerically handled by means of a suitable Monte Carlo sampling of the one-fermion-per-site Hilbert subspace. 

 Here, we compute the static spin structure factor 
\begin{equation}
S(\mathbf{k})=\frac{1}{N_s}\sum_{i,\,j} e^{\dot{\iota} \mathbf{k}\cdot (\mathbf{r}_i-\mathbf{r}_j)} \langle \Psi| \hat{\mathbf{S}}_i \cdot \hat{\mathbf{S}}_j |\Psi\rangle,
\end{equation}
for the Gutzwiller-projected $SU(2)$ flux states, $SA3$ and $SB1$. The calculations are performed on finite clusters of $N_s=2(L\times L \times L)$ lattice sites, whose position are denoted by the vectors $\mathbf{r}_i$. In Fig.~\ref{fig:sq_path}, we show the $S(\mathbf{k})$ profile along high-symmetry lines of the extended Brillouin zone [see Fig.~\ref{fig:fig3}(b)], which is dominated by the sharp maxima located at the $H=(0,0,4\pi)$ points, and by subleading peaks at the $P=(2\pi,2\pi,2\pi)$ points. The location of these Bragg peaks corresponds to that expected from the two-sublattice N\'eel order. The finite-size scaling analysis in Fig.~\ref{fig:sq_scaling} shows that $S(\mathbf{k}=H)/N_s=m^2$ extrapolates to a finite value in the thermodynamic limit, indicating that both $SA3$ and $SB1$ exhibit a finite (antiferro)magnetic order after Gutzwiller-projection, with order parameter ${m=0.0656(6)}$ and ${m=0.0812(4)}$, respectively. The value of $m$ for the $\pi$-flux state is similar to that obtained in an earlier work~\cite{Albuquerque-2012} employing large-scale quantum Monte Carlo simulations of the NN-Valence Bond wave function on the diamond lattice. We note that since the Fermi surface in the unprojected wave function is nested, the development of long-range magnetic order may be viewed as arising from its magnetic instability once correlations are incorporated via the Gutzwiller projector~\cite{Zhang-2011}. It is therefore plausible that a similar scenario may occur on other 3D bipartite lattices, such as the body centered cubic lattice. On this lattice, the Fermi surface of the uniform 0-flux fermionic \textit{Ansatz} consists of (almost) parallel planes forming
a cube in momentum space~\cite{Sonnenschein-2020}. The two opposite planes are connected by a nesting vector $q=(2\pi,0,0)$, which is precisely the ordering wave vector of the two sublattice N\'eel order on the body centered cubic lattice~\cite{Ghosh-2019}. A similar situation has been observed on bipartite lattices in 2D, e.g., on the square lattice, where the projection of the $0$-flux Fermi sea leads to a wave function displaying finite N\'eel order~\cite{Zhang-2011,Li_2013,Grover-2013}.

\section{Conclusions and Outlook}
\label{sec:discussion}
We performed a projective symmetry group classification of fermionic mean-field \textit{Ans\"atze} of fully symmetric $S=1/2$ quantum spin liquids on the diamond lattice obtaining 8 $SU(2)$, 62 $U(1)$ and 80 $\mathds{Z}_2$ algebraic PSGs. Upon restricting the \textit{Ans\"atze} to third nearest neighbor mean-field amplitudes, we find that only $2$ $SU(2)$, $7$ $U(1)$, and $8$ $\mathds{Z}_{2}$ distinct PSGs can be realized. The symmetry allowed singlet and triplet fields of the \textit{Ans\"atze} are obtained, with the latter being of relevance to spin-orbit coupled systems. We performed a self-consistent mean-field analysis at a point of high frustration in the $J_1$-$J_2$-$J_3$ Heisenberg antiferromagnet and presented the spinon band structures and dynamical spin structure factors for the different \textit{Ans\"atze}. The $SU(2)$ $0$-flux state is shown to host three fourfold degenerate nodal loops which are robust at the mean-field level, being protected by the projective realization of rotoinversion and screw symmetries. These stable nodal loops are also present in two of its $U(1)$ and one $\mathds{Z}_2$ descendant spin liquids. We reveal a nontrivial feature of our parton mean-field theory by showing that one of the PSG-classified quadratic spinon Hamiltonians, namely, the nonprojective $U(1)$ \textit{Ansatz}, yields the Fu-Kane-Mele model for a three-dimensional topological insulator on the diamond lattice. Finally, we investigate the effects of fluctuations beyond mean-field via Gutzwiller projection, and show that the projected 0-flux and $\pi$-flux $SU(2)$ \textit{Ans\"atze} exhibit long-range N\'eel magnetic order.

Our work sets the stage for future studies aimed at investigating the energetic competition of the corresponding Gutzwiller projected variational wave functions for the $S=1/2$ frustrated $J_{1}$-$J_{2}$ Heisenberg model on the diamond lattice. In particular, given the conflicting claims concerning the presence~\cite{Oitmaa-2019,Oitmaa_2018} or absence~\cite{Buessen-2018} of long-range magnetic order in this model, it will be of interest to assess the competition of these spin liquids with magnetically ordered states. In the scenario that the ground state is nonmagnetic, a further exploration of the static structure factors of these Gutzwiller projected states to identify signatures of approximate spiral surfaces with weak modulations due to quantum fluctuations~\cite{Bernier-2008,Buessen-2018} would constitute a worthwhile endeavor. These \textit{Ans\"atze} also serve as starting point for explorations of valence bond crystal ground states which can be viewed as dimerization instabilities of these quantum spin liquids, and the energetic viability of such symmetry broken dimer orders as ground states for the $J_{1}$-$J_{2}$ model remains to be explored.  

\section*{Acknowledgements}
Y.I. thanks Leon Balents, Ganapathy Baskaran, Subhro Bhattacharjee, Lasse Gresista, Tarun Grover, Sid Parameswaran, Johannes Reuther, Ronny Thomale, and Simon Trebst for valuable discussions. Y.I. acknowledges financial support by the Science and Engineering Research Board (SERB), DST, India through the MATRICS Grant No.~MTR/2019/001042, and the Indo-French Centre for the Promotion of Advanced Research (CEFIPRA) Project No. 64T3-1. The research of Y.I. was supported in part by the National Science Foundation under Grant No.~NSF~PHY-1748958 during a visit to the Kavli Institute for Theoretical Physics (KITP), UC Santa Barbara, USA for participating in the program ``A Quantum Universe in a Crystal: Symmetry and Topology across the Correlation Spectrum,'' ICTP through the Associates Programme and from the Simons Foundation through grant number 284558FY19, IIT Madras through the Institute of Eminence (IoE) program for establishing the QuCenDiEM CoE (Project No. SP22231244CPETWOQCDHOC), the International Centre for Theoretical Sciences (ICTS), Bengaluru, India during a visit for participating in the program “Frustrated Metals and Insulators” (Code: ICTS/frumi2022/9). The work of Y. I. was performed in part and completed at the Aspen Center for Physics, which is supported by National Science Foundation grant PHY-2210452. The participation of Y. I. at the Aspen Center for Physics was supported by the Simons Foundation. Y.I. acknowledges the use of the computing resources at HPCE, IIT Madras. F. F. acknowledges financial support from the Deutsche Forschungsgemeinschaft (DFG, German Research Foundation) for funding through TRR 288 -- 422213477 (project A05). F.\,F. thanks IIT Madras for a visiting Postdoctoral fellowship position under the IoE program which facilitated the completion of this research work. C.L. was supported by the EPiQS program of the Gordon and Betty Moore Foundation. J.S. received financial support from the Theory of Quantum Matter Unit of the Okinawa Institute of Science and Technology Graduate University (OIST). 

\appendix

\section{Symmetries of the diamond lattice}
\label{app:symmetry_properties}
\begin{figure}[b]
\includegraphics[width=0.75\linewidth]{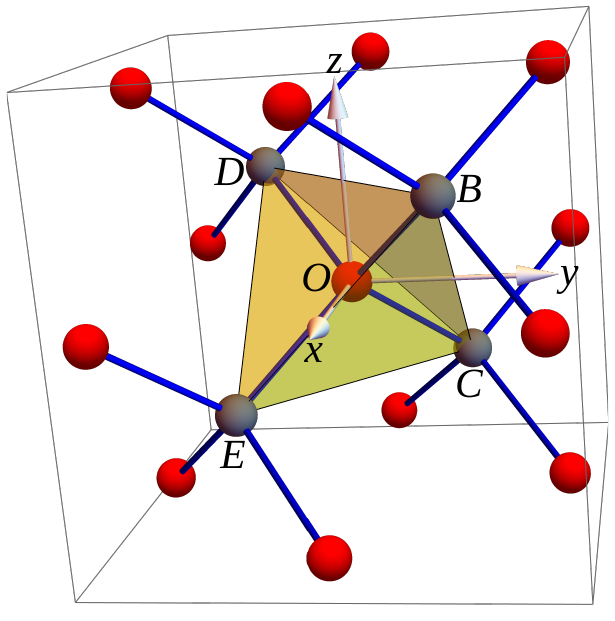}
\caption{One of the tetrahedrons with origin ``O'' as the center of the tetrahedron and the four vertices are denoted by ``B'', ``C'', ``D'', and ``A'' respectively.}
\label{fig:tetrahedron}
\end{figure}
\subsection{Space group}
\label{app:space_group}
   The space group of the diamond lattice is given by $Fd\bar{3}m$. This space group is comprised of the $24$ elements constituting the tetrahedral point group (T$_d$) and another $24$ nonsymmorphic elements. To define the action of the tetrahedral point group elements, we fix a tetrahedron with its centre at the origin of the Cartesian coordinate system $(x,y,z)$ [see Fig.~\ref{fig:tetrahedron}]. The different group elements and their action on the coordinates $(x,y,z)$ is then given by\\
 
      (1) The identity $\mathds{1}$.\\
      
      (2) $8C_3$: The symmetry elements $C_{31}$, $C_{32}$, $C_{33}$ and $C_{34}$ correspond to the four three fold rotation axes passing through the origin `O' and the centre of the triangular faces ECD, ECB, BCD and EBD, respectively. Their actions on a point $(x,y,z)$ are given by
       \begin{equation}
        \left.\begin{aligned}
        &C_{31}:(x,y,z)\rightarrow(z,x,y),\\
        &C_{32}:(x,y,z)\rightarrow(-z,x,-y),\\
        &C_{33}:(x,y,z)\rightarrow(-z,-x,y),\\
        &C_{34}:(x,y,z)\rightarrow(z,-x,-y).\\
        \end{aligned}\right.
        \end{equation}
        These four symmetry elements generate 8 symmetry operations: $C_{31}$, $C_{32}$, $C_{33}$, $C_{34}$, $C^2_{31}$, $C^2_{32}$, $C^2_{33}$, and $C^2_{34}$.\\

     (3) $6\sigma_d$: There are six elements that correspond to mirror reflections about planes bisecting the triangles OBD, OBC, OCD, OBE, OCE, and ODE, respectively. We thus have six distinct symmetry operations.
    Their action on coordinates $(x,y,z)$ is given by 
    
    \begin{equation}
        \left.\begin{aligned}
        &\sigma_{d1}:(x,y,z)\rightarrow(y,x,z),\\
        &\sigma_{d2}:(x,y,z)\rightarrow(-y,-x,z),\\
        &\sigma_{d3}:(x,y,z)\rightarrow(x,z,y),\\
        &\sigma_{d4}:(x,y,z)\rightarrow(x,-z,-y),\\
        &\sigma_{d5}:(x,y,z)\rightarrow(-z,y,-x),\\
        &\sigma_{d6}:(x,y,z)\rightarrow(z,y,x).\\
        \end{aligned}\right.
        \end{equation}\\

    (4) $3C_2$: We have three two fold rotation axes which pass through the origin and coincide with the Cartesian $x$, $y$, and $z$ axes. These three rotation axes generate three symmetry operations with the transformation of coordinates given by
        \begin{equation}
        \left.\begin{aligned}
        &C_{21}:(x,y,z)\rightarrow(-x,-y,z),\\
        &C_{22}:(x,y,z)\rightarrow(x,-y,-z),\\
        &C_{23}:(x,y,z)\rightarrow(-x,y,-z).\\
        \end{aligned}\right.
        \end{equation}\\

    (5) $6C'_4$: There are three improper rotation axes. These correspond to $C_4$ rotations about the Cartesian $x$, $y$, and $z$ axes, followed by reflections about planes perpendicular to the axes and passing through the origin. They act as
        \begin{equation}
        \left.\begin{aligned}
        &C'_{41}:(x,y,z)\rightarrow(-y,x,-z),\\
        &C'_{42}:(x,y,z)\rightarrow(-x,-z,y),\\
        &C'_{43}:(x,y,z)\rightarrow(z,-y,-x).\\
        \end{aligned}\right.
        \end{equation}    
    These three elements generate six symmetry operations $C'_{41},C'_{42},C'_{43},C'^3_{41},C'^3_{42},C'^3_{43}$. We note that the operations $C'^2_{41},C'^2_{42},C'^2_{43}$ are the same as $C_{21},C_{22},C_{23}$.\\

    We now proceed to discuss the remaining $24$ nonsymmorphic elements and their action on coordinates. We distinguish them from the symmorphic ones by labeling a generic element $O$ with a bar, as $\bar{O}$.\\ 

    (1) $\bar{I}$: Inversion with respect to the point $(1/8,1/8,1/8)$. The transformation is given by 
    \begin{equation}
        \bar{I}:(x,y,z)\rightarrow(1/4-x,1/4-y,1/4-z).
    \end{equation}
    This can be viewed as an inversion about the origin $O(0,0,0)$ followed by a translation $(1/4,1/4,1/4)$.\\

    (2) $6\bar{S}_4$: There exist six screw symmetry operations which transform the Cartesian coordinates as
    \begin{equation}
    \left.\begin{aligned}
    &\bar{S}_{41}:(x,y,z)\rightarrow(1/4+z,1/4+y,1/4-x),\\
    &\bar{S}_{42}:(x,y,z)\rightarrow(1/4-y,1/4+x,1/4+z),\\
    &\bar{S}_{43}:(x,y,z)\rightarrow(1/4+x,1/4-z,1/4+y),\\
    &\bar{S}_{44}:(x,y,z)\rightarrow(1/4-z,1/4+y,1/4+x),\\
    &\bar{S}_{45}:(x,y,z)\rightarrow(1/4+y,1/4-x,1/4+z),\\
    &\bar{S}_{46}:(x,y,z)\rightarrow(1/4+x,1/4+z,1/4-y).\\
    \end{aligned}\right.
    \end{equation}
    Here, $\bar{S}_{41}$, $\bar{S}_{42}$, and $\bar{S}_{43}$ correspond to a $C_4$ clockwise rotation about the $y$-, $z$- and $x$-axes anchored at $(1/4,0,0)$, $(0,1/4,0)$, and $(0,0,1/4)$ points followed by a translation $(0,1/4,0)$, $(0,0,1/4)$, and $(1/4,0,0)$, respectively. The remaining $\bar{S}_{44}$, $\bar{S}_{45}$ and $\bar{S}_{46}$ correspond to $C_4$ counter clockwise rotations about the $y$, $z$, and $x$ axes centered at the $(0,0,1/4)$, $(1/4,0,0)$, and $(1/4,0,0)$ points followed by a translation given by $(0,1/4,0)$, $(0,0,1/4)$, and $(1/4,0,0)$, respectively.\\

    (3) $6\bar{C}_2$: There are six nonsymmorphic $C_2$ operations whose actions are given by
    \begin{equation}
    \left.\begin{aligned}
    &\bar{C}_{21}:(x,y,z)\rightarrow(1/4-x,1/4-z,1/4-y),\\
    &\bar{C}_{22}:(x,y,z)\rightarrow(1/4-z,1/4-y,1/4-x),\\
    &\bar{C}_{23}:(x,y,z)\rightarrow(1/4-y,1/4-x,1/4-z),\\
    &\bar{C}_{24}:(x,y,z)\rightarrow(1/4-x,1/4+z,1/4+y),\\
    &\bar{C}_{25}:(x,y,z)\rightarrow(1/4+z,1/4-y,1/4+x),\\
    &\bar{C}_{26}:(x,y,z)\rightarrow(1/4+y,1/4+x,1/4-z).\\
    \end{aligned}\right.
    \end{equation}
    The operations $\bar{C}_{21}$, $\bar{C}_{22}$, and $\bar{C}_{23}$ are composed of a $C_2$ rotation about the $[0,-1,1]$, $[-1,0,1]$, and $[1,-1,0]$ axes anchored at $(1/8,1/4,0)$, $(1/4,1/8,0)$, and $(0,1/4,1/8)$ points, respectively. The remaining three operations $\bar{C}_{24}$, $\bar{C}_{25}$, and $\bar{C}_{26}$ are composed of a $C_2$ rotation about $[0,1,1]$, $[1,0,1]$, and $[1,1,0]$ axes centered at $(1/8,0,0)$, $(0,1/8,0)$, and $(0,0,1/8)$ points followed by a translation $(0,1/4,1/4)$, $(1/4,0,1/4)$, and $(1/4,1/4,0)$, respectively.\\
    
    (4) $3\bar{\sigma}_{d}$: The three nonsymmorphic operations $\bar{\sigma}_{d1}$, $\bar{\sigma}_{d2}$, and $\bar{\sigma}_{d3}$ are given by reflections about the planes $y$-$z$, $x$-$z$, and $x$-$y$ passing through the points $(1/8,0,0)$, $(0,1/8,0)$, and $(0,0,1/8)$ followed by translations $(0,1/4,1/4)$, $(1/4,0,1/4)$, and $(1/4,1/4,0)$, respectively. 
    \begin{equation}
    \left.\begin{aligned}
    &\bar{\sigma}_{d1}:(x,y,z)\rightarrow(1/4-x,1/4+y,1/4+z),\\
    &\bar{\sigma}_{d2}:(x,y,z)\rightarrow(1/4+x,1/4-y,1/4+z),\\
    &\bar{\sigma}_{d3}:(x,y,z)\rightarrow(1/4+x,1/4+y,1/4-z).\\
    \end{aligned}\right.
    \end{equation}\\

    (5) $8\bar{C}_3$: There are eight rotoinversion symmetry elements composed of $C_3$ operation and an inversion, with their action on the coordinates given by
    \begin{equation}
    \left.\begin{aligned}
    &\bar{C}_{31}:(x,y,z)\rightarrow(1/4-z,1/4-x,1/4-y),\\
    &\bar{C}_{32}:(x,y,z)\rightarrow(1/4+z,1/4+x,1/4-y),\\
    &\bar{C}_{33}:(x,y,z)\rightarrow(1/4-z,1/4+x,1/4+y),\\
    &\bar{C}_{34}:(x,y,z)\rightarrow(1/4+z,1/4-x,1/4+y),\\
    &\bar{C}_{35}:(x,y,z)\rightarrow(1/4+y,1/4-z,1/4+x),\\
    &\bar{C}_{36}:(x,y,z)\rightarrow(1/4+y,1/4+z,1/4-x),\\
    &\bar{C}_{37}:(x,y,z)\rightarrow(1/4-y,1/4+z,1/4+x),\\
    &\bar{C}_{38}:(x,y,z)\rightarrow(1/4-y,1/4-z,1/4-x).\\
    \end{aligned}\right.
    \end{equation}
    The first four correspond to clockwise $C_3$ rotations around the axes $[1, 1, 1]$, $[1, -1, -1]$, $[-1, 1, -1]$, and $[-1, -1, 1]$ passing through the points $(0,0,0)$, $(0,1/2,0)$, $(-1/2,1/2,0)$, and $(1/2,0,0)$ followed by inversion with respect to the points $(1/8, 1/8, 1/8)$, $(1/8, 3/8, -1/8)$, $(-1/8, 1/8, 3/8)$, and $(3/8, -1/8, 1/8)$, respectively. The last four correspond to counterclockwise $C_3$ rotations around the axes $[1, -1, -1]$, $[-1, 1, -1]$, $[-1, -1, 1]$, and $[1, 1, 1]$ passing through the points $(1/2,-1/2,0)$, $(1/2,0,0)$, $(0,1/2,0)$ and $(0,0,0)$ followed by inversion with respect to the points $(1/8, -1/8, 3/8)$, $(3/8, 1/8, -1/8)$, $(-1/8, 3/8, 1/8)$, and $(1/8, 1/8, 1/8)$, respectively. 
    
\subsection{Minimal set of symmetry generators}
\label{app:minimal_set}
Here we show there exist a minimal set of generators given by a screw rotation and rotoinversion in terms of which all aforementioned elements of the $Fd\bar{3}m$ space group can be expressed. We choose $\bar{C}_{31}$ and $\bar{S}_{42}$ and henceforth in the ensuing discussion assign them a new notation $\bar{C}_{3}$ and $S_{4}$, respectively. The $24$ tetrahedral point group elements can be expressed as
\begin{itemize}
    \item $\mathds{1}$: $\mathds{1}$=$\bar{C}^6_3$.
    \item $8C_3$: $C_{31}=\bar{C}^4_3$, $C_{32}$=$\bar{C}_3$$S_4\bar{C}_3$$S_4\bar{C}^{-1}_3$$S_4$$\bar{C}^{-3}_3$$S_4$$\bar{C}^{-4}_3$, $C_{33}=C^2_{32}C_{31}C_{32}$ and $C_{34}=(C_{31}C_{32})^2$.
    \item $6\sigma_d$: $\sigma_{d1}=\Bar{C}_3S_4\Bar{C}^{-1}_3S_4\Bar{C}^{-3}_3S_4\Bar{C}^{-4}_3$, $\sigma_{d2}=\sigma^{-1}_{d1}C_{31}C_{32}C_{31}$, $\sigma_{d3}=\sigma^{-1}_{d1}C_{31}$, $\sigma_{d4}=\sigma^{-1}_{d2}C_{34}$, $\sigma_{d5}=C^{-1}_{33}\sigma_{d2}C_{33}$ and $\sigma_{d6}=C_{31}\sigma_{d1}$.
    \item $3C_2$: $C_{21}=\sigma_{d1}\sigma_{d2}=C_{31}C_{32}C_{31}$, $C_{22}=\sigma_{d3}\sigma_{d4}$ and $C_{23}=\sigma_{d5}\sigma_{d6}$.
    \item $6C'_4$: $C'_{41}=\sigma_{d2}C_{23}$, $C'_{42}={C'}^{-1}_{41}C_{34}$ and $C'_{43}={C'}^{-1}_{42}C_{32}$.
\end{itemize}

Similarly, the $24$ nonsymmorphic elements can be represented in terms of $\bar{C}_{3}$ and $S_{4}$ as follows:
\begin{itemize}
    \item $\bar{I}$: $\bar{I}=\bar{C}^3_3$.
    \item $6\bar{S}_4$: $\bar{S}_{41}=\bar{I}C_{23}C'_{43}$, $\bar{S}_{42}=\bar{S}_4$, $\bar{S}_{43}=\bar{I}C_{23}\sigma_{d3}$, $\bar{S}_{44}=\bar{I}C_{22}\sigma_{d6}$, $\bar{S}_{45}=\bar{I}C_{22}\sigma_{d2}$ and $\bar{S}_{46}=\bar{I}C'_{42}$.
    \item $6\bar{C}_2$: $\bar{C}_{21}=\bar{I}\sigma_{d3}$, $\bar{C}_{22}=\bar{I}\sigma_{d6}$, $\bar{C}_{23}=\bar{I}\sigma_{d1}$, $\bar{C}_{24}=\bar{I}C_{22}\sigma_{d3}$, $\bar{C}_{25}=\bar{I}\sigma_{d5}$ and $\bar{C}_{26}=\bar{I}\sigma_{d2}$.
    \item $3\bar{\sigma}_{d}$: $\bar{\sigma}_{d1}=\bar{I}C_{22}$, $\bar{\sigma}_{d2}=\bar{I}C_{23}$ and $\bar{\sigma}_{d1}=\bar{I}C_{21}$.
    \item $8\bar{C}_3$: $\bar{C}_{31}=\bar{C}_3$, $\bar{C}_{32}=\bar{I}C_{33}$, $\bar{C}_{33}=\bar{I}C_{34}$, $\bar{C}_{34}=\bar{I}C_{32}$, $\bar{C}_{35}=\bar{I}C^2_{33}$, $\bar{C}_{36}=\bar{I}C^2_{34}$, $\bar{C}_{37}=\bar{I}C^2_{32}$ and $\bar{C}_{38}=\bar{I}C^2_{31}$.
\end{itemize}

Thus, $\bar{C}_3$ and $S_4$ serve as the ``fundamental'' generators of the diamond lattice.

\section{Generic gauge conditions}
\label{sec:genric_gauge_con}
Generalizing Eqs.~\eqref{eq:id_relation} to incorporate the gauge transformations corresponding to lattice symmetry operations yields
\begin{subequations}
\label{eq:id_gauge_relation}
\begin{align}
   (G_{T_i} T_i)(G_{T_{i+1}}T_{i+1}) (G_{T_i}T_i)^{-1}(G_{T_{i+1}}T_{i+1})^{-1}&\in {\rm IGG}\ , \; \\
  (G_{T_i} T_i)(G_{\bar{C}_3}\bar{C}_3)^{-1}(G_{T_{i+1}}T_{i+1})(G_{\bar{C}_3}\bar{C}_3)&\in {\rm IGG}\ , \; \\
  (G_{\bar{C_3}}\bar{C_3})^6 & \in IGG \ , \; \\
(G_{T_1}T_1)^{-1}(G_{S_4}S_4)(G_{T_2}T_2)(G_{S_4}S_4)^{-1}& \in {\rm IGG} \ , \; \\
(G_{T_2}T_2)^{-1}(G_{S_4}S_4)(G_{T_2}T_2)(G_{T_3}T_3)^{-1}(G_{S_4}S_4)^{-1}& \in {\rm IGG}\ , \; \\
   (G_{S_4}S_4)(G_{T_1}T_1)^{-1}(G_{T_2}T_2)(G_{S_4}S_4)^{-1}(G_{T_3}T_3)^{-1}&\in {\rm IGG}\ , \; \\
  (G_{S_4}S_4)(G_{T_1}T_1)(G_{S_4}S_4)^{-1}(G_{T_1}T_1)^{-1}(G_{T_3}T_3)&\in {\rm IGG}\ , \; \\
  (G_{S_4}S_4)(G_{T_3}T_3)^{-1}(G_{S_4}S_4)^{-1}(G_{T_2}T_2)^{-1}(G_{T_1}T_1)&\in {\rm IGG}\ , \; \\
  (G_{T_3}T_3)(G_{T_1}T_1)^{-1}(G_{T_2}T_2)^{-1}(G_{S_4}S_4)^4&\in {\rm IGG}\ , \; \\
((G_{S_4}S_4) (G_{\bar{C_3}}\bar{C_3}))^2 & \in {\rm IGG} \ , \; \\
((G_{\bar{C_3}}\bar{C_3})^2 (G_{S_4}S_4)^2)^3 & \in {\rm IGG}\ , \; \\
((G_{\bar{C_3}}\bar{C_3})^3 (G_{S_4}S_4))^4 & \in {\rm IGG} \ , \; \\
((G_{\bar{C_3}}\bar{C_3}) (G_{S_4}S_4)^2)^6 & \in {\rm IGG} \ , \; \\
(G_\mathcal{T}\mathcal{T})^2 & \in {\rm IGG} \ , \; \\
(G_\mathcal{T}\mathcal{T}) (G_\mathcal{O}\mathcal{O}) (G_\mathcal{T}\mathcal{T})^{-1} (G_\mathcal{O}\mathcal{O})^{-1} \in {\rm IGG}  \ , \; \mathcal{O} \in &\{T_i,S_4, \bar{C_3}\} \ .  
\end{align}
\end{subequations}

\section{$\mathbf{SU(2)}$ PSG}
\label{app:su2_psg_derivation}
\subsection{Canonical form of {\it{Ansatz}} and corresponding gauges}
The canonical form of a $SU(2)$ \textit{Ansatz} is given by
\begin{equation}
    \label{eq:canonical_su2}
    u_{ij}=\dot{\iota}\chi_{ij}\tau^0,
\end{equation}
and the loop operators must be trivial, i.e., $P_C\propto\tau^0$. The structure of the gauge that keeps the above canonical form intact must have the following form:
\begin{equation}
    \label{eq:canonical_su2_gauge structrure}
    G_\mathcal{O}(r,\mu)=\eta_\mathcal{O}(r)g_\mathcal{O}(\mu),
\end{equation}
where $\eta_\mathcal{O}(r)=\pm1$ and $g_\mathcal{O}(\mu)\in SU(2)$.
\subsection{Derivation of $\mathbf{SU(2)}$ PSG corresponding to space group generators}
The relations~\eqref{eq:canonical_su2_gauge structrure} for $\mathcal{O}\in\{T_1,T_2,T_3\}$ can be written as
\begin{equation}
\left.\begin{aligned}
&G_{T_1}(r_1,r_2,r_3,\mu)=\eta_{T_1}(r_1,r_2,r_3)g_{T_1}(\mu),\\
&G_{T_2}(r_1,r_2,r_3,\mu)=\eta_{T_2}(r_1,r_2,r_3)g_{T_2}(\mu),\\
&G_{T_3}(r_1,r_2,r_3,\mu)=\eta_{T_3}(r_1,r_2,r_3)g_{T_3}(\mu).\\
\end{aligned}\right.
\end{equation}
Now, the \textit{Ans\"atze} on the bonds which connect different sublattices will take the form~\eqref{eq:canonical_su2} only if we consider $g_{T_i}(\mu)=\eta_{\mu,i}g_{T_i}$ with $\eta_{\mu,i}=\pm1$ and thus,
\begin{equation}
\label{eq:su2_tran}
\left.\begin{aligned}
&G_{T_1}(r_1,r_2,r_3,\mu)=\eta_{T_1}(r_1,r_2,r_3)\eta_{\mu,1}g_{T_1},\\
&G_{T_2}(r_1,r_2,r_3,\mu)=\eta_{T_2}(r_1,r_2,r_3)\eta_{\mu,2}g_{T_2},\\
&G_{T_3}(r_1,r_2,r_3,\mu)=\eta_{T_3}(r_1,r_2,r_3)\eta_{\mu,3}g_{T_3}.\\
\end{aligned}\right.
\end{equation}
A local gauge transformation with the form $W(r,\mu)=\eta(r)\tau^0$ maintains the explicit $SU(2)$ form~\eqref{eq:canonical_su2} of $u_{ij}$. Exploiting this gauge freedom we can set $\eta_{T_1}(r_1,r_2,r_3)\eta_{\mu,1}=\eta_{T_2}(0,r_2,r_3)\eta_{\mu,2}=\eta_{T_3}(0,0,r_3)\eta_{\mu,3}=1$ and accordingly the relations~\eqref{eq:su2_tran} can be expressed as
\begin{equation}
\left.\begin{aligned}
&G_{T_1}(r_1,r_2,r_3,\mu)=g_{T_1},\\
&G_{T_2}(r_1,r_2,r_3,\mu)=\eta_{T_2}(r_1)g_{T_2},\\
&G_{T_3}(r_1,r_2,r_3,\mu)=\eta_{T_3}(r_1,r_2)g_{T_3},\\
\end{aligned}\right.
\end{equation}
which upon imposing the constraint~\eqref{eq:id_gauge_relation}(a) reads
\begin{equation}
\label{eq:tran_1_su2}
\left.\begin{aligned}
&G_{T_1}(r_1,r_2,r_3,\mu)=g_{T_1},\\
&G_{T_2}(r_1,r_2,r_3,\mu)=\eta^{r_1}_1g_{T_2},\\
&G_{T_3}(r_1,r_2,r_3,\mu)=\eta^{r_1}_{3}\eta^{r_2}_{2}g_{T_3},\\
\end{aligned}\right.
\end{equation}
where $g_{T_1},g_{T_2},g_{T_3}\in SU(2)$. Now, we seek the form of $G_{\bar{C}_3}(r,\mu)$. From Eq.~\eqref{eq:id_gauge_relation}(b) we have,
\begin{equation}
\label{eqn:gtgc3}
\left.\begin{aligned}
G_{T_i}[\bar{C}^{-1}_3T^{-1}_{i+1}(r,\mu)]G^{-1}_{\bar{C}_3}(r,\mu)&G_{T_{i+1}}(r,\mu)\\
&G_{\bar{C}_3}[T^{-1}_{i+1}(r,\mu)]\in SU(2).
\end{aligned}\right.
\end{equation}
From Eq.~\eqref{eq:canonical_su2_gauge structrure}, we see that the projective gauge for rotoinversion takes the general form $G_{\bar{C}_3}(r,\mu)=\eta_{\bar{C}_3}(r)g_{\bar{C}_3}(\mu)$. Substituting this expression for $G_{\bar{C}_3}(r,\mu)$ and those of $G_{T_{i}}$ [Eq.~\eqref{eq:tran_1_su2}] in Eq.~\eqref{eqn:gtgc3} and adopting the convention $\xi_i[\eta_{\bar{C}_3}(r)]=\eta_{\bar{C}_3}(r)\eta_{\bar{C}_3}[T^{-1}_i(r)]$ yields
\begin{equation}
\label{eq:cb3_su21}
\left.\begin{aligned}
&\xi_1[\eta_{\bar{C}_3}(r,\mu)]=\eta_{\bar{C}_3T_1}\eta^{r_2}_{3}\eta^{r_3}_{2},\\
&\xi_2[\eta_{\bar{C}_3}(r,\mu)]=\eta_{\bar{C}_3T_2}\eta^{r_1}_1,\\
&\xi_3[\eta_{\bar{C}_3}(r,\mu)]=\eta_{\bar{C}_3T_3}\eta^{r_2}_2\eta^{r_1}_{3}\eta^{r_2}_{2},\\\
\end{aligned}\right.
\end{equation}
and
\begin{equation}
\label{}
g_{T_i}g^{-1}_{\bar{C}_3}(\mu)g_{T_{I+1}}g_{\bar{C}_3}(\mu)\in SU(2).
\end{equation}
The consistency equation can be written as,
\begin{equation}
\label{eq:consistent_su2}
\left.\begin{aligned}
\xi_i[\eta_{\bar{C}_3}(r)]&\xi_{i+1}[\eta_{\bar{C}_3}[T^{-1}_i(r)]]\\
&=\xi_{i+1}[\eta_{\bar{C}_3}(r)]\xi_{i}[\eta_{\bar{C}_3}[T^{-1}_{i+1}(r)]],\\
\end{aligned}\right.
\end{equation}
which requires $\eta_1=\eta_{2}=\eta_{3}$. With this fixing, $G_{\bar{C}_3}$ has the following solution:
\begin{equation}
\label{eq:cb3_su22}
 G_{\bar{C}_3}(r,\mu)=\eta^{r_1}_{\bar{C}_3T_3}\eta^{r_2}_{\bar{C}_3T_1}\eta^{r_3}_{\bar{C}_3T_2}\eta^{r_1(r_2+r_3)}_1g_{\bar{C}_3}(\mu).
\end{equation}

Now plugging this result in Eq.~\eqref{eq:id_gauge_relation}(c), one gets
\begin{equation}
	\label{eq:cb3_su2}
	[g_{\bar{C}_3}(\mu)g_{\bar{C}_3}(\bar{\mu})]^3\in SU(2).
\end{equation}

Now, let us consider the gauge transformation corresponding to screw symmetry $G_{S_4}$. Eqs.~\eqref{eq:id_gauge_relation}(d),~\eqref{eq:id_gauge_relation}(e), and~\eqref{eq:id_gauge_relation}(f) give
\begin{equation}
\label{eqn:gS4_T}
\left.\begin{aligned}
&G^{-1}_{T_1}(r_1+1,r_2,r_3,\mu)G_{S_4}(r_1+1,r_2,r_3,\mu)\\
&G_{T_2}(-r_3,r_1+r_2+r_3-\bar{\mu}+1,-r_2,\bar{\mu})\\
&G^{-1}_{S_4}(r_1,r_2,r_3,\mu)\in SU(2),\\
&G^{-1}_{T_2}(r_1,r_2,r_3,\mu)G_{S_4}(r_1,r_2,r_3,\mu)\\
&G_{T_2}(-r_3,r_1+r_2+r_3-\bar{\mu},-r_2,\bar{\mu})\\
&G^{-1}_{T_3}(-r_3,r_1+r_2+r_3-\bar{\mu}-1,-r_2+1,\bar{\mu})\\
&G^{-1}_{S_4}(r_1,r_2-1,r_3,\mu)\in SU(2),\\
&G_{S_4}(r_1,r_2,r_3,\mu)G^{-1}_{T_1}(-r_3+1,r_1+r_2+r_3-\bar{\mu},-r_2,\bar{\mu})\\
&G_{T_2}(-r_3+1,r_1+r_2+r_3-\bar{\mu},-r_2,\bar{\mu})\\
&G^{-1}_{S_4}(r_1,r_2,r_3-1,\mu)G^{-1}_{T_3}(r_1,r_2,r_3,\mu)\in SU(2).\\
\end{aligned}\right.
\end{equation}
From Eq.~\eqref{eq:canonical_su2_gauge structrure}, the projective gauge for $S_4$ has the general form 
$G_{S_4}(r,\mu)=\eta_{S_4}(r)g_{S_4}(\mu)$. Substituting this expression together with Eq.~\eqref{eq:tran_1_su2} in Eq.~\eqref{eqn:gS4_T}, and defining $\xi_i[\eta_{S_4}(r)]=\eta_{S_4}(r)\eta_{S_4}[T^{-1}_i(r)]$, we have
\begin{equation}
\label{eq:s4_su21}
\left.\begin{aligned}
&\xi_1[\eta_{S_4}(r)]=\eta_{ST_1}\eta^{r_3}_1,\\
&\xi_2[\eta_{S_4}(r)]=\eta_{ST_2}\eta^{r_2+r_3-1-\bar{\mu}}_1,\\
&\xi_3[\eta_{S_4}(r)]=\eta_{ST_3}\eta^{r_1+r_2-r_3+1}_1,\\
\end{aligned}\right.
\end{equation}
and
\begin{equation}
\label{}
\left.\begin{aligned}
&g^{-1}_{T_1}g_{S_4}(\mu)g_{T_2}g^{-1}_{S_4}(\mu)\in SU(2),\\
&g^{-1}_{T_2}g_{S_4}(\mu)g_{T_2}g_{T_3}g^{-1}_{S_4}(\mu)\in SU(2),\\
&g_{S_4}(\mu)g^{-1}_{T_1}g_{T_2}g^{-1}_{S_4}(\mu)g^{-1}_{T_3}\in SU(2).\\
\end{aligned}\right.
\end{equation}
The resulting consistency conditions equivalent to Eq.~\eqref{eq:consistent_su2}, do not impose any restriction on $\eta_1$. Hence, the solution of $G_{S_4}(r,\mu)$ is given by 
 \begin{equation}
\label{eq:s4_su22}
\left.\begin{aligned}
	G_{S_4}(r,\mu)&=\eta^{r_1}_{ST_1}\eta^{r_2}_{ST_2}\eta^{r_3}_{ST_3}\\
	&\eta^{r_3(r_1+r_2)-r_2\bar{\mu}+r_2\left(\frac{r_2-1}{2}\right)-r_3\left(\frac{r_3-1}{2}\right)}_1g_{S_4}(\mu).\\
\end{aligned}\right.
\end{equation}
Equations~\eqref{eq:id_gauge_relation}(g) and ~\eqref{eq:id_gauge_relation}(h) give
\begin{equation}
\label{}
\left.\begin{aligned}
&g_{S_4}(\mu)g^{-1}_{T_3}g^{-1}_{S_4}(\mu)g^{-1}_{T_2}g_{T_1}\eta_{ST_1}\eta_{ST_2}\in SU(2),\\
&g_{S_4}(\mu)g_{T_1}g^{-1}_{S_4}(\mu)g^{-1}_{T_1}g_{T_3}\eta_{ST_1}\eta_{ST_3}\eta_1\in SU(2),\\
\end{aligned}\right.
\end{equation}
and Eq.~\eqref{eq:id_gauge_relation}(i) gives
\begin{subequations}
\label{eq:s4_tran_su21}
\begin{align}
&\mu = 0:\Rightarrow [g_{S_4}(0)g_{S_4}(1)]^2\eta_{ST_1}\eta_{ST_2}\in SU(2) \; \notag \\
&\Rightarrow  [g_{S_4}(0)g_{S_4}(1)]^2\eta_{ST_1}\eta_{ST_2}\in SU(2)\; , \\ 
&\mu = 1:\Rightarrow [g_{S_4}(1)g_{S_4}(0)]^2\eta_{ST_3}\eta_{1}\in SU(2) \; \notag \\
&\Rightarrow  [g_{S_4}(1)g_{S_4}(0)]^2\eta_{ST_3}\eta_1\in SU(2)\;. 
\end{align}
\end{subequations}
From the above we thus get
\begin{equation}
	\label{eq:s4_trans_su44}
\left.\begin{aligned}
	&[g_{S_4}(0)g_{S_4}(1)]^2\eta_{ST_1}\eta_{ST_2}=[g_{S_4}(1)g_{S_4}(0)]^2\eta_{ST_3}\eta_1\\
	&\Rightarrow \eta_{ST_1}\eta_{ST_2}\eta_{ST_3}=\eta_1.\\
\end{aligned}\right.
\end{equation}
A further simplification is possible by effecting the local gauge transformation,
\begin{equation}
\label{eq:local_su2}
    W(r,\mu)=\eta^{r_1}_x\eta^{r_2}_y\eta^{r_3}_z\tau^0.
\end{equation}

We find that the above transformation does not change the structure of the translation gauge $G_{T_{1}}$, $G_{T_{2}}$, and $G_{T_{3}}$ up to a possible global sign, which is unimportant. Also, it can be seen that the gauge transformation given by Eq.~\eqref{eq:local_su2} modulates the sign of $\eta_{ST_i}$ and $\eta_{\bar{C}_3T_i}$, however, by a proper choice of the constants 
$\eta_x$, $\eta_y$ and $\eta_z$, together with Eq.~\eqref{eq:s4_trans_su44}, we can fix $\eta_{ST_1},\eta_{ST_2}$ and $\eta_{\bar{C}_3T_3}$ to be positive, i.e.,
\begin{equation}
\label{eq:s4_fixing_su2}
    \eta_{ST_1}=\eta_{ST_2}=\eta_{\bar{C}_3T_3}=+1,\;\eta_{ST_3}=\eta_{1}.
\end{equation}
Thus Eq.~\eqref{eq:s4_su22} and Eq.~\eqref{eq:s4_tran_su21} take the forms
 \begin{equation}
\label{eq:s4_su23}
G_{S_4}(r,\mu)=\eta^{r_3(r_1+r_2+1)+r_2\left(\frac{r_2-1}{2}-\bar{\mu}\right)-r_3\left(\frac{r_3-1}{2}\right)}_1g_{S_4}(\mu)
\end{equation}
and
\begin{equation}
\label{eq:s4_tran_su22}
[g_{S_4}(0)g_{S_4}(1)]^2=[g_{S_4}(1)g_{S_4}(0)]^2\in SU(2).
\end{equation}

Further constraints are imposed by the four conditions given by Eqs.~\eqref{eq:id_gauge_relation}(j),~\eqref{eq:id_gauge_relation}(k),~\eqref{eq:id_gauge_relation}(l), and ~\eqref{eq:id_gauge_relation}(m). 
From Eq.~\eqref{eq:id_gauge_relation}(j), we have
\begin{subequations}
	\label{eq:cs1_su2}
	\begin{align}
		&\mu = 0:\Rightarrow [g_{S_4}(0) g_{\bar{C_3}}(1)]^2 \in SU(2)\;\text{with} \hspace{2mm} \eta_{\bar{C}_3T_1}  =+1 \; , \\ 
		&\mu = 1: \Rightarrow [g_{S_4}(1) g_{\bar{C_3}}(0)]^2  \in SU(2),\; \text{with} \hspace{2mm} \eta_{\bar{C}_3T_1}  =+1 \;. 
	\end{align}
\end{subequations}
From Eq.~\eqref{eq:id_gauge_relation}(k), we have
\begin{subequations}
	\label{eq:cs2_su2}
	\begin{align}
		&\mu = 0: \Rightarrow [g_{\bar{C_3}}(0)g_{\bar{C_3}}(1)g_{S_4}(0)g_{S_4}(1)]^3 \in SU(2)\;, \\
		&\mu = 1:\Rightarrow [g_{\bar{C_3}}(1)g_{\bar{C_3}}(0)g_{S_4}(1)g_{S_4}(0)]^3  \eta_{\bar{C}_3T_2}\in SU(2)  \;.
	\end{align}
\end{subequations}
From Eq.~\eqref{eq:id_gauge_relation}(l), we have
\begin{subequations}
	\label{eq:cs3_su2}
	\begin{align}
		&\mu = 0: \Rightarrow [g_{\bar{C_3}}(0)g_{\bar{C_3}}(1)g_{\bar{C_3}}(0)g_{S_4}(1)]^4 \in SU(2) \; , \\
		&\mu = 1: \Rightarrow [g_{\bar{C_3}}(1)g_{\bar{C_3}}(0)g_{\bar{C_3}}(1)g_{S_4}(0)]^4 \eta_{\bar{C}_3T_2}\in SU(2) \;.
	\end{align}
\end{subequations}

From Eq.~\eqref{eq:id_gauge_relation}(m), we have
\begin{equation}
\label{eq:cs4_su2}
[g_{\bar{C_3}}(\mu)g_{S_4}(\bar{\mu})g_{S_4}(\mu)g_{\bar{C_3}}(\bar{\mu})g_{S_4}(\mu)g_{S_4}(\bar{\mu})]^3 \in SU(2) .
\end{equation}

Let us now find the PSG for time-reversal symmetry. Using Eq.~\eqref{eq:id_gauge_relation}(o) for $\mathcal{O}\in T_i$, we get
\begin{equation}
\label{eq:tr_su2_1}
G_{\mathcal{T}}(r,\mu)G_{T_i}(r,\mu)G^{-1}_{\mathcal{T}}[T^{-1}_i(r,\mu)]G^{-1}_{T_i}(r,\mu)\in SU(2).
\end{equation} 
From Eq.~\eqref{eq:canonical_su2_gauge structrure}, the projective gauge for $\mathcal{T}$ has the general form 
$G_{\mathcal{T}}(r,\mu)=\eta_{\mathcal{T}}(r)g_{\mathcal{T}}(\mu)$. 
Replacing this in the above equation with the definition 
$\xi_i[\eta_{\mathcal{T}}(r)]=\eta_{\mathcal{T}}(r)\eta_{\mathcal{T}}[T^{-1}_i(r)]$, we have
\begin{equation}
\label{eq:tr_z1_su2}
f_i[G_{\mathcal{T}}(r,\mu)]=\eta_{\mathcal{T}T_i}\tau^0.
\end{equation} 
The consistency condition does not impose any constraint. Thus, from Eq.~\eqref{eq:tr_z1_su2}, the solution for $G_{\mathcal{T}}(r,\mu)$ can be written as
\begin{equation}
\label{eq:tr_su2}
G_{\mathcal{T}}(r,\mu)=\eta^{r_1}_{\mathcal{T}T_1}\eta^{r_2}_{\mathcal{T}T_2}\eta^{r_3}_{\mathcal{T}T_3}g_\mathcal{T}(\mu).
\end{equation}
Using Eq.~\eqref{eq:id_gauge_relation}(o) for $\mathcal{O}\in S_4,\bar{C}_3$, one can set
\begin{equation}
    \eta_{\mathcal{T}T_1}=\eta_{\mathcal{T}T_2}=\eta_{\mathcal{T}T_3}=1.
\end{equation}

Thus the PSG solutions for IGG$\in$ $SU(2)$ can be gathered and rewritten as follows:

\begin{subequations}
\label{eq:psg_sol_su2_1}
\begin{align}
&G_{T_1}(r,\mu)=g_{T_1},\;G_{T_2}(r,\mu)=\eta^{r_1}_1g_{T_2},\;G_{T_3}(r,\mu)=\eta^{r_1+r_2}_1g_{T_3},\\
&G_{\mathcal{T}}(r,\mu)=g_\mathcal{T}(\mu),\\
&G_{S_4}(r,\mu)=\eta^{r_3(r_1+r_2+1)+r_2\left(\frac{r_2-1}{2}-\bar{\mu}\right)-r_3\left(\frac{r_3-1}{2}\right)}_1g_{S_4}(\mu),\\
&G_{\bar{C}_3}(r,\mu)=\eta^{r_3}_{\bar{C}_3T_2}\eta^{r_1(r_2+r_3)}_1g_{\bar{C}_3}(\mu),
\end{align}
\end{subequations}
where the unit cell representation matrices must satisfy the following constraints:
\begin{subequations}
\label{eq:psg_sol_su2_2}
\begin{align}
&[g_\mathcal{T}(\mu)]^2\in SU(2),\\
&[g_{\bar{C}_3}(\mu)g_{\bar{C}_3}(\bar{\mu})]^3\in SU(2),\\
&[g_{S_4}(\mu)g_{S_4}(\bar{\mu}) ]^2\in SU(2),\\
&[g_{S_4}(\mu) g_{\bar{C}_3}(\bar{\mu})]^2  \in SU(2),\\
&[g_{\bar{C_3}}(0)g_{\bar{C_3}}(1)g_{S_4}(0)g_{S_4}(1)]^3 \in SU(2),\\
&[g_{\bar{C_3}}(1)g_{\bar{C_3}}(0)g_{S_4}(1)g_{S_4}(0)]^3  \eta_{\bar{C}_3T_2}\in SU(2),\\
&[g_{\bar{C_3}}(0)g_{\bar{C_3}}(1)g_{\bar{C_3}}(0)g_{S_4}(1)]^4 \in SU(2),\\
&[g_{\bar{C_3}}(1)g_{\bar{C_3}}(0)g_{\bar{C_3}}(1)g_{S_4}(0)]^4 \eta_{\bar{C}_3T_2}\in SU(2),\\
&[g_{\bar{C_3}}(\mu)g_{S_4}(\bar{\mu})g_{S_4}(\mu)g_{\bar{C_3}}(\bar{\mu})g_{S_4}(\mu)g_{S_4}(\bar{\mu})]^3 \in SU(2),\\
&g_{\mathcal{T}}(\mu)g_{\bar{C_3}}(\mu)g^{-1}_{\mathcal{T}}(\bar{\mu})g^{-1}_{\bar{C_3}}(\mu) \in SU(2),\\
&g_{\mathcal{T}}(\mu)g_{S_4}(\mu)g^{-1}_{\mathcal{T}}(\bar{\mu})g^{-1}_{S_4}(\mu) \in SU(2) \;.
\end{align}
\end{subequations}
The time-reversal symmetry condition for non vanishing \textit{Ans\"atze} on 1NN bonds requires $g_\mathcal{T}(\mu)=(-1)^\mu g_\mathcal{T}$. Also, on the diamond lattice, the nearest neighbor bonds connect different sublattices, and thus, to maintain the canonical form~\eqref{eq:canonical_su2} of $SU(2)$ \textit{Ans\"atze} we need to consider $g_{\bar{C_3}}(\mu)=\eta_{c,\mu}g_{\bar{C_3}}$ and $g_{S_4}(\mu)=\eta_{s,\mu}g_{S_4}$. With this fixing, Eqs.~\eqref{eq:psg_sol_su2_2} can be recast as
\begin{subequations}
\label{eq:psg_sol_su2_3}
\begin{align}
&[g_\mathcal{T}]^2\in SU(2),\\
&\eta_{c,\mu}\eta_{c,\bar{\mu}}[g_{\bar{C}_3}]^3\in SU(2),\\
&[g_{S_4}]^4\in SU(2),\\
&[g_{S_4}g_{\bar{C}_3}]^2  \in SU(2),\\
&\eta_{c,0}\eta_{c,1}\eta_{s,0}\eta_{s,1}[g_{\bar{C_3}}g_{\bar{C_3}}g_{S_4}g_{S_4}]^3 \in SU(2),\\
&\eta_{c,0}\eta_{c,1}\eta_{s,0}\eta_{s,1}[g_{\bar{C_3}}g_{\bar{C_3}}g_{S_4}g_{S_4}]^3  \eta_{\bar{C}_3T_2}\in SU(2),\\
&[g_{\bar{C_3}}g_{\bar{C_3}}g_{\bar{C_3}}g_{S_4}]^4 \in SU(2),\\
&[g_{\bar{C_3}}g_{\bar{C_3}}g_{\bar{C_3}}g_{S_4}]^4 \eta_{\bar{C}_3T_2}\in SU(2),\\
&\eta_{c,\mu}\eta_{c,\bar{\mu}}[g_{\bar{C_3}}g_{S_4}g_{S_4}g_{\bar{C_3}}g_{S_4}g_{S_4}]^3 \in SU(2),\\
&-g_{\mathcal{T}}g_{\bar{C_3}}g^{-1}_{\mathcal{T}}g^{-1}_{\bar{C_3}} \in SU(2),\\
&-g_{\mathcal{T}}g_{S_4}g^{-1}_{\mathcal{T}}g^{-1}_{S_4} \in SU(2) \;.
\end{align}
\end{subequations}
From Eqs.~\eqref{eq:psg_sol_su2_3}(e) and ~\eqref{eq:psg_sol_su2_3}(f), it can be readily seen that $\eta_{\bar{C}_3T_2}=+1$. 
In total, we find there are eight gauge inequivalent $SU(2)$ PSGs, which are listed in Table~\ref{table:su2_psg} and the solutions given by Eqs.~\eqref{eq:psg_sol_su2_1} take the form of Eqs.~\eqref{eq:su2_gauge_sol} in the main text.

\section{$\mathbf{U(1)}$ PSG}
\label{app:u1_psg_derivation}
\subsection{Canonical {\it{Ansatz}} form and corresponding gauges}
The canonical form of an U(1) \textit{Ansatz} is given by
\begin{equation}
    \label{eq:canonical_u1}
    u_{ij}=\dot{\iota}\text{Im}h_{ij} \tau^0 + \text{Re}h_{ij} \tau^3=\dot{\iota} A_{ij}g_3(f(ij)\phi(ij)),
\end{equation}
where $g_3(\theta)=e^{\dot{\iota}\theta\tau^3}$. Correspondingly, the loop operators take the form $P_C=g_3(\xi)$. The structure of the gauges that keep the above canonical form intact are
\begin{equation}
    \label{eq:canonical_u1_gauge structrure}
    G_\mathcal{O}(r,\mu)=g_3(\phi_\mathcal{O}(r,\mu))(\dot{\iota}\tau^1)^{w_\mathcal{O}},
\end{equation}
where $w_\mathcal{O}$ can take values 0,1 and $\mathcal{O}\in\{T_i,\bar{C}_3,S_4\}$. 

\subsection{Derivation of PSG corresponding to space group generators}
The relations~\eqref{eq:canonical_u1_gauge structrure} for $\mathcal{O}\in\{T_1,T_2,T_3\}$ can be written as
\begin{equation}
\left.\begin{aligned}
&G_{T_1}(r_1,r_2,r_3,\mu)=g_3(\phi_{T_1}(r_1,r_2,r_3,\mu))(\dot{\iota}\tau^1)^{w_{T_1}},\\
&G_{T_2}(r_1,r_2,r_3,\mu)=g_3(\phi_{T_2}(r_1,r_2,r_3,\mu))(\dot{\iota}\tau^1)^{w_{T_2}},\\
&G_{T_3}(r_1,r_2,r_3,\mu)=g_3(\phi_{T_3}(r_1,r_2,r_3,\mu))(\dot{\iota}\tau^1)^{w_{T_3}}.\\
\end{aligned}\right.
\end{equation}
Here, it is worth mentioning that the loop operators connected by translations $T_i$ differ by a sign $(-)^{w_{T_i}}$ in order to fulfill translational invariance, and four cases may arise
\begin{subequations}
\begin{align}
  &w_{T_1}=0,\; w_{T_2}=0,\; w_{T_3}=0\ , \; \label{eq:w_T_000}\\
  &w_{T_1}=1,\; w_{T_2}=0,\; w_{T_3}=0\ , \; \label{eq:w_T_100}\\
  &w_{T_1}=1,\; w_{T_2}=1,\; w_{T_3}=0\ , \; \label{eq:w_T_110}\\
  &w_{T_1}=1,\; w_{T_2}=1,\; w_{T_3}=1\ . \; \label{eq:w_T_111} \   
\end{align}
\end{subequations}
It can be readily inferred that cases given by Eqs.~\eqref{eq:w_T_100} and \eqref{eq:w_T_110} need to be excluded as these do not satisfy the condition~\eqref{eq:id_gauge_relation}(b). Similarly, one must also exclude the case given by Eq.~\eqref{eq:w_T_111} otherwise conditions~\eqref{eq:id_gauge_relation}(e),~\eqref{eq:id_gauge_relation}(f),~\eqref{eq:id_gauge_relation}(g),~\eqref{eq:id_gauge_relation}(h), and \eqref{eq:id_gauge_relation}(i) cannot be satisfied. Hence, one needs to consider only the first [Eq.~\eqref{eq:w_T_000}] case, i.e., $w_{T_i}=0$\;$\forall~i$. This is known as the uniform gauge~\cite{Wen-2002} since the loop operators connected by translations are the same in both sign and direction. Now, one can always choose a gauge transformation of the form $G(r_1,r_2,r_3,\mu)=g_3(\theta(r_1,r_2,r_3,\mu))$ such that it results in
\begin{equation}
G_{T_1}(r_1,r_2,r_3,\mu)=G_{T_2}(0,r_2,r_3,\mu)=G_{T_3}(0,0,r_3,\mu)=\tau^0.\\
\end{equation}
Upon fixing this, the symmetry condition~\eqref{eq:id_gauge_relation}(a) leads to the following solution for the translation gauges 
\begin{equation}
\label{eq:tran_1}
\left.\begin{aligned}
&G_{T_1}(r_1,r_2,r_3,\mu)=\tau^0,\\
&G_{T_2}(r_1,r_2,r_3,\mu)=g_3(-r_1\chi_1),\\
&G_{T_3}(r_1,r_2,r_3,\mu)=g_3(r_1\chi_3-r_2\chi_2).\\
\end{aligned}\right.
\end{equation}
Now, with the definition $\Delta_i\phi(r,\mu)=\phi(r,\mu)-\phi[T^{-1}_i(r,\mu)]$, Eq.~\eqref{eq:id_gauge_relation}(b) gives
\begin{equation}
\label{eq:cb3_1}
\left.\begin{aligned}
\Delta_i\phi_{\bar{C}_3}(r,\mu)=(-1)^{w_{\bar{C}_3}}(-\chi_{\bar{C}_3T_{i-1}}&+\phi_{T_{i-1}}[\bar{C}^{-1}_3T^{-1}_i(r,\mu)]),\\
&+\phi_{T_i}(r,\mu)\\
\end{aligned}\right.
\end{equation}
in which we substitute Eq.~\eqref{eq:tran_1}, resulting in
\begin{equation}
\label{eq:cb3_2}
\left.\begin{aligned}
&\Delta_1\phi_{\bar{C}_3}(r,\mu)=(-1)^{w_{\bar{C}_3}}(-\chi_{\bar{C}_3T_3}-r_2\chi_3+r_3\chi_2),\\
&\Delta_2\phi_{\bar{C}_3}(r,\mu)=(-1)^{w_{\bar{C}_3}}(-\chi_{\bar{C}_3T_1})-r_1\chi_1,\\
&\Delta_3\phi_{\bar{C}_3}(r,\mu)=(-1)^{w_{\bar{C}_3}}(-\chi_{\bar{C}_3T_2}+r_2\chi_1)+r_1\chi_3-r_2\chi_2,\\
\end{aligned}\right.
\end{equation}
which must obey the following consistency relations:
\begin{equation}
\label{eq:consistent}
\left.\begin{aligned}
\Delta_i\phi_{\bar{C}_3}(r,\mu)&+\Delta_{i+1}\phi_{\bar{C}_3}[T^{-1}_i(r,\mu)]\\
&=\Delta_{i+1}\phi_{\bar{C}_3}(r,\mu)+\Delta_i\phi_{\bar{C}_3}[T^{-1}_{i+1}(r,\mu)].\\
\end{aligned}\right.
\end{equation}
For $i=1,2,3$ in Eq.~\eqref{eq:consistent}, we use Eq.~\eqref{eq:cb3_2} which imposes the following constraints:
\begin{equation}
    \chi_1=(-1)^{w_{\bar{C}_3}}\chi_3,\;    \chi_1=(-1)^{w_{\bar{C}_3}}\chi_2,\;    \chi_3=(-1)^{w_{\bar{C}_3}}\chi_2
\end{equation}
implying
\begin{equation}
\label{eq:trans_phase}
\left.\begin{aligned}
    &w_{\bar{C}_3}=0\Rightarrow\chi_1=\chi_2=\chi_3,\\ 
    &w_{\bar{C}_3}=1\Rightarrow\chi_1=\chi_2=\chi_3=0,\pi.\\
\end{aligned}\right.
\end{equation}
Using this, the solution of $\phi_{\bar{C}_3}(r,\mu)$ can be found from Eq.~\eqref{eq:cb3_2} and is given by 
\begin{equation}
\label{eq:cb3_3}
\left.\begin{aligned}
   \phi_{\bar{C}_3}(r,\mu)&=-r_1(r_2-r_3)\chi_1\\
   &-(-1)^{w_{\bar{C}_3}}(r_1\chi_{\bar{C}_3T_3}+r_2\chi_{\bar{C}_3T_1}+r_3\chi_{\bar{C}_3T_2})+\rho_{\mu},\\
\end{aligned}\right.
\end{equation}
where $\rho_{\mu}=\phi_{\bar{C}_3}(0,0,0,\mu)$. Now, plugging this result in Eq.~\eqref{eq:id_gauge_relation}(c), we obtain
\begin{equation}
\label{eq:cb3_4}
\left.\begin{aligned}
w_{\bar{C}_3}=0:\;& g_3(3(\rho_{\mu}+\rho_{\bar{\mu}}))=g_3(\chi_{\bar{C}_3}), \\
w_{\bar{C}_3}=1:\;& g_3(3(\rho_{\mu}-\rho_{\bar{\mu}}))=g_3(\chi_{\bar{C}_3}),\\
& g_3(\chi_{\bar{C}_3T_3}+\chi_{\bar{C}_3T_1}+\chi_{\bar{C}_3T_2})=1 ,\\
\end{aligned}\right.
\end{equation}
where $\rho_{\bar{0}/\bar{1}}=\rho_{1/0}$. 

Now, let us consider the screw symmetry gauge $G_{S_4}$. Eqs.~\eqref{eq:id_gauge_relation}(d),~\eqref{eq:id_gauge_relation}(e), and~\eqref{eq:id_gauge_relation}(f) give
\begin{equation}
\label{eq:s4_1}
\left.\begin{aligned}
\Delta_1\phi_{S_4}(r,\mu)&=\chi_{ST_1}-(-1)^{w_{S_4}}r_3\chi_1,\\
\Delta_2\phi_{S_4}(r,\mu)&=\chi_{ST_2}-r_1\chi_1\\
&-(-1)^{w_{S_4}}(r_1+r_2+r_3-1-\bar{\mu})\chi_1,\\
\Delta_3\phi_{S_4}(r,\mu)&=\chi_{ST_3}+(r_1-r_2)\chi_1-(-1)^{w_{S_4}}(r_3-1)\chi_1,\\
\end{aligned}\right.
\end{equation}
which must obey the consistency condition
\begin{equation}
\label{eq:consistent_s4}
\left.\begin{aligned}
\Delta_i\phi_{S_4}(r,\mu)&+\Delta_{i+1}\phi_{S_4}[T^{-1}_i(r,\mu)]\\
&=\Delta_{i+1}\phi_{S_4}(r,\mu)+\Delta_i\phi_{S_4}[T^{-1}_{i+1}(r,\mu)].\\
\end{aligned}\right.
\end{equation}
For $i=1,2,3$ in Eq.~\eqref{eq:consistent_s4}, we use Eq.~\eqref{eq:s4_1} which imposes the following constraints:
\begin{equation}
\label{eq:trans_phase_s4}
\left.\begin{aligned}
 &(1+(-1)^{w_{S_4}})\chi_1=0,\;    \chi_1=(-1)^{w_{S_4}}3\chi_1,\\ 
 &(-1)^{w_{S_4}}\chi_1=-\chi_1.\\
 \end{aligned}\right.
\end{equation}
This has the consequence that if $w_{S_4}=0$, then $\chi_1=0,\pi$, and if $w_{S_4}=1$, then $\chi_1=0,\pi/2,\pi,3\pi/2$. We combine the results from the consistency conditions for $\phi_{\bar{C}_3}$ and $\phi_{S_4}$ in Table~\ref{table:chi_constraint}.

\begin{table}[b]
\begin{ruledtabular}
\begin{tabular}{ ccc } 
 $w_{\bar{C}_3}$ & $w_{S_4}$ & $\chi_1=\chi_2=\chi_3$ \\ 
 \hline
 0 & 0 & $0,\pi$ \\ 
 0 & 1 & $0,\pi/2,\pi,3\pi/2$ \\ 
 1 & 0 & $0,\pi$ \\ 
 1 & 1 & $0,\pi$ \\ 
\end{tabular}
\end{ruledtabular}
\caption{Allowed values of $\chi_1$ depending on different $(w_{\bar{C}_3},w_{S_4})$.}
\label{table:chi_constraint}
\end{table}

Using Eq.~\eqref{eq:trans_phase_s4}, the solution of $\phi_{S_4}(r,\mu)$ can be found from Eq.~\eqref{eq:s4_1} and is given by 
 \begin{equation}
\label{eq:s4_0}
\left.\begin{aligned}
   \phi_{S_4}&(r,\mu)=r_1\chi_{ST_1}+r_2\chi_{ST_2}+r_3\chi_{ST_3}+\theta_{\mu}-(-1)^{w_{S_4}}\\
   &\times\left[r_2\left(\frac{r_2-1}{2}+3r_3-\bar{\mu}\right)+r_3\left(r_1+\frac{r_3-1}{2}\right)\right]\chi_1,\\
    \end{aligned}\right.
\end{equation}
where $\theta_{\mu}=\phi_{S_4}(0,0,0,\mu)$. In the expression of $\phi_{S_4}(r,\mu)$, the relation $3r_2r_3\chi_1=-r_2r_3\chi_1$ holds for any choice of $w_{S_4}$. Thus we can rewrite Eq.~\eqref{eq:s4_0} as
 \begin{equation}
	\label{eq:s4_2}
	\left.\begin{aligned}
		\phi_{S_4}&(r,\mu)=r_1\chi_{ST_1}+r_2\chi_{ST_2}+r_3\chi_{ST_3}+\theta_{\mu}-(-1)^{w_{S_4}}\\
		&\times\left[r_2\left(\frac{r_2-1}{2}-r_3-\bar{\mu}\right)+r_3\left(r_1+\frac{r_3-1}{2}\right)\right]\chi_1.\\
	\end{aligned}\right.
\end{equation}
Now plugging the results in Eqs.~\eqref{eq:id_gauge_relation}(g) and ~\eqref{eq:id_gauge_relation}(h), we obtain 
\begin{equation}
    \label{eq:s4_trans_1}
    \chi_{ST_1}-\chi_{ST_3}-(-1)^{w_{S_4}}\chi_1=\chi_{ST_{4}},
\end{equation}
\begin{equation}
    \label{eq:s4_trans_2}
    \chi_{ST_2}-\chi_{ST_1}=\chi_{ST_{5}}.
\end{equation}
Now, Eq.~\eqref{eq:id_gauge_relation}(i) gives
\begin{equation}
\left.\begin{aligned}
\phi_{S_4}[S^4_4(r,\mu)]&+\phi_{S_4}[S^2_4(r,\mu)]+(-1)^{w_{S_4}}(\phi_{S_4}[S^3_4(r,\mu)]\\
&+\phi_{S_4}[S_4(r,\mu)])=\chi_{ST_6}+\phi_{T_1}[T_1T^{-1}_3(r,\mu)]\\
&+\phi_{T_2}[T_1T_2T^{-1}_3(r,\mu)]-\phi_{T_1}(r,\mu).\\
\end{aligned}\right.   
\end{equation}
On substituting Eq.~\eqref{eq:s4_2} and Eq.~\eqref{eq:tran_1}, we get the following constraints:
\begin{enumerate}
    \item For $w_{S_4}=0$ using $2\chi_{1}=0$ 
    \begin{equation}
    \label{eq:s4_trans_3}
        \chi_{ST_1}+\chi_{ST_2}-\chi_{ST_3}=\chi_1=0,\pi.
    \end{equation}
    \item For $w_{S_4}=1$ using $4\chi_1=0$, we find
    \begin{equation}
    \label{eq:s4_trans_4}
        \chi_{ST_1}+\chi_{ST_2}+\chi_{ST_3}+4(\theta_{0}-\theta_{1})=\chi_1.
    \end{equation}
\end{enumerate}
It is worth stating that all the above-mentioned $U(1)$ phases may not be gauge independent. Some of them can be fixed by the use of local gauge transformations. We know that under a local gauge transformation $W(r,\mu)$, the projective gauge $G_\mathcal{O}$ transforms as $G_\mathcal{O}(r,\mu)\rightarrow W^\dagger(r,\mu)G_\mathcal{O}(r,\mu)W[\mathcal{O}^{-1}(r,\mu)]$. Let us choose a gauge
\begin{equation}
\label{eq:local}
    W(r,\mu)=g_3(r_1\theta_x+r_2\theta_y+r_3\theta_z+\phi_\mu).
\end{equation}
It is easy to see that this transformation does not change the structure of the translational gauges except they acquire three global phases along the three translation directions, i.e., $G_{T_1}(r,\mu)=g_3(\theta_x)$, $G_{T_2}(r,\mu)=g_3(-r_1\chi_1+\theta_y)$ and  $G_{T_3}(r,\mu)=g_3((r_1-r_2)\chi_1+\theta_z)$. These global phases do not have any consequence for the \textit{Ans\"atze} and thus can be neglected. However, with the gauge transformation~\eqref{eq:local}, the gauges corresponding to the screw rotation and rotoinversion are now given by
\begin{widetext}
	\begin{equation}
	\left.\begin{aligned}
	\tilde{\phi}_{S_4}(r,\mu)=&r_1(-\theta_x+\chi_{ST_1}+(-1)^{w_{S_4}}\theta_y)+r_2(-\theta_y+\chi_{ST_2}+(-1)^{w_{S_4}}(\theta_y-\theta_z))+r_3(-\theta_z+\chi_{ST_3}+(-1)^{w_{S_4}}(\theta_y-\theta_x))\\
	&-(-1)^{w_{S_4}}\left[r_2\left(\frac{r_2-1}{2}+3r_3-\bar{\mu}\right)+r_3\left(r_1+\frac{r_3-1}{2}\right)\right]\chi_1,\\
	\tilde{\phi}_{\bar{C}_3}(r,\mu)=&r_1(-\theta_x-(-1)^{w_{\bar{C}_3}}(\chi_{\bar{C}_3T_3}+\theta_z))+r_2(-\theta_y-(-1)^{w_{\bar{C}_3}}(\chi_{\bar{C}_3T_1}+\theta_x))+r_3(-\theta_z-(-1)^{w_{\bar{C}_3}}(\chi_{\bar{C}_3T_2}+\theta_y))\\
	&-r_1(r_2-r_3)\chi_1,\\
\end{aligned}\right. 	
	\end{equation}
\end{widetext}
where $\theta_{x}$, $\theta_{y}$, and $\theta_{z}$ can be chosen such that 
\begin{equation}
\label{eq:s4_fixing}
\chi_{ST_1}=\chi_{ST_2}=\chi_{\bar{C}_3T_3}=0\,.
\end{equation}
Accordingly, $\chi_{ST_3}$, $\chi_{\bar{C}_3T_1}$, and $\chi_{\bar{C}_3T_2}$ will also be modified, however, in the new definition, we still denote them by $\chi_{ST_3}$, $\chi_{\bar{C}_3T_1}$, and $\chi_{\bar{C}_3T_2}$. Using the fixing in Eqs.~\eqref{eq:s4_fixing}, \eqref{eq:s4_trans_3}, and~\eqref{eq:s4_trans_4} give
\begin{enumerate}
    \item For $w_{S_4}=0$
    \begin{equation}
    \label{eq:s4_trans_5}
      \chi_{ST_3}=-\chi_1.
    \end{equation}
    \item For $w_{S_4}=1$
    \begin{equation}
    \label{eq:s4_trans_6}
        \chi_{ST_3}-4(\theta_{1}-\theta_0)=\chi_1.
    \end{equation}
\end{enumerate}

There are four more conditions given by Eqs.~\eqref{eq:id_gauge_relation}(j), \eqref{eq:id_gauge_relation}(k), \eqref{eq:id_gauge_relation}(l), and \eqref{eq:id_gauge_relation}(m). We will now consider these for all the four cases $(w_{\bar{C}_3},w_{S_4})=(0,0)$, $(w_{\bar{C}_3},w_{S_4})=(0,1)$, $(w_{\bar{C}_3},w_{S_4})=(1,0)$, and $(w_{\bar{C}_3},w_{S_4})=(1,1)$ separately in the following.

\subsubsection{$w_{\bar{C}_3}=0,\;w_{S_4}=0$}
Using Eq.~\eqref{eq:id_gauge_relation}(j), we can write
\begin{equation}
\left.\begin{aligned}
    \phi_{S_4}(r,\mu)+\phi_{\bar{C}_3}[S^{-1}_4(r,\mu)]&+\phi_{S_4}[S_4\bar{C}_3(r,\mu)]\\
    &+\phi_{\bar{C}_3}[\bar{C}_3(r,\mu)]=\chi_{cs1}.\\
    \end{aligned}\right. 
\end{equation}
This imposes the following constraints:
\begin{equation}
\label{eq:cs1_00}
\left.\begin{aligned}
    &2\chi_{\bar{C}_3T_2}=\chi_{\bar{C}_3T_1}=0\, ,\\
    &2(\rho_{1}-\rho_{0}+\theta_0-\theta_1)=0.\\
    \end{aligned}\right. 
\end{equation}

Now, Eq.~\eqref{eq:id_gauge_relation}(k) leads to the following constraint:
\begin{equation}
\left.\begin{aligned}
    \label{eq:cs2_00}
    &\chi_{\bar{C}_3T_2}=0\, .\\
\end{aligned}\right. 
\end{equation}
Equations~\eqref{eq:id_gauge_relation}(l) and \eqref{eq:id_gauge_relation}(m) do not impose any further constraints. Exploiting IGG gauge freedom we can set $\rho_0=\theta_0=0$. The sublattice dependent gauge transformation does not give any further fixing. From Eq.~\eqref{eq:cs1_00}, we have $\rho_1=\theta_1+m_1\pi$.

\subsubsection{$w_{\bar{C}_3}=0,\;w_{S_4}=1$}
Equation~\eqref{eq:id_gauge_relation}(j) gives
\begin{equation}
\left.\begin{aligned}
    \phi_{S_4}(r,\mu)-\phi_{\bar{C}_3}[S^{-1}_4(r,\mu)]&-\phi_{S_4}[S_4\bar{C}_3(r,\mu)]\\
    &+\phi_{\bar{C}_3}[\bar{C}_3(r,\mu)]=\chi_{cs1},\\
    \end{aligned}\right. 
\end{equation}
\begin{equation}
\label{eq:cs1_01}
\implies    \chi_{\bar{C}_3T_1}=0\,.
\end{equation}
Equation~\eqref{eq:id_gauge_relation}(k) together with Eq.~\eqref{eq:s4_trans_6} gives
\begin{subequations}
	\label{eq:cs2_01}
	\begin{align}
		& \chi_{\bar{C}_3T_2}-\chi_{ST_3}+\chi_1+6(\theta_{1}-\theta_0)=0\ , \; \\
		& \chi_{\bar{C}_3T_2}+2(\theta_{1}-\theta_0)=0\ , \; \\
		&4\chi_{\bar{C}_3T_2}+2\chi_{ST_3}=2\chi_1\, .
	\end{align}
\end{subequations}
Equation~\eqref{eq:id_gauge_relation}(l) yields
\begin{equation}
    \label{eq:cs3_01}
    \chi_{\bar{C}_3T_2}+\chi_{ST_3}+\chi_1=0\, .
\end{equation}
From Eqs.~\eqref{eq:cs2_01}(c) and \eqref{eq:cs3_01}, we get $2\chi_{\bar{C}_3T_2}=0$. Using this in Eq.~\eqref{eq:cs2_01}(b), we get $4(\theta_1-\theta_0)=0$ and hence from Equation~\eqref{eq:s4_trans_6}, we have $\chi_{ST_3}=\chi_1$.
Eq.~\eqref{eq:id_gauge_relation}(m) does not give any further constraints.

\subsubsection{$w_{\bar{C}_3}=1,\;w_{S_4}=0$}
For $w_{\bar{C}_3}=1,\;w_{S_4}=0$, Eq.~\eqref{eq:id_gauge_relation}(j) imposes a constraint similar to that given by Eq.~\eqref{eq:cs1_01}.
Equation~\eqref{eq:id_gauge_relation}(k) in conjunction with Eq.~\eqref{eq:cb3_4}, i.e., $6(\rho_1-\rho_0)=0$ and $\chi_{\bar{C}_3T_1}+\chi_{\bar{C}_3T_2}+\chi_{\bar{C}_3T_3}=0$, gives
\begin{equation}
\label{eq:cs2_10}
\left.\begin{aligned}
&\chi_{\bar{C}_3T_2}=0\\
    \end{aligned}\right. 
\end{equation}
Equations~\eqref{eq:id_gauge_relation}(l) and \eqref{eq:id_gauge_relation}(m) do not impose any further constraints.

\subsubsection{$w_{\bar{C}_3}=1,\;w_{S_4}=1$}
Equation~\eqref{eq:id_gauge_relation}(j) gives 
\begin{equation}
\label{eq:cs1_11}
\left.\begin{aligned}
&2\chi_{\bar{C}_3T_2}=2\chi_{ST_3}=\chi_{\bar{C}_3T_1},\\
&\chi_{\bar{C}_3T_1}=2(\theta_{1}-\theta_0+\rho_1-\rho_{0}).\\
    \end{aligned}\right. 
\end{equation}
Using Eqs.~\eqref{eq:s4_trans_6} and \eqref{eq:cb3_4}, Eq.~\eqref{eq:id_gauge_relation}(k) gives
\begin{equation}
\label{eq:cs2_11}
\left.\begin{aligned}
&2\chi_{\bar{C}_3T_1}=2(\chi_{\bar{C}_3T_2}+\chi_{ST_3}),\\
&\chi_{\bar{C}_3T_2}-\chi_{ST_3}+6(\theta_1-\theta_0)=\chi_1,\\
&\chi_{\bar{C}_3T_1}=-\chi_{\bar{C}_3T_2}=2(\theta_1-\theta_0)\, .\\
    \end{aligned}\right. 
\end{equation}
Equation~\eqref{eq:id_gauge_relation}(l) does not lead to any new constraint while 
Eq.~\eqref{eq:id_gauge_relation}(m) gives
\begin{equation}
    \label{eq:cs4_11}
\left.\begin{aligned}
   & 3\chi_{\bar{C}_3T_1}=0,\\
   & 2\chi_{\bar{C}_3T_2}+\chi_{ST_3}=\chi_1.
\end{aligned}\right. 
\end{equation}
Further simplification gives
\begin{equation}
	\label{eq:cs_psg_11}
	\left.\begin{aligned}
		& 3\chi_{\bar{C}_3T_1}=3\chi_{\bar{C}_3T_2}=6(\theta_1-\theta_0)=0\, ,\\
		& \chi_{ST_3}=\chi_{\bar{C}_3T_2}+\chi_1.
	\end{aligned}\right. 
\end{equation}
Furthermore, using IGG gauge freedom we can set $\rho_0=0,\;\rho_1=\rho_c$. With the help of these gauge fixings, we can simplify all the gauge equations and we have listed the simplified forms in Sec.~\ref{sec:classification}.

\subsection{Derivation of time-reversal PSG}
Now, we proceed to find the PSG solutions for time-reversal symmetry. Using Eq.~\eqref{eq:id_gauge_relation}(o) for $\mathcal{O}\in T_i$, we get
\begin{equation}
\label{eq:tr_1}
\left.\begin{aligned}
&\Delta_1\phi_{\mathcal{T}}(r,\mu)=\chi_{\mathcal{T}T_1},\\
&\Delta_2\phi_{\mathcal{T}}(r,\mu)=\chi_{\mathcal{T}T_2}-[1-(-1)^{w_{\mathcal{T}}}]2r_1\chi_1,\\
&\Delta_3\phi_{\mathcal{T}}(r,\mu)=\chi_{\mathcal{T}T_3}+[1-(-1)^{w_{\mathcal{T}}}]2(r_1-r_2)\chi_1.\\
\end{aligned}\right.
\end{equation}
We consider the $w_{\mathcal{T}}=0$ and $w_{\mathcal{T}}=1$ separately in the following.
\subsubsection{$w_{\mathcal{T}}=0$}
Let us first consider the case $w_{\mathcal{T}}=0$. Here, the consistency condition similar to Eq.~\eqref{eq:consistent} does not impose any new constraint on $\chi_1$ and the form of the solution can be written readily as
\begin{equation}
    \label{eq:tr_2}
    \phi_\mathcal{T}(r,\mu)=r_1\chi_{\mathcal{T}T_1}+r_2\chi_{\mathcal{T}T_2}+r_3\chi_{\mathcal{T}T_3}+\phi_\mathcal{T}(0,\mu)\, .
\end{equation}
 Now Eq.~\eqref{eq:id_gauge_relation}(n) requires
\begin{equation}
    \label{eq:tr_3}
    \chi_{\mathcal{T}T_1},\chi_{\mathcal{T}T_2},\chi_{\mathcal{T}T_3}=0,\pi.
\end{equation}
Now, the condition~\eqref{eq:id_gauge_relation}(o) for $\mathcal{O}=\bar{C}_3$ can be written as 
\begin{equation}
    \phi_\mathcal{T}(r,\mu)-(-)^{w_{\bar{C}_3}}\phi_\mathcal{T}[\bar{C}^{-1}_3(r,\mu)]=\chi_{\mathcal{T}\bar{C}_3}\, .
\end{equation}
This results in 
\begin{equation}
\label{eq:tr_4}
\left.\begin{aligned}
&\chi_{\mathcal{T}T_1}=\chi_{\mathcal{T}T_2}=\chi_{\mathcal{T}T_3}=0,\pi;\\
&\phi_\mathcal{T}(0,\mu)-(-1)^{w_{\bar{C}_3}}\phi_\mathcal{T}(0,\bar{\mu})=\chi_{\mathcal{T}\bar{C}_3}.\\
\end{aligned}\right.
\end{equation}
The condition~\eqref{eq:id_gauge_relation}(o) for $\mathcal{O}=S_4$ takes the following form:
\begin{equation}
    \phi_\mathcal{T}(r,\mu)-(-)^{w_{S_4}}\phi_\mathcal{T}[S^{-1}_4(r,\mu)]=\chi_{\mathcal{T}S}.
\end{equation}
Using $\chi_{\mathcal{T}T_1}=\chi_{\mathcal{T}T_2}=\chi_{\mathcal{T}T_3}$, from Eq.~\eqref{eq:tr_4} this results in 
\begin{equation}
\label{eq:tr_5}
\left.\begin{aligned}
&\chi_{\mathcal{T}T_1}=\chi_{\mathcal{T}T_2}=\chi_{\mathcal{T}T_3}=0,\\
&\phi_\mathcal{T}(0,\mu)-(-1)^{w_{S_4}}\phi_\mathcal{T}(0,\bar{\mu})=\chi_{\mathcal{T}S}\, .\\
\end{aligned}\right.
\end{equation}
Therefore, in this case, the projective gauge can be written as
\begin{equation}
    \label{eq:tr_6}
    G_\mathcal{T}(r,\mu)=g_3(\phi_\mathcal{T}(0,\mu)).
\end{equation}
To have non vanishing \textit{Ans\"atze} on the 1NN bonds, we need to set
\begin{equation}
    \label{eq:tr_7}
    \phi_\mathcal{T}(0,0)=0,\;\phi_\mathcal{T}(0,1)=\pi.
\end{equation}
It is easy to check that with this setting the \textit{Ans\"atze} on 2NN bonds vanish. 

\subsubsection{$w_{\mathcal{T}}=1$}
Now, we proceed with the case $w_\mathcal{T}=1$. Employing consistency condition equivalent to Eq.~\eqref{eq:consistent} we obtain $2\chi_1=0$. Therefore the class with $\chi_1=\pi/2$ is excluded. Also, we find Eq.~\eqref{eq:id_gauge_relation}(n) does not impose any new restriction. In this case, also the projective gauge corresponding to time reversal can be written as
\begin{equation}
    \label{eq:tr_8}
    \phi_\mathcal{T}(r,\mu)=r_1\chi_{\mathcal{T}T_1}+r_2\chi_{\mathcal{T}T_2}+r_3\chi_{\mathcal{T}T_3}+\phi_\mathcal{T}(0,\mu).
\end{equation}
The Eq.~\eqref{eq:id_gauge_relation}(o) for $\mathcal{O}=S_4$ and $\mathcal{O}=\bar{C}_3$ gives the following two relations:
\begin{equation}
\label{eq:tr_9}
\left.\begin{aligned}
&\phi_\mathcal{T}(r,\mu)-(-)^{n_{S_4}}\phi_\mathcal{T}[S^{-1}_4(r,\mu)]-2\theta_{S_4}(r,\mu)=\chi_{\mathcal{T}S_4},\\
&\phi_\mathcal{T}(r,\mu)-(-)^{n_{\bar{C}_3}}\phi_\mathcal{T}[\bar{C}^{-1}_3(r,\mu)]-2\theta_{\bar{C}_3}(r,\mu)=\chi_{\mathcal{T}\bar{C}_3}\, .\\
\end{aligned}\right.
\end{equation}
Using the above relations, $\chi_{\mathcal{T}T_i}$ can be fixed as $\chi_{\mathcal{T}T_i}=0$. The PSGs for $w_\mathcal{T}=0,1$ are summarized in Table~\ref{table:u1_psg}.

\section{$\mathbf{\mathds{Z}_2}$ PSG}
\label{app:z2_psg_derivation}
\subsection{Canonical forms of {\it{Ansatz}} and corresponding gauges.}
When the $SU(2)$ IGG is completely broken down to $\mathds{Z}_2$, the $SU(2)$ flux operators must be non collinear. Therefore the \textit{Ans\"atze} admit pairing terms along with the hopping terms. Thus the canonical form of the $\mathds{Z}_{2}$ \textit{Ans\"atze} is given by
\begin{equation}
	\label{eq:canonical_z2}
	u_{ij}=\dot{\iota}\text{Im}h_{ij}\tau^0+\text{Re}h_{ij}\tau^3+\text{Re}p_{ij}\tau^1+\text{Im}p_{ij}\tau^2,
\end{equation}
where the first two terms represent the imaginary and real hoppings, respectively, while the last two term represent real and imaginary pairing, respectively. For a $\mathds{Z}_2$ IGG, the identity element of the symmetry group is defined up to a sign parameter. The different possible choices of sign parameters then correspond to the different PSGs.
\subsection{Derivation of $\mathbf{\mathds{Z}_2}$ PSG corresponding to space group generators}
Similar to the $U(1)$ case, using local gauge redundancy, the relation~\eqref{eq:id_gauge_relation}(a) leads to the following solution for the projective gauges of $\mathcal{O}\in\{T_1,T_2,T_3\}$
\begin{equation}
	\label{eq:tran_1_z2}
	\left.\begin{aligned}
		&G_{T_1}(r_1,r_2,r_3,\mu)=\tau^0,\\
		&G_{T_2}(r_1,r_2,r_3,\mu)=\eta^{r_1}_1\tau^0,\\
		&G_{T_3}(r_1,r_2,r_3,\mu)=\eta^{r_1}_{3}\eta^{r_2}_{2}\tau^0.\\
	\end{aligned}\right.
\end{equation}
Now, Eq.~\eqref{eq:id_gauge_relation}(b) gives
\begin{equation}
	\label{}
	\left.\begin{aligned}
		G_{T_i}(r,\mu)G^{-1}_{\bar{C}_3}[T_{i+1}\bar{C}_3(r,\mu)]&G_{T_{i+1}}[T_{i+1}\bar{C}_3(r,\mu)]\\
		&G_{\bar{C}_3}[\bar{C}_3(r,\mu)]=\eta_{\bar{C}_3T_i}\tau^0,
	\end{aligned}\right.
\end{equation}
where $\eta_{\bar{C}_3T_i}=\pm1$. With the definition $f_i[G_{\bar{C}_3}(r,\mu)]=G_{\bar{C}_3}(r,\mu)G^{-1}_{\bar{C}_3}[T^{-1}_i(r,\mu)]$, the above equation after substituting Eq.~\eqref{eq:tran_1_z2} results in
\begin{equation}
	\label{eq:cb3_z2}
	\left.\begin{aligned}
		&f_1[G_{\bar{C}_3}(r,\mu)]=\eta_{\bar{C}_3T_3}\eta^{r_2}_{3}\eta^{r_3}_{2}\tau^0,\\
		&f_2[G_{\bar{C}_3}(r,\mu)]=\eta_{\bar{C}_3T_1}\eta^{r_1}_1\tau^0,\\
		&f_3[G_{\bar{C}_3}(r,\mu)]=\eta_{\bar{C}_3T_2}\eta^{r_2}_1\eta^{r_1}_{3}\eta^{r_2}_{2}\tau^0.\\
	\end{aligned}\right.
\end{equation}
The consistency equation reads as
\begin{equation}
	\label{eq:consistent_z2}
	\left.\begin{aligned}
		f_i[G_{\bar{C}_3}(r,\mu)]&f_{i+1}[G_{\bar{C}_3}[T^{-1}_i(r,\mu)]]\\
		&=f_{i+1}[G_{\bar{C}_3}(r,\mu)]f_{i}[G_{\bar{C}_3}[T^{-1}_{i+1}(r,\mu)]],\\
	\end{aligned}\right.
\end{equation}
which requires that $\eta_1=\eta_{2}=\eta_{3}$ and therefore $G_{\bar{C}_3}$ has the following solution:
\begin{equation}
	\label{eq:cb3_z3}
	G_{\bar{C}_3}(r,\mu)=\eta^{r_1}_{\bar{C}_3T_3}\eta^{r_2}_{\bar{C}_3T_1}\eta^{r_3}_{\bar{C}_3T_2}\eta^{r_1(r_2+r_3)}_1g_{\bar{C}_3}(\mu).
\end{equation}
Here, $g_{\bar{C}_3}(\mu)= G_{\bar{C}_3}(0,0,0,\mu)$. Now, plugging this result in Eq.~\eqref{eq:id_gauge_relation}(c), we obtain
\begin{equation}
	\label{eq:cb3_z4}
	[g_{\bar{C}_3}(\mu)g_{\bar{C}_3}(\bar{\mu})]^3=\eta_{\bar{C}_3}\tau^0.
\end{equation}
Now, let us consider the screw symmetry gauge $G_{S_4}$, where, Eqs.~\eqref{eq:id_gauge_relation}(d),~\eqref{eq:id_gauge_relation}(e), and~\eqref{eq:id_gauge_relation}(f) give
\begin{equation}
	\label{eq:s4_z1}
	\left.\begin{aligned}
		&f_1[G_{S_4}(r,\mu)]=\eta_{ST_1}\eta^{r_3}_1\tau^0,\\
		&f_2[G_{S_4}(r,\mu)]=\eta_{ST_2}\eta^{r_2+r_3-1-\bar{\mu}}_1\tau^0,\\
		&f_3[G_{S_4}(r,\mu)]=\eta_{ST_3}\eta^{r_1+r_2-r_3+1}_1\tau^0.\\
	\end{aligned}\right.
\end{equation}
The consistency condition similar to Eq.~\eqref{eq:consistent_z2} does not impose any restriction on $\eta_1$ and the solution of $G_{S_4}(r,\mu)$ is given by
\begin{equation}
	\label{eq:s4_z2}
	\left.\begin{aligned}
		G_{S_4}(r,\mu)&=\eta^{r_1}_{ST_1}\eta^{r_2}_{ST_2}\eta^{r_3}_{ST_3}\\
		&\times\eta^{r_3(r_1+r_2)-r_2\bar{\mu}+r_2\left(\frac{r_2-1}{2}\right)-r_3\left(\frac{r_3-1}{2}\right)}_1g_{S_4}(\mu).\\
	\end{aligned}\right.
\end{equation}
We find that plugging this result in Eq.~\eqref{eq:id_gauge_relation}(g) and Eq.~\eqref{eq:id_gauge_relation}(h) does not impose any new constraints.
Now, Eq.~\eqref{eq:id_gauge_relation}(i) gives
\begin{equation}
	\left.\begin{aligned}
		&G_{T_3}(r_1,r_2,r_3,\mu)G^{-1}_{T_1}(r_1+1,r_2,r_3-1,\mu)\\
		&G^{-1}_{T_2}(r_1+1,r_2+1,r_3-1,\mu)G_{S_4}(r_1+1,r_2+1,r_3-1,\mu)\\
		&G_{S_4}(-r_3+1,r_1+r_2+r_3+\mu,-r_2-1,\bar{\mu},\bar{\mu})\\
		&G_{S_4}(r_2+1,r_1,-r_1-r_2-r_3-\mu,\mu)\\
		&G_{S_4}(r_1+r_2+r_3+\mu,-r_3,-r_1,\bar{\mu})=\eta_{s3}\tau^0.\\
	\end{aligned}\right.   
\end{equation}
On substituting Eqs.~\eqref{eq:s4_z2} and \eqref{eq:tran_1_z2} in the above equation, we get the following constraints.
\begin{enumerate}
	\item For $\mu=0$
	\begin{equation}
		\label{eq:s4_trans_z3}
		[g_{S_4}(0)g_{S_4}(1)]^2=\eta_1\eta_{ST_1}\eta_{ST_2}\eta_{s3}\tau^0.
	\end{equation}
	\item For $\mu=1$
	\begin{equation}
		\label{eq:s4_trans_z4}
		[g_{S_4}(1)g_{S_4}(0)]^2=\eta_{ST_3}\eta_{s3}\tau^0.
	\end{equation}
\end{enumerate}
As $[g_{S_4}(0)g_{S_4}(1)]^2=[g_{S_4}(1)g_{S_4}(0)]^2$ irrespective of whether $g_{S_4}(0)$ and $g_{S_4}(1)$ commute or not, we get
\begin{equation}
	\label{eq:s4_trans_z44}
	\eta_{ST_3}\eta_{ST_1}\eta_{ST_2}=\eta_1\, .
\end{equation}

Further simplification can be achieved by considering the following gauge transformation:
\begin{equation}
	\label{eq:local_z2}
	W(r,\mu)=\eta^{r_1}_x\eta^{r_2}_y\eta^{r_3}_z\tau^0.
\end{equation}

We find that the above transformation does not change the structure of the translation gauges except a global sign modification and these can be put to their original form by a redefinition of these modified signs. Similar to $SU(2)$ and $U(1)$, these signs can be neglected. Also, it can be seen that the gauge transformation given by Eq.~\eqref{eq:local_z2} modulates the signs $\eta_{ST_i},\eta_{\bar{C}_3T_i}$. By a proper choice of constants $\eta_x$, $\eta_y$ and $\eta_z$ and using Eq.~\eqref{eq:s4_trans_z44}, we can fix $\eta_{ST_1}$,  $\eta_{ST_2}$, and $\eta_{\bar{C}_3T_3}$ as
\begin{equation}
	\label{eq:s4_fixing_z2}
	\eta_{ST_1}=\eta_{ST_2}=\eta_{\bar{C}_3T_3}=+1,\;\eta_{ST_3}=\eta_{1}\, .
\end{equation}
Also, we have another gauge freedom which is independent of lattice coordinates but depends on the sublattice indices, i.e., $W(r,\mu)=g(\mu)$. With this transformation $g_{S_4}(\mu)$ transforms as
\begin{equation}
	\left.\begin{aligned}
		&g_{S_4}(0)\rightarrow g^\dagger(0)g_{S_4}(0)g(1),\\
		&g_{S_4}(1)\rightarrow g^\dagger(1)g_{S_4}(1)g(0).\\
	\end{aligned}\right.   
\end{equation}
By a proper choice of $\{g(0),g(1)\}$ we can do the following simplification:
\begin{equation}
	\label{eq:s4_fixing_z2_sublat}
	g_{S_4}(0)=\tau^0,\;\;g_{S_4}(1)=g_s,\;\;g_s\in SU(2)\, .
\end{equation}
More constraints are found from the remaining four conditions given by Eqs.~\eqref{eq:id_gauge_relation}(j), \eqref{eq:id_gauge_relation}(k), \eqref{eq:id_gauge_relation}(l), and \eqref{eq:id_gauge_relation}(m). 
From Eq.~\eqref{eq:id_gauge_relation}(j), we have
\begin{subequations}
	\label{eq:cs1_z2}
	\begin{align}
		&\mu = 0 : G_{S_4}(r_1,r_2,r_3,0) G_{\bar{C_3}}(-r_3,r_1+r_2+r_3-1,-r_2,1) \; \notag \\
		&G_{S_4}(-r_1-r_2-r_3+1,r_2,r_3,0) \; \notag \\
		&G_{\bar{C_3}}(-r_3,-r_1,-r_2,1) =\eta_{cs1}\tau^0 \; \notag \\
		&\Rightarrow [g_{S_4}(0) g_{\bar{C_3}}(1)]^2 = \eta_{cs1} \tau^0,\;\text{with} \hspace{2mm} \eta_{\bar{C}_3T_1}  =+1 \; , \\ 
		&\mu = 1 : G_{S_4}(r_1,r_2,r_3,1) G_{\bar{C_3}}(-r_3,r_1+r_2+r_3,-r_2,0) \; \notag \\
		&G_{S_4}(-r_1-r_2-r_3,r_2,r_3,1) G_{\bar{C_3}}(-r_3,-r_1,-r_2,0)=\eta_{cs1}\tau^0 \; \notag \\
		&\Rightarrow [g_{S_4}(1) g_{\bar{C_3}}(0)]^2 = \eta_{cs1} \tau^0,\; \text{with} \hspace{2mm} \eta_{\bar{C}_3T_1}  =+1 \;. 
	\end{align}
\end{subequations}

From Eq.~\eqref{eq:id_gauge_relation}(k), we have
\begin{subequations}
	\label{eq:cs2_z2}
	\begin{align}
		&\mu = 0:
		G_{\bar{C_3}}(r_1,r_2,r_3,0) G_{\bar{C_3}}(-r_2,-r_3,-r_1,1)  \; \notag \\
		& G_{S_4}(r_3,r_1,r_2,0)G_{S_4}(-r_2,r_1+r_2+r_3-1,-r_1,1) \; \notag \\
		& G_{\bar{C_3}}(r_1,r_3-1,-r_1-r_2-r_3+1,0) \; \notag \\
		& G_{\bar{C_3}}(-r_3+1,r_1+r_2+r_3-1,-r_1,1) \; \notag \\
		& G_{S_4}(-r_1-r_2-r_3+1,r_1,r_3-1,0) \; \notag \\
		& G_{S_4}(-r_3+1,-r_2-1,-r_1,1) \; \notag \\
		& G_{\bar{C_3}}(r_1,-r_1-r_2-r_3,r_2+1,0) \; \notag \\
		& G_{\bar{C_3}}(r_1+r_2+r_3,-r_2-1,-r_1,1) \; \notag \\
		& G_{S_4}(r_2+1,r_1,-r_1-r_2-r_3,0) \; \notag \\
		&G_{S_4}(r_1+r_2+r_3,-r_3,-r_1,1)=\eta_{cs2}\tau^0 \; \notag \\
		&\Rightarrow [g_{\bar{C_3}}(0)g_{\bar{C_3}}(1)g_{S_4}(0)g_{S_4}(1)]^3 = \eta_{cs2} \tau^0 \;, \\
		&\mu = 1: G_{\bar{C_3}}(r_1,r_2,r_3,1) G_{\bar{C_3}}(-r_2,-r_3,-r_1,0)  \; \notag \\
		& G_{S_4}(r_3,r_1,r_2,1)G_{S_4}(-r_2,r_1+r_2+r_3,-r_1,0) \; \notag \\
		&G_{\bar{C_3}}(r_1,r_3-1,-r_1-r_2-r_3,1) \; \notag \\
		&G_{\bar{C_3}}(-r_3+1,r_1+r_2+r_3,-r_1,0) \; \notag \\
		& G_{S_4}(-r_1-r_2-r_3,r_1,r_3-1,1) \; \notag \\
		& G_{S_4}(-r_3+1,-r_2-1,-r_1,0) \; \notag \\
		& G_{\bar{C_3}}(r_1,-r_1-r_2-r_3-1,r_2+1,1) \; \notag \\
		& G_{\bar{C_3}}(r_1+r_2+r_3+1,-r_2-1,-r_1,0) \; \notag \\
		& G_{S_4}(r_2+1,r_1,-r_1-r_2-r_3-1,1) \; \notag \\
		&G_{S_4}(r_1+r_2+r_3+1,-r_3,-r_1,0)=\eta_{cs2}\tau^0  \; \notag \\
		&\Rightarrow [g_{\bar{C_3}}(1)g_{\bar{C_3}}(0)g_{S_4}(1)g_{S_4}(0)]^3 = \eta_{\bar{C}_3T_2}\eta_{cs2} \tau^0  \;.
	\end{align}
\end{subequations}
Similarly from Eq.~\eqref{eq:id_gauge_relation}(l), we have
\begin{subequations}
	\label{eq:cs3_z2}
	\begin{align}
		&\mu = 0 : 
		 [g_{\bar{C_3}}(0)g_{\bar{C_3}}(1)g_{\bar{C_3}}(0)g_{S_4}(1)]^4 =\eta_{cs3} \tau^0 \; , \\
		&\mu = 1 : 
		[g_{\bar{C_3}}(1)g_{\bar{C_3}}(0)g_{\bar{C_3}}(1)g_{S_4}(0)]^4 =\eta_{\bar{C}_3T_2}\eta_{cs3} \tau^0 \;.
	\end{align}
\end{subequations}

From Eq.~\eqref{eq:id_gauge_relation}(m), we have

\begin{equation}
	\label{eq:cs4_z2}
	[g_{\bar{C_3}}(\mu)g_{S_4}(\bar{\mu})g_{S_4}(\mu)g_{\bar{C_3}}(\bar{\mu})g_{S_4}(\mu)g_{S_4}(\bar{\mu})]^3 =\eta_{cs4} \tau^0 \, .
\end{equation}
Let us now find the PSG for time-reversal symmetry. Using Eq.~\eqref{eq:id_gauge_relation}(o) for $\mathcal{O}\in T_i$, we get
\begin{equation}
	\label{eq:tr_z1}
	f_i[G_{\mathcal{T}}(r,\mu)]=\eta_{\mathcal{T}T_i}\tau^0.
\end{equation} 
The consistency condition does not impose any constraint. Thus, from Eq.~\eqref{eq:tr_z1}, the solution for $G_{\mathcal{T}}(r,\mu)$ can be written as
\begin{equation}
	\label{eq:tr_z2}
	G_{\mathcal{T}}(r,\mu)=\eta^{r_1}_{\mathcal{T}T_1}\eta^{r_2}_{\mathcal{T}T_2}\eta^{r_3}_{\mathcal{T}T_3}g_\mathcal{T}(\mu).
\end{equation} 
From Eq.~\eqref{eq:id_gauge_relation}(o) for $\mathcal{O}\in \bar{C}_3$, we have
\begin{subequations}
	\label{eq:ct_z}
	\begin{align}
		&G_{\mathcal{T}}(r_1,r_2,r_3,\mu)G_{\bar{C}_3}(r_1,r_2,r_3,\mu) \; \notag \\
		&\times G^{-1}_{\mathcal{T}}(-r_2,-r_3,-r_1,\bar{\mu})\;\notag \\
  &\times G^{-1}_{\bar{C}_3}(r_1,r_2,r_3,\mu)=\eta_{\mathcal{T}\bar{C}_3}\tau^0 \\
		& \Rightarrow \eta_{\mathcal{T}T_1}=\eta_{\mathcal{T}T_2}=\eta_{\mathcal{T}T_3}\; ; \\
		& g_{\mathcal{T}}(\mu)g_{\bar{C_3}}(\mu)g^{-1}_{\mathcal{T}}(\bar{\mu})g^{-1}_{\bar{C_3}}(\mu) =\eta_{\mathcal{T}\bar{C}_3} \tau^0 \;.
	\end{align}
\end{subequations}
Similarly, for $\mathcal{O}\in {S}_4$, we have
\begin{subequations}
	\label{eq:st_z}
	\begin{align}
		&G_{\mathcal{T}}(r_1,r_2,r_3,\mu)G_{S_4}(r_1,r_2,r_3,\mu) \; \notag \\
		&\times G^{-1}_{\mathcal{T}}(r_1+r_2+r_3+\mu,-r_3,-r_1,\bar{\mu}),\;\notag\\
  & \times G^{-1}_{S_4}r_1,r_2,r_3,\mu)=\eta_{\mathcal{T}S}\tau^0 \; \\
		& \Rightarrow \eta_{\mathcal{T}T_1}=\eta_{\mathcal{T}T_2},\;\;\eta_{\mathcal{T}T_3}=1\; ; \\
		& g_{\mathcal{T}}(\mu)g_{S_4}(\mu)g^{-1}_{\mathcal{T}}(\bar{\mu})g^{-1}_{S_4}(\mu) =\eta_{\mathcal{T}S} \tau^0 \;.
	\end{align}
\end{subequations}
From Eqs.~\eqref{eq:ct_z}(b) and \eqref{eq:st_z}(b), we get $\eta_{\mathcal{T}T_1}=\eta_{\mathcal{T}T_2}=\eta_{\mathcal{T}T_3}=+1$. Also, Eq.~\eqref{eq:id_gauge_relation}(n) gives $[g_\mathcal{T}(\mu)]^2=\eta'_{\mathcal{T}}\tau^0$.\\
Hence, the $\mathds{Z}_2$ PSG solutions can be summarized as
	\begin{subequations}
		\label{eq:psg_sol_z2}
		\begin{align}
			&G_{T_1}(r,\mu)=\eta^{r_1}_1G_{T_2}(r,\mu)=\eta^{r_1+r_2}_1G_{T_3}(r,\mu)=\tau^0,\\
			&G_{\mathcal{T}}(r,\mu)=g_\mathcal{T}(\mu),\\
			&G_{S_4}(r,\mu)=\eta^{r_3}_{ST_3}\nonumber\\
			&\times\eta^{r_3(r_1+r_2)-r_2\bar{\mu}+r_2\left(\frac{r_2-1}{2}\right)-r_3\left(\frac{r_3-1}{2}\right)}_1g_{S_4}(\mu),\\
			&G_{\bar{C}_3}(r,\mu)=\eta^{r_2}_{\bar{C}_3T_1}\eta^{r_3}_{\bar{C}_3T_2}\eta^{r_1(r_2+r_3)}_1g_{\bar{C}_3}(\mu),\\
			&[g_\mathcal{T}(\mu)]^2=\eta'_{\mathcal{T}}\tau^0,\\
			&[g_{\bar{C}_3}(\mu)g_{\bar{C}_3}(\bar{\mu})]^3=\eta_{\bar{C}_3}\tau^0,\\
			&[g_{S_4}(\mu)g_{S_4}(\bar{\mu}) ]^2=\eta_{S}\tau^0,\\
			&[g_{S_4}(\mu) g_{\bar{C_3}}(\bar{\mu})]^2 = \eta_{cs1} \tau^0,\\
			&[g_{\bar{C_3}}(0)g_{\bar{C_3}}(1)g_{S_4}(0)g_{S_4}(1)]^3 = \eta_{cs2} \tau^0,\\
			&[g_{\bar{C_3}}(1)g_{\bar{C_3}}(0)g_{S_4}(1)g_{S_4}(0)]^3 = \eta_{\bar{C}_3T_2}\eta_{cs2} \tau^0,\\
			&[g_{\bar{C_3}}(0)g_{\bar{C_3}}(1)g_{\bar{C_3}}(0)g_{S_4}(1)]^4 =\eta_{cs3} \tau^0,\\
			&[g_{\bar{C_3}}(1)g_{\bar{C_3}}(0)g_{\bar{C_3}}(1)g_{S_4}(0)]^4 =\eta_{\bar{C}_3T_2}\eta_{cs3} \tau^0,\\
			&[g_{\bar{C_3}}(\mu)g_{S_4}(\bar{\mu})g_{S_4}(\mu)g_{\bar{C_3}}(\bar{\mu})g_{S_4}(\mu)g_{S_4}(\bar{\mu})]^3 =\eta_{cs4} \tau^0,\\
			&g_{\mathcal{T}}(\mu)g_{\bar{C_3}}(\mu)g^{-1}_{\mathcal{T}}(\bar{\mu})g^{-1}_{\bar{C_3}}(\mu) =\eta_{\mathcal{T}\bar{C}_3} \tau^0,\\
			&g_{\mathcal{T}}(\mu)g_{S_4}(\mu)g^{-1}_{\mathcal{T}}(\bar{\mu})g^{-1}_{S_4}(\mu) =\eta_{\mathcal{T}S} \tau^0 \;.
		\end{align}
	\end{subequations}
	
	The different sign combinations of $\eta$ parameters enables us to classify different PSGs. Substitution of $g_{S_4}(0)=\tau^0$ in Eq.~\eqref{eq:psg_sol_z2}(g) gives $[g_{S_4}(1)]^2=\eta_{S}\tau^0$. The following two cases are possible:
	\begin{subequations}
		\label{eq:gs4_z2}
		\begin{align}
			&\text{(I):}\hspace{2mm} \eta_S=+1\Rightarrow g_{S_4}(1)=\eta_{S_4}\tau^0 \hspace{2mm}\text{with}\hspace{2mm}\eta_{S_4}=\pm1 \; ,\\
			&\text{(II):}\hspace{2mm} \eta_S=-1\Rightarrow g_{S_4}(1)=\dot{\iota}(\mathbf{a}\cdot{\boldsymbol\tau}) \hspace{2mm}\text{with}\hspace{2mm}|\mathbf{a}|^2=1\;.
		\end{align}
	\end{subequations}
	Now, for Eq.~\eqref{eq:psg_sol_z2}(f), the following four cases are possible. For $\eta_{\bar{C}_3}=+1$,
	\begin{subequations}
		\label{eq:gc3_z2_1}
		\begin{align}
			&\text{(a)}\hspace{2mm} g_{\bar{C_3}}(0)g_{\bar{C_3}}(1)=\tau^0, \; \notag \\
			&\Rightarrow g_{\bar{C_3}}(0)=g^\dagger_{\bar{C_3}}(1)=e^{\dot{\iota}\theta(\mathbf{b}\cdot\boldsymbol{\tau})} \hspace{2mm}\text{with}\hspace{2mm}|\mathbf{b}|^2=1,\\
			&\text{(b)}\hspace{2mm} g_{\bar{C_3}}(0)g_{\bar{C_3}}(1)=e^{\dot{\iota}\frac{2\pi}{3}\tau^\alpha}, \; \notag \\
			&\Rightarrow g_{\bar{C_3}}(0)=e^{\dot{\iota}\zeta\tau^\alpha},\; g_{\bar{C_3}}(1)=e^{\dot{\iota}(\frac{2\pi}{3}-\zeta)\tau^\alpha},\;
		\end{align}
	\end{subequations}
	and for $\eta_{\bar{C}_3}=-1$,
	\begin{subequations}
		\label{eq:gc3_z2_2}
		\begin{align}
			&\text{(c)}\hspace{2mm} g_{\bar{C_3}}(0)g_{\bar{C_3}}(1)=-\tau^0, \; \notag \\
			&\Rightarrow g_{\bar{C_3}}(\mu)=(-1)^\mu\tau^0\;\text{or}\;\dot{\iota}(\mathbf{b}\cdot\boldsymbol{\tau}) \hspace{2mm}\text{with}\hspace{2mm}|\mathbf{b}|^2=1,\\
			&\text{(d)}\hspace{2mm} g_{\bar{C_3}}(0)g_{\bar{C_3}}(1)=e^{\dot{\iota}\frac{\pi}{3}\tau^\alpha}, \; \notag \\
			&\Rightarrow g_{\bar{C_3}}(0)=e^{\dot{\iota}\zeta\tau^\alpha},\; g_{\bar{C_3}}(1)=e^{\dot{\iota}(\frac{\pi}{3}-\zeta)\tau^\alpha}\; ,
		\end{align}
	\end{subequations}
	where $\alpha=1,2,3$ and $0\leq \zeta,\theta\leq2\pi$. Let us first consider the case (I)(a). For this case, Eq.~\eqref{eq:psg_sol_z2}(h) gives
		\begin{align}
			&[g_{\bar{C_3}}(1)]^2=[g_{\bar{C_3}}(0)]^2=\eta_{cs1}\tau^0 \; \notag \\
			&\Rightarrow g_{\bar{C_3}}(0)=g^\dagger_{\bar{C_3}}(1)=\tau^0,\dot{\iota}(\mathbf{b}\cdot\boldsymbol{\tau})\; \label{eq:Ia_z2_cs1} .
		\end{align}
	Equations~\eqref{eq:psg_sol_z2}(i) and~\eqref{eq:psg_sol_z2}(j) give $\eta_{\bar{C_3}T_2}=1$. For both the choices in Eq.~\eqref{eq:Ia_z2_cs1}, Eq.~\eqref{eq:psg_sol_z2}(m) is also satisfied. For the case (I)(c) we also find $\eta_{\bar{C_3}T_2}=1$. Now, let us look into the case (I)(b). Equation~\eqref{eq:psg_sol_z2}(h), for this case reads as follows:
		\begin{align}
			&[g_{\bar{C_3}}(1)]^2=[g_{\bar{C_3}}(0)]^2=\eta_{cs1}\tau^0 \; \notag \\
			&\Rightarrow e^{2\dot{\iota}(\frac{2\pi}{3}-\zeta)\tau^\alpha}=e^{2\dot{\iota}\zeta\tau^\alpha}=\eta_{cs1}\tau^0\;, \label{eq:Ib_z2_cs1}
		\end{align}
	which cannot be satisfied for any choice of $\zeta$ and $\eta_{cs1}$. This happens also for the case (I)(d). Therefore we need to exclude these two cases. Now, let us consider the case (II)(a). It can readily be seen that the substitution of $g_{\bar{C_3}}(0)g_{\bar{C_3}}(1)=\tau^0$ and $g_{S_4}(1)=\dot{\iota}(\mathbf{a}\cdot\boldsymbol{\tau})$ do not satisfy 
	Eqs.~\eqref{eq:psg_sol_z2}(i) and \eqref{eq:psg_sol_z2}(j). A similar observation can be made for (II)(a). Therefore, one needs to exclude these two cases. Now, let us proceed with the case (II)(b). In this scenario, Eqs.~\eqref{eq:psg_sol_z2}(i) and \eqref{eq:psg_sol_z2}(j) give
	\begin{subequations}
		\label{eq:IIb_z2_cs1}
		\begin{align}
			&[e^{\dot{\iota}\frac{2\pi}{3}\tau^\alpha}\dot{\iota}(\mathbf{a}\cdot\boldsymbol{\tau})]^3=\eta_{cs2}\tau^0 , \\
			&[e^{\dot{\iota}\frac{2\pi}{3}\tau^\alpha}\dot{\iota}(\mathbf{a}\cdot\boldsymbol{\tau})]^3=\eta_{\bar{C_3}T_2}\eta_{cs2}\tau^0\;, 
		\end{align}
	\end{subequations}
	which cannot be satisfied for any choice of $\mathbf{a}$. Similarly, the case (II)(d) does not satisfy Eqs.~\eqref{eq:psg_sol_z2}(i) and \eqref{eq:psg_sol_z2}(j). This implies that in order to find screw rotation ($S_4$) and rotoinversion ($\bar{C}_3$) symmetric \textit{Ans\"atze}, one needs to consider only two cases, (I)(a) and (I)(c). Our next task is to find out possible sets of $\{\mathbf{a},\mathbf{b},\eta_{S_4}\}$ which satisfy Eqs.~\eqref{eq:psg_sol_z2}(e), \eqref{eq:psg_sol_z2}(n), and \eqref{eq:psg_sol_z2}(o).
	For Eq.~\eqref{eq:psg_sol_z2}(e), the following two cases are possible:
	\begin{subequations}
		\label{eq:eta_time_z2}
		\begin{align}
			&\text{(A):}\hspace{2mm} \eta'_{\mathcal{T}}=+1\Rightarrow g_{\mathcal{T}}(\mu)=\eta_{\mathcal{T},\mu}\tau^0,\\
			&\text{(B):}\hspace{2mm} \eta'_{\mathcal{T}}=-1\Rightarrow g_{\mathcal{T}}(\mu)=\dot{\iota}(\mathbf{c}_\mu\cdot\boldsymbol{\tau}) \hspace{2mm}\text{with}\hspace{2mm}|\mathbf{c}|^2=1\;.
		\end{align}
	\end{subequations}
	Now, the Eqs.~\eqref{eq:psg_sol_z2}(n) and \eqref{eq:psg_sol_z2}(o) can be rewritten as
	\begin{subequations}
		\label{eq:time_psg_z2_unit cell}
		\begin{align}
			&g_{\mathcal{T}}(\mu)=\eta_{\mathcal{T}\bar{C}_3} g_{\bar{C_3}}(\mu)g_{\mathcal{T}}(\bar{\mu})g^{-1}_{\bar{C_3}}(\mu) , \\
			&g_{\mathcal{T}}(\mu)=\eta_{\mathcal{T}S} g_{S_4}(\mu)g_{\mathcal{T}}(\bar{\mu})g^{-1}_{S_4}(\mu)\; .
		\end{align}
	\end{subequations}
	Upon substituting the case (I), i.e., $g_{S_4}(0)=\tau^0$ and $g_{S_4}(1)=\pm\tau^0$, Eq.~\eqref{eq:time_psg_z2_unit cell}(b) implies $g_{\mathcal{T}}(\mu)=\pm g_{\mathcal{T}}(\bar{\mu})$. We have summarized these results and the solutions given by Eqs.~\eqref{eq:psg_sol_z2}, Eqs.~\eqref{eq:z2_gauge_sol}, and Table~\ref{table:z2_psg}, where we find a total of 80 $\mathds{Z}_2$ PSGs.
	
\section{Definitions of reference bonds}
\label{app:ref_bond}
To represent the mean-field \textit{Ans\"atze}, the point $O(0,0,0,0)$ has been considered as the reference site. There are four first nearest neighbors (1NN) surrounding the reference site, namely, $A(0,0,0,1)$, $B(-1,0,0,1)$, $C(0,-1,0,1)$ and $D(0,0,-1,1)$. We have chosen $u_{OA}\equiv u_1$ as the \textit{Ansatz} on the reference 1NN bond and the \textit{Ansatz} on the other 1NN bonds are obtained by employing various strings of space group operations $\{T_i,S_4,\bar{C}_3\}$ on the reference bond. Around the $O$ point, there are twelve second-nearest neighbors (2NN) among which six sites are given by $(1,0,0,0)$, $(0,1,-1,0)$, $(0,1,0,0)$, $(-1,0,1,0)$, $(0,0,1,0)$, and $(1,-1,0,0)$, and other six can be found from these by translation operations. Let us denote the former six sites by $i=1,2,\cdots,6$ and the corresponding bonds can be denoted as $u^0_{Oi}$. ``$0$'' stands for sublattice $\mu=0$. Similarly, the second NN bond for the second sublattice ``$1$'' can be denoted as $u^1_{Oi}$ and we consider the reference 2NN bond to be $u^0_{O1}=u_2$. Similarly, the twelve third-nearest neighbours (3NN) bonds can be defined by connecting $O(0,0,0,0)$ with the twelve 3NN sites given by $a(1,-1,0,1)$, $b(-1,1,0,1)$, $c(-1,-1,0,1)$, $d(0,1,-1,1)$, $e(0,-1,1,1)$, $f(0,-1,-1,1)$, $g(-1,0,1,1)$, $h(1,0,-1,1)$, $i(-1,0,-1,1)$, $j(1,-1,-1,1)$, $k(-1,1,-1,1)$, and $l(-1,-1,1,1)$, and denoting the bonds by $u_{O\alpha}$ with $\alpha=a,b,c,\cdots,l$. We choose $u_{Oa}=u_3$ to be the reference bond for 3NNs. With reference to these notations, we define the \textit{Ans\"atze} in Table~\ref{table:ansatz} and Sec.~\ref{sec:mean_field_amplitude}.

The pattern of spatial modulation of the \textit{Ans\"atze} are given below for the 1NN, 2NN, and 3NN bonds
\begin{equation}
	\label{eq:sign_str_1nn}
	\left.\begin{aligned}
 &\text{For 1NN bonds,}\\
		&\langle(r_1,r_2,r_3,0)(r_1,r_2,r_3,1)\rangle=u_{OA}=u_1,\\
		&\langle(r_1,r_2,r_3,0)(r_1-1,r_2,r_3,1)\rangle=g_3((-r_2+r_3)\chi_1)u_{OB},\\
		&\langle(r_1,r_2,r_3,0)(r_1,r_2-1,r_3,1)\rangle=g_3((-r_3)\chi_1)u_{OC},\\
		&\langle(r_1,r_2,r_3,0)(r_1,r_2,r_3-1,1)\rangle=u_{OD}.\\
	\end{aligned}\right.
\end{equation}

\begin{equation}
	\label{eq:sign_str_2nn}
	\left.\begin{aligned}
 &\text{For 2NN bonds,}\\
		&\langle(r_1,r_2,r_3,\mu)(r_1+1,r_2,r_3,\mu)\rangle=g_3((r_2-r_3)\chi_1)u^\mu_{O1},\\
		&\langle(r_1,r_2,r_3,\mu)(r_1,r_2+1,r_3-1,\mu)\rangle=g_3(r_3\chi_1)u^\mu_{O2},\\
		&\langle(r_1,r_2,r_3,\mu)(r_1,r_2+1,r_3,\mu)\rangle=g_3(r_3\chi_1)u^\mu_{O3},\\
		&\langle(r_1,r_2,r_3,\mu)(r_1-1,r_2,r_3+1,\mu)\rangle=g_3((-r_2+r_3)\chi_1)u^\mu_{O4},\\
		&\langle(r_1,r_2,r_3,\mu)(r_1,r_2,r_3+1,\mu)\rangle=u^\mu_{O5},\\
		&\langle(r_1,r_2,r_3,\mu)(r_1+1,r_2-1,r_3,\mu)\rangle=g_3((r_2-2r_3)\chi_1)u^\mu_{O6}.\\
	\end{aligned}\right.
\end{equation}

\begin{equation}
	\label{eq:sign_str_3nn}
	\left.\begin{aligned}
 \text{For 3NN bon} &\text{ds ,}\\
		\langle(r_1,r_2,r_3,0)&(r_1+1,r_2-1,r_3,1)\rangle=g_3((r_2-2 r_3)\chi_1)u_{Oa},\\
		\langle(r_1,r_2,r_3,0)&(r_1-1,r_2+1,r_3,1)\rangle=g_3((-r_2+2 r_3)\chi_1)u_{Ob},\\
        \langle(r_1,r_2,r_3,0)&(r_1-1,r_2-1,r_3,1)\rangle=g_3(-r_2\chi_1)u_{Oc},\\
        \langle(r_1,r_2,r_3,0)&(r_1,r_2+1,r_3-1,1)\rangle=g_3(r_3\chi_1)u_{Od},\\
        \langle(r_1,r_2,r_3,0)&(r_1,r_2-1,r_3+1,1)\rangle=g_3(-r_3\chi_1)u_{Oe},\\
        \langle(r_1,r_2,r_3,0)&(r_1,r_2-1,r_3-1,1)\rangle=g_3(-r_3\chi_1)u_{Of},\\
        \langle(r_1,r_2,r_3,0)&(r_1-1,r_2,r_3+1,1)\rangle=g_3((-r_2+r_3)\chi_1)u_{Og},\\
        \langle(r_1,r_2,r_3,0)&(r_1+1,r_2,r_3-1,1)\rangle=g_3((r_2-r_3)\chi_1)u_{Oh},\\
        \langle(r_1,r_2,r_3,0)&(r_1-1,r_2,r_3-1,1)\rangle=g_3((-r_2+r_3)\chi_1)u_{Oi},\\
        \langle(r_1,r_2,r_3,0)&(r_1+1,r_2-1,r_3-1,1)\rangle\\
        &\;\;\;\;\;\;\;\;\;\;\;\;\;\;\;\;\;\;=g_3((r_2-2r_3)\chi_1)u_{Oj},\\
        \langle(r_1,r_2,r_3,0)&(r_1-1,r_2+1,r_3-1,1)\rangle\\
        &\;\;\;\;\;\;\;\;\;\;\;\;\;\;\;\;\;\;=g_3((-r_2+2r_3)\chi_1)u_{Ok},\\
        \langle(r_1,r_2,r_3,0)&(r_1-1,r_2-1,r_3+1,1)\rangle=g_3(-r_2\chi_1)u_{Ol}.\\
	\end{aligned}\right.
\end{equation}

\section{Symmetry conditions on the short-range mean field {\it{Ans\"atze}}}
\label{app:sym_cond_mfa}

In Sec.~\ref{app:ref_bond}, we have chosen some reference bonds and the others can be found through symmetry operations. The \textit{Ans\"atze} on these bonds cannot be chosen arbitrarily, and must follow all the symmetry conditions. In the following, those symmetry conditions are summarized. 

The conditions on the reference bond $u_{OA}=u_1$ are
\begin{subequations}
\label{eq:cond_1nn}
\begin{align}
 \bar{C}_3T_3T_2\bar{C}_3T_3S^4_4:\;&u_{1}\rightarrow u_{1}\ , \; \\
 T_3T_2\bar{C}_3T_3S^4_4:\;&u_{1}\rightarrow u^\dagger_{1}\ , \; \\
 \bar{C}_3:\;&u_{1}\rightarrow u^\dagger_{1}\ , \; \\
 S_4\bar{C}_3S_4:\;&u_{1}\rightarrow u^\dagger_{1}\ , \; \\
 S_4\bar{C}^2_3S_4\bar{C}_3S^2_4:\;&u_{1}\rightarrow u^\dagger_{1}\ , \; \\
 S_4\bar{C}^4_3S_4\bar{C}^3_3S^2_4\bar{C}_3S^3_4:\;&u_{1}\rightarrow u^\dagger_{1} \
.  
\end{align}
\end{subequations}
 $u_{OB}$, $u_{OC}$ and $u_{OD}$ can be found from $u_1$ by using the symmetry operations $ T^{-1}_1S_4:\;u_{1}\rightarrow u^\dagger_{OB}$, $ \bar{C}^2_3S_4\bar{C}_3S^2_4:\;u_{1}\rightarrow u_{OC}$ and $S_4\bar{C}_3S^2_4:\;u_{1}\rightarrow u_{OD}$, respectively.
 
The conditions on the 2NN reference bond $u^0_{O1}=u_2$ are given by
\begin{subequations}
\label{eq:cond_2nn}
\begin{align}
 \bar{C}_3T_3T_2\bar{C}_3T_3S^4_4:\;&u_{2}\rightarrow u_{2}\ , \; \\
 T_1\bar{C}^2_3T^{-1}_1S_4\bar{C}^2_3S_4:\;&u_{2}\rightarrow u^\dagger_{2}\ , \; \\
 S^{-1}_4T_2\bar{C}_3S^2_4:\;&u_{2}\rightarrow u_{2}\ , \; \\
 S_4\bar{C}_3:\;&u_{2}\rightarrow u^\dagger_{2}\ , \; \\
 T_1\bar{C}^{-1}_3S_4\bar{C}^3_3S_4\bar{C}^3_3S_4:\;&u_{2}\rightarrow u^\dagger_{2}\
.  
\end{align}
\end{subequations}
Using $u_2$, $u^\mu_{Oi}$ ($i=2,3,4,5,6$) can be found by the symmetry conditions given by $\bar{C}^5_3S_4 :u_2\rightarrow u^0_{O2}$, $\bar{C}^{-2}_3 :u_2\rightarrow u^0_{O3}$, $\bar{C}^3_3S_4 :u_2\rightarrow u^0_{O4}$, $\bar{C}^2_3 :u_2\rightarrow u^0_{O5}$, $\bar{C}_3S_4 :u_2\rightarrow u^0_{O6}$. 

The conditions on the third NN reference bond $u_{Oa}=u_3$ are given by
\begin{subequations}
\label{eq:cond_3nn}
\begin{align}
 \bar{C}_3S_4:\;&u_{3}\rightarrow u_{3}\ , \; \\
 T^{-1}_2T_1\bar{C}^3_3:\;&u_{3}\rightarrow u^\dagger_{3}\ , \; \\
 \bar{C}^2_3S_4\bar{C}_3S^2_4:\;&u_{3}\rightarrow u_{3}\; .  
\end{align}
\end{subequations}

Using $u_3$, $u_{O\gamma}$ ($\gamma=a,b,..,l$) can be found by the symmetry conditions given by $\bar{C}^2_3 :u_3\rightarrow u_{Og}$, $\bar{C}^4_3 :u_3\rightarrow u_{Od}$, $T^{-1}_1S^2_4 :u_3\rightarrow u_{Ok}$, $\bar{C}^2_3T^{-1}_1S^2_4 :u_3\rightarrow u_{Oj}$, $\bar{C}^4_3T^{-1}_1S^2_4 :u_3\rightarrow u_{Ol}$, $\bar{C}_3S_4\bar{C}^2_3T^{-1}_1S^2_4 :u_3\rightarrow u_{Oh}$, $\bar{C}^3_3S_4\bar{C}^2_3T^{-1}_1S^2_4 :u_3\rightarrow u_{Oe}$, $\bar{C}^5_3S_4\bar{C}^2_3T^{-1}_1S^2_4 :u_3\rightarrow u_{Ob}$, $T^{-1}_1T^{-1}_2S^3_4 :u_3\rightarrow u^\dagger_{Of}$, $\bar{C}^2_3T^{-1}_1T^{-1}_2S^3_4 :u_3\rightarrow u^\dagger_{Oc}$ and $T^{-1}_1T^{-1}_3\bar{C}_3T^{-1}_1T^{-1}_2S^3_4 :u_3\rightarrow u^\dagger_{Oi}$.

\section{Singlet and triplet bond parameters}\label{app:singlet_trplet}

\subsection{SU(2) {\it{Ans\"{a}tze}}}

We write the onsite terms
\begin{equation}
u_{OO} \!=\! \dot{\iota} h\,\tau^0,\! \quad\! 
u^{(x,y,z)}_{OO} = 0,
\end{equation}
the first nearest neighbor terms
\begin{equation}
u_{OA} \!=\! \dot{\iota} h_1\,\tau^0 ,\!\quad\! 
u^{(x)}_{OA} \!=\! h_1^x\,\tau^0 ,\!\quad\! 
u^{(y)}_{OA} \!=\! h_1^y\,\tau^0 ,\!\quad\! 
u^{(z)}_{OA} \!=\! h_1^z\,\tau^0,
\end{equation}
and the second-nearest neighbor terms
\begin{equation}
u_{O1} \!=\! \dot{\iota} h_2\,\tau^0 ,\!\quad\! 
u^{(x)}_{O1} \!=\! h_2^x\,\tau^0 ,\!\quad\! 
u^{(y)}_{O1} \!=\! h_2^y\,\tau^0 ,\!\quad\! 
u^{(z)}_{O1} \!=\! h_2^z\,\tau^0,
\end{equation}
where all the bond parameters are real, and we have introduced the notation $u_{O1} \equiv u_{O1}^0$ in relation to the definition in Appendix~\ref{app:ref_bond}.

The results are summarized in Table \ref{tab:su2_triplet_singlet}.

\begin{table*}
\centering
\begin{tabular}{l|ccc|l}
\hline
\multirow{ 2}{*}{SU(2)-Class ($\eta_1,\eta_{S_4},\eta_{\bar{C}_3}$)}
&\multicolumn{3}{c|}{\multirow{ 1}{*}{Independent nonzero parameters}}&\multirow{ 2}{*}{Constraints (1NN and 2NN)}\\
\cline{2-4}
&Onsite&1NN&2NN&\\
\hline
{{$(+,+,+)$  or $(-,-,+)$}}       & {$h$} & ---& ---&--- \\
{{$(+,+,-)$  or $(-,-,-)$}}       & {$h$} &  ---& ---& ---\\
{{$(+,-,+)$  or $(-,+,+)$}}       & {$h$} & {$h_1^x$} & --- & $h_1^x\!=\!h_1^y\!=\!h_1^z$\\
{{$(+,-,-)$  or $(-,+,-)$}}       & {$h$} & {$h_1$} &--- & ---\\
\hline
\end{tabular}\caption{Summary of the bond parameters for the $SU(2)$ Ans\"{a}tze. All bond parameters not explicitly mentioned have zero values.}\label{tab:su2_triplet_singlet}
\end{table*}

\subsection{U(1) {\it{Ans\"{a}tze}}}

We write the onsite, first nearest neighbor and second nearest neighbor terms as
\begin{equation}
u_{OO} = \dot{\iota} \mathrm{Im} h\, \tau^0+\mathrm{Re}h \,\tau^3,\!\quad \!
u^{(x,y,z)}_{OO} = 0,
\end{equation}
\begin{equation}
\begin{aligned}
u_{OA} &\!=\! \dot{\iota} \mathrm{Im} h_1 \,\tau^0+\mathrm{Re}h_1 \,\tau^3,\!\quad \! u^{(x)}_{OA} \!=\! \mathrm{Re}h_1^x \,\tau^0 +\dot{\iota} \mathrm{Im}h_1^x\, \tau^3,\\
u^{(y)}_{OA} &\!=\! \mathrm{Re}h_1^y \,\tau^0 +\dot{\iota} \mathrm{Im}h_1^y \,\tau^3,\!\quad\!
u^{(z)}_{OA} \!=\! \mathrm{Re}h_1^z \,\tau^0 +\dot{\iota} \mathrm{Im}h_1^z \,\tau^3,
\end{aligned}
\end{equation} 
\begin{equation}
\begin{aligned}
u_{O1} &\!=\! \dot{\iota} \mathrm{Im} h_2 \,\tau^0+\mathrm{Re} h_2 \,\tau^3,\!\quad \! 
u^{(x)}_{O1} \!=\! \mathrm{Re}h_2^x \,\tau^0 +\dot{\iota} \mathrm{Im}h_2^x \,\tau^3,\\
u^{(y)}_{O1} &\!=\! \mathrm{Re}h_2^y \,\tau^0 +\dot{\iota} \mathrm{Im}h_2^y \,\tau^3,\!\quad \!
u^{(z)}_{O1} \!=\! \mathrm{Re}h_2^z \,\tau^0 +\dot{\iota} \mathrm{Im}h_2^z \,\tau^3,
\end{aligned}
\end{equation}

\begin{itemize}
\item For class $(\omega_{\bar{C}_3},\omega_{S_4})=(0,0)$: we have $\theta_1 =\rho_1 +m_1 \pi$, and
\begin{subequations}
\begin{align}
h\text{ arbitrary},\\
h_1={h_1^*} e^{\dot{\iota} (-{\theta_1}-{\chi_1})}={h_1^*} e^{-\dot{\iota}\rho_1},\\
h_1^x=-{h_1^{x*}} e^{\dot{\iota} (-{\theta_1}-{\chi_1})}=h_1^z=h_1^y={h_1^{x*}} e^{ -\dot{\iota}{\rho_1}},\\h_2=h_2^*,\!\quad\!
h_2^x=0,\!\quad\!
h_2^y=-h_2^{y*}=h_2^{z*}
.
\end{align}
\end{subequations}

\item For class $(\omega_{\bar{C}_3},\omega_{S_4})=(0,1)$: this is the class with $\chi_1=0,\pi,\pi/2$. We have \begin{subequations}
\begin{align}
h= -h^*,\\
h_1=-h_1 e^{\dot{\iota} (-3 {\theta_1}+{\chi_1})}={h_1^*} e^{-\dot{\iota}{\rho_1}},\\
h_1^x=-h_1^x e^{\dot{\iota} (-3 {\theta_1}+{\chi_1})}=h_1^z=h_1^y={h_1^{x*}} e^{ -\dot{\iota}{\rho_1}},\\h_2=
h_2^x=0,\quad
h_2^y=-h_2^{y*} e^{2\dot{\iota} \chi_1}=h_2^z.
\end{align}
\end{subequations}

\item For class $(\omega_{\bar{C}_3},\omega_{S_4})=(1,0)$: we have $6 \rho_1=0$, and $\theta_1$ is arbitrary.
\begin{subequations}
\begin{align}
h=-h^*,\\
h_1={h_1^*} e^{\dot{\iota} (-{\theta_1}- {\chi_1})}=-h_1 e^{-\dot{\iota} {\rho_1}},\\
h_1^x\!=\!-{h_1^{x*}} e^{\dot{\iota} (-{\theta_1}+4 {\rho_1}- {\chi_1})}\!=\!h_1^z e^{2\dot{\iota} {\rho_1}}\notag \\
\!=\!h_1^y e^{-2 \dot{\iota} {\rho_1}}\!=\!h_1^x e^{3 \dot{\iota} {\rho_1}},\\
h_2=h_2^x=0,\quad
h_2^y=-h_2^{y*}=h_2^z
.
\end{align}
\end{subequations}
\item For class $(\omega_{\bar{C}_3},\omega_{S_4})=(1,1)$: $2\theta_1=0$ and $2\rho_1=0$. 
\begin{subequations}
\begin{align}
h\text{ arbitrary},\\
h_1=-h_1 e^{\dot{\iota} ({\theta_1}+ {\chi_1})} =-h_1 e^{\dot{\iota} {\rho_1}},\\
h_1^x=-h_1^x e^{\dot{\iota} ({\theta_1}+ {\chi_1})}=h_1^z=h_1^y =h_1^x e^{ \dot{\iota} {\rho_1}},\\
h_2=h_2^*,\!\quad\!
h_2^x=0,\!\quad\!
h_2^y=-h_2^{y*}=h_2^{z*}.
\end{align}
\end{subequations}
\end{itemize}

The results are summarized in Table~\ref{tab:u1_triplet_singlet}.

\begin{table*}
\centering
\begin{tabular}{l|lll|l}
\hline
\multirow{ 2}{*}{U(1)-Classes ($w_{\bar{C}_3},w_{S_4}$)}
&\multicolumn{3}{c|}{\multirow{ 1}{*}{Independent nonzero parameters}}&\multirow{ 2}{*}{Constraints (1NN and 2NN)}\\
\cline{2-4}
&Onsite&1NN&2NN&\\
\hline
$(0,0)$, $m_1 \pi=\chi_1, \rho_1=0$ & $h$ & $\mathrm{Re}h_1$ & $\mathrm{Re}h_2$, $\mathrm{Im}h_2^y$ & $h_2^{z*}=h_2^y$\\
$(0,0)$, $m_1 \pi=\chi_1+\pi, \rho_1=0$ & $h$ & $\mathrm{Re}h_1^x$ &
$\mathrm{Re}h_2$, $\mathrm{Im}h_2^y$ & $h_1^x=h_1^y=h_1^z$, $h_2^{z*}=h_2^y$\\
$(0,1)$, $-3\theta_1+\chi_1=0$, $\chi_1=0$ or $\pi$ & $\mathrm{Re}h$&& $\mathrm{Im}h_2^y$ & $h_2^y=h_2^z$\\
$(0,1)$, $-3\theta_1+\chi_1=\pi$, $\chi_1=0$ or $\pi$ & $\mathrm{Im}h$& $\mathrm{Re}h_1$, $\mathrm{Re}h_1^x$& $\mathrm{Im}h_2^y$ & $h_1^x=h_1^y=h_1^z$, $h_2^y=h_2^z$\\
$(0,1)$, $-3\theta_1+\chi_1=0$, $\chi_1=\frac{\pi}{2}$ & $\mathrm{Im}h$& & $\mathrm{Re}h_2^y$ & $h_2^y=h_2^z$\\
$(0,1)$, $-3\theta_1+\chi_1=0$, $\chi_1=\frac{\pi}{2}$ & $\mathrm{Im}h$& $\mathrm{Re}h_1$, $\mathrm{Re}h_1^x$& $\mathrm{Re}h_2^y$ & $h_1^x=h_1^y=h_1^z$, $h_2^y=h_2^z$\\
$(1,0)$, $\theta_1+\chi_1=0$, $\rho_1=0,\frac{2\pi}{3},\frac{4\pi}{3}$ & $\mathrm{Im}h$& $\mathrm{Re}h_1^x$ & $\mathrm{Im}h_2^y$ & $\mathrm{Arg}(h_1^x) = 2\rho_1$, $h_1^x=h_1^z e^{2i \rho_1}=h_1^y e^{-2i \rho_1}$, $h_2^z=h_2^y$\\
$(1,0)$, $\theta_1+\chi_1=0$, $\rho_1=\frac{\pi}{3},\frac{5\pi}{3}$ & $\mathrm{Im}h$& & $\mathrm{Im}h_2^y$ & $h_2^z=h_2^y$\\
$(1,0)$, $\theta_1+\chi_1=0$, $\rho_1=\pi$ & $\mathrm{Im}h$& $\mathrm{Re}h_1$ & $\mathrm{Im}h_2^y$ & $h_2^z=h_2^y$\\
$(1,1)$, $\theta_1+\chi_1=0$, $\rho_1=0$ or $\pi$ & $h$ && $\mathrm{Re}h_2$, $\mathrm{Im}h_2^y$ & $h_2^y = h_2^{z*}$\\
$(1,1)$, $\theta_1+\chi_1=\pi$, $\rho_1=0$ & $h$ & $h_1^x$ & $\mathrm{Re}h_2$, $\mathrm{Im}h_2^y$ & $h_1^x=h_1^y=h_1^z$, $h_2^y = h_2^{z*}$\\
$(1,1)$, $\theta_1+\chi_1=\pi$, $\rho_1=\pi$ & $h$ & $h_1$ & $\mathrm{Re}h_2$, $\mathrm{Im}h_2^y$ & $h_2^y = h_2^{z*}$\\
\hline
\end{tabular}
\caption{Summary of bond parameters for the $U(1)$ \textit{Ans\"{a}tze}. All bond parameters not explicitly mentioned have zero values.}\label{tab:u1_triplet_singlet}
\end{table*}

\subsection{$\mathbb{Z}_2$ {\it{Ans\"{a}tze}}}

For $\mathbb{Z}_2$ PSG, we always have $\omega_{S_4}=0$, while  $g_{\bar{C}_3}(0)$ can take values of either $\tau^0$, $\dot{\iota}\tau^2$, or $\dot{\iota}\tau^3$. We write the onsite, first nearest neighbor and second nearest neighbor terms as
\begin{equation}
u_{OO}\!=\! \dot{\iota}\mathrm{Im}h\,\tau^0 \!+\!\mathrm{Re}p \,\tau^1 \!+\! \mathrm{Im}p \,\tau^2 \!+\! \mathrm{Re}h \,\tau^3,\quad u^{(x,y,z)}_{OO} = 0,
\end{equation}
\begin{equation}
\begin{aligned}
u_{OA} &\!=\! \dot{\iota}\mathrm{Im}h_1\,\tau^0 \!+\!\mathrm{Re}p_1 \,\tau^1 \!+\! \mathrm{Im}p_1 \,\tau^2 \!+\! \mathrm{Re}h_1 \,\tau^3,\\
u^{(x)}_{OA} &\!=\! \mathrm{Re}h_1^x\,\tau^0 + \dot{\iota}(\mathrm{Re}p_1^x \,\tau^1 \!+\! \mathrm{Im}p_1^x \,\tau^2 \!+\! \mathrm{Im}h_1^x \,\tau^3),\\
u^{(y)}_{OA} &\!=\! \mathrm{Re}h_1^y\,\tau^0 \!+\! \dot{\iota}(\mathrm{Re}p_1^y \,\tau^1 \!+\! \mathrm{Im}p_1^y \,\tau^2 \!+\! \mathrm{Im}h_1^y \,\tau^3),\\
u^{(z)}_{OA} &\!=\! \mathrm{Re}h_1^z \,\tau^0 \!+\! \dot{\iota}(\mathrm{Re}p_1^z \,\tau^1 \!+\! \mathrm{Im}p_1^z \,\tau^2 \!+\! \mathrm{Im}h_1^z \,\tau^3),
\end{aligned}
\end{equation}
\begin{equation}
\begin{aligned}
u_{O1} &\!=\! \dot{\iota}\mathrm{Im}h_2\,\tau^0 \!+\!\mathrm{Re}p_2 \,\tau^1 \!+\! \mathrm{Im}p_2 \,\tau^2 \!+\! \mathrm{Re}h_2 \,\tau^3,\\
u^{(x)}_{O1} &\!=\! \mathrm{Re}h_2^x\,\tau^0 \!+\! \dot{\iota}(\mathrm{Re}p_2^x \,\tau^1 \!+\! \mathrm{Im}p_2^x \,\tau^2 \!+\! \mathrm{Im}h_2^x \,\tau^3),\\
u^{(y)}_{O1} &\!=\! \mathrm{Re}h_2^y\,\tau^0 \!+\! \dot{\iota}(\mathrm{Re}p_2^y \,\tau^1 \!+\! \mathrm{Im}p_2^y \,\tau^2 \!+\! \mathrm{Im}h_2^y \,\tau^3),\\
u^{(z)}_{O1} &\!=\! \mathrm{Re}h_2^z\,\tau^0 \!+\! \dot{\iota}(\mathrm{Re}p_2^z \,\tau^1 \!+\! \mathrm{Im}p_2^z \,\tau^2 \!+\! \mathrm{Im}h_2^z \,\tau^3),
\end{aligned}
\end{equation}

We have
\begin{itemize}
\item $g_{\bar{C}_3}(0)=\tau^0$: 
\begin{subequations}
\begin{align}
h\text{ arbitrary},\quad 
p\text{ arbitrary},\\
h_1={(-1)}^{(n_1+n_{S_4})} h_1^*={(-1)}^{n_{\bar{C}_3}} h_1^*,\\
h_1^x\!=\!-{(-1)}^{(n_1+n_{S_4})} h_1^{x*}\!=\!h_1^z\!=\!h_1^y\!=\!{(-1)}^{n_{\bar{C}_3}} h_1^{x*},\\
p_1={(-1)}^{(n_1+n_{S_4})} p_1={(-1)}^{n_{\bar{C}_3}} p_1,\\
p_1^x={(-1)}^{(n_1+n_{S_4})} p_1^x=p_1^z=p_1^y=-{(-1)}^{n_{\bar{C}_3}} p_1^x,\\
h_2=h_2^*,\!\quad \!
h_2^x=0,\!\quad \!
h_2^y=-h_2^{y*}=h_2^{z*}
,\\
{p_2\text{ arbitrary}},\quad
p_2^x=0,\quad
p_2^y=-p_2^z.
\end{align}
\end{subequations}
\item $g_{\bar{C}_3}(0)=\dot{\iota}\tau^2$: 
\begin{subequations}
\begin{align}
h = -h^*,\quad 
p = -p^*,\\
h_1={(-1)}^{(n_1+n_{S_4})} h_1^*=-{(-1)}^{n_{\bar{C}_3}} h_1,\\
h_1^x\!=\!-{(-1)}^{(n_1+n_{S_4})} h_1^{x*}\!=\!h_1^z\!=\!h_1^y\!=\!{(-1)}^{n_{\bar{C}_3}} h_1^x,\\
p_1={(-1)}^{(n_1+n_{S_4})} p_1=-{(-1)}^{n_{\bar{C}_3}} p_1^*,\\
p_1^x={(-1)}^{(n_1+n_{S_4})} p_1^x=p_1^z=p_1^y={(-1)}^{n_{\bar{C}_3}} p_1^{x*},\\
h_2=h_2^x=0,\quad
h_2^y=-h_2^{y*}=h_2^z,\\
p_2=-p_2^*,\quad
p_2^x=0,\quad
p_2^y=p_2^{z*}.
\end{align}
\end{subequations}
\item $g_{\bar{C}_3}(0)=\dot{\iota}\tau^3$:
\begin{subequations}
\begin{align}
h\text{ arbitrary},\quad p=0,\\
h_1={(-1)}^{(n_1+n_{S_4})} h_1^*={(-1)}^{n_{\bar{C}_3}} h_1^*,\\
h_1^x\!=\!-{(-1)}^{(n_1+n_{S_4})} h_1^{x*}\!=\!h_1^z\!=\!h_1^y\!=\!{(-1)}^{n_{\bar{C}_3}} h_1^{x*},\\
p_1={(-1)}^{(n_1+n_{S_4})} p_1=-{(-1)}^{n_{\bar{C}_3}} p_1,\\
p_1^x={(-1)}^{(n_1+n_{S_4})} p_1^x=p_1^z=p_1^y={(-1)}^{n_{\bar{C}_3}} p_1^x,\\
h_2=h_2^*,\!\quad \!
h_2^x=0,\!\quad \!
h_2^y=-h_2^{y*}=h^{z*},\\
p_2=p_2^x=0,\quad
p_2^y=p_2^z.
\end{align}
\end{subequations}
\end{itemize}

\begin{table*}[!thb]
\centering
\begin{tabular}{l|lll|l}
\hline
\multirow{ 2}{*}{$\mathbb{Z}_2$-Class ($g_{\bar{C}_3}(0),\eta_1,\eta_{S_4},\eta_{\bar{C}_3}$)}
&\multicolumn{3}{c|}{\multirow{ 1}{*}{Independent nonzero parameters}}&\multirow{ 2}{*}{Constraints (1NN and 2NN)}\\
\cline{2-4}
&Onsite&1NN&2NN&\\
\hline
{{$(\tau^0,+,+,+)$  or $(\tau^0,-,-,+)$}}       & {$h$, $p$} & {$\mathrm{Re}h_1$, $p_1$} &{{$\mathrm{Re}h_2$, $\mathrm{Im}h_2^y$, $p_2$, $p_2^y$}} & {$h_2^{z*}\!=\!h_2^y$, $p_2^z\!=\!-p_2^y$}\\
{{$(\tau^0,+,+,-)$  or $(\tau^0,-,-,-)$}}       & {$h$, $p$} & {$\mathrm{Im}h_1^x$, $p_1^x$} &{{$\mathrm{Re}h_2$, $\mathrm{Im}h_2^y$, $p_2$, $p_2^y$}} & {$h_1^x\!=\!h_1^y\!=\!h_1^z$, $p_1^x\!=\!p_1^y\!=\!p_1^z$, $h_2^{z*}\!=\!h_2^y$, $p_2^z\!=\!-p_2^y$}\\
{{$(\tau^0,+,-,+)$  or $(\tau^0,-,+,+)$}}       & {$h$, $p$} & {$\mathrm{Re}h_1^x$} &{{$\mathrm{Re}h_2$, $\mathrm{Im}h_2^y$, $p_2$, $p_2^y$}} & {$h_1^x\!=\!h_1^y\!=\!h_1^z$, $h_2^{z*}\!=\!h_2^y$, $p_2^z\!=\!-p_2^y$}\\
{{$(\tau^0,+,-,-)$  or $(\tau^0,-,+,-)$}}       & {$h$, $p$} & {$\mathrm{Re}h_1$} &{{$\mathrm{Re}{h}_2$, $\mathrm{Im}{h_2^y}$, $p_2$, $p_2^y$}} & {$h_2^{z*}\!=\! h_2^y$, $p_2^z\!=\!-p_2^y$}\\
{{$(\dot{\iota}\tau^2,+,+,+)$  or $(\dot{\iota}\tau^2,-,-,+)$}}       & {{$\mathrm{Im}h$,  $\mathrm{Im}p$}} & {{$\mathrm{Im}h_1^x$, $\mathrm{Im}p_1$, $\mathrm{Re}p_1^x$}} &{{$\mathrm{Im}h_2^y$, $p_2^y$}} & {$h_1^x\!=\!h_1^y\!=\!h_1^z$, $p_1^x\!=\!p_1^y\!=\!p_1^z$, $h_2^z\!=\!h_2^y$, $p_2^z\!=\!p_2^{y*}$}\\
{{$(\dot{\iota}\tau^2,+,+,-)$  or $(\dot{\iota}\tau^2,-,-,-)$}}       & {{$\mathrm{Im}h$,  $\mathrm{Im}p$}} & {{$\mathrm{Re}h_1$, $\mathrm{Re}p_1$, $\mathrm{Im}p_1^x$}} &{{$\mathrm{Im}h_2^y$, $p_2^y$}} & {$p_1^x\!=\!p_1^y\!=\!p_1^z$, $h_2^z\!=\! h_2^y$, $p_2^z\!=\!p_2^{y*}$}\\
{{$(\dot{\iota}\tau^2,+,-,+)$  or $(\dot{\iota}\tau^2,-,+,+)$}}       & {{$\mathrm{Im}h$, $\mathrm{Im}p$}} & {$\mathrm{Re}h_1^x$} &
{{$\mathrm{Im}h_2^y$, $p_2^y$}} & {$h_1^x\!=\!h_1^y\!=\!h_1^z$, $h_2^z\!=\!h_2^y$, $p_2^z\!=\!p_2^{y*}$}\\
{{$(\dot{\iota}\tau^2,+,-,-)$  or $(\dot{\iota}\tau^2,-,+,-)$}}       & {{$\mathrm{Im}h$, $\mathrm{Im}p$}} &{$\mathrm{Im}h_1$} &{{$\mathrm{Im}h_2^y$, $p_2^y$}} & {$h_2^z\!=\!h_2^y$, $p_2^z\!=\!p_2^{y*}$}\\
{{$(\dot{\iota}\tau^3,+,+,+)$  or $(\dot{\iota}\tau^3,-,-,+)$}}       & {$h$} & {{$\mathrm{Re}h_1$, $\mathrm{Im}p_1$,  $p_1^x$}} &{{$\mathrm{Re}h_2$, $\mathrm{Im}h_2^y$,  $p_2^y$}} & {$p_1^x\!=\!p_1^y\!=\!p_1^z$, $h_2^z\!=\!h_2^{y*}$, $p_2^z=p_2^y$}\\
{{$(\dot{\iota}\tau^3,+,+,-)$  or $(\dot{\iota}\tau^3,-,-,-)$}}& {$h$} & {$\mathrm{Im}h_1^x$, $p_1$} &{{$\mathrm{Re}h_2$, $\mathrm{Im}h_2^y$, $p_2^y$}} & {$h_1^x\!=\!h_1^y\!=\!h_1^z$, $h_2^z\!=\!h_2^{y*}$, $p_2^z\!=\!p_2^y$}\\
{{$(\dot{\iota}\tau^3,+,-,+)$  or $(\dot{\iota}\tau^3,-,+,+)$}}       & {$h$} & {$\mathrm{Re}h_1^x$} &
{{$\mathrm{Re}h_2$, $\mathrm{Im}h_2^y$, $p_2^y$}} & {$h_1^x\!=\!h_1^y\!=\!h_1^z$, $h_2^z\!=\!h_2^{y*}$, $p_2^z\!=\!p_2^y$}\\
{{$(\dot{\iota}\tau^3,+,-,-)$  or $(\dot{\iota}\tau^3,-,+,-)$}}       & {$h$} & {$\mathrm{Im}h_1$} &{{$\mathrm{Re}h_2$, $\mathrm{Im}h_2^y$, $p_2^y$}} & {$h_2^z\!=\!h_2^{y*}$, $p_2^z\!=\!p_2^y$}\\
\hline
\end{tabular}\caption{Summary of bond parameters for $\mathbb{Z}_2$ \textit{Ans\"{a}tze}. All bond parameters not explicitly mentioned have zero values.}\label{tab:z2_triplet_singlet}
\end{table*}

The results are summarized in Table \ref{tab:z2_triplet_singlet}. If we further take time-reversal symmetry into account, we have the following constraints:
\begin{itemize}
\item $(g_{\mathcal{T}}(0),\eta_{\mathcal{T}}) = (\tau^0,+)$: vanishing \textit{Ans\"{a}tze};
\item $(g_{\mathcal{T}}(0),\eta_{\mathcal{T}}) = (\tau^0,-)$: vanishing on even 1NN bonds; no additional constraints on odd 1NN bonds;
\item $(g_{\mathcal{T}}(0),\eta_{\mathcal{T}}) = (\dot{\iota}\tau^2,+)$: only $\tau^{1,3}$, i.e. $\mathrm{Re}p^{x,y,z}$, $\mathrm{Im}h^{x,y,z}$, $\mathrm{Re}p$ and $\mathrm{Re}h$ can be nonvanishing;
\item $(g_{\mathcal{T}}(0),\eta_{\mathcal{T}}) = (\dot{\iota}\tau^2,-)$: for odd 1NN bonds: only $\tau^{0,2}$, i.e. $\mathrm{Re}h^{x,y,z}$, $\mathrm{Im}p^{x,y,z}$, $\mathrm{Im}h$ and $\mathrm{Im}p$ can be nonvanishing; for even 1NN bonds: only $\tau^{1,3}$, i.e. $\mathrm{Re}p^{x,y,z}$, $\mathrm{Im}h^{x,y,z}$, $\mathrm{Re}p$ and $\mathrm{Re}h$ can be nonvanishing.
\end{itemize}

\section{Quadratic Spinon Hamiltonian}
\label{app:quadratic_ham}
Here, we discuss the generalized structure of the quadratic spinon Hamiltonian. To begin with, let us consider the following Bogoliubov–de Gennes(BdG) basis:
\begin{equation}
    \hat{\psi}_\mathbf{k}=(\hat{f}_{\mathbf{k},\uparrow},\hat{f}^\dagger_{-\mathbf{k},\downarrow})^T
    \label{eq:bdg_basis}
\end{equation}
where $\hat{f}_{\mathbf{k},\uparrow}=(\hat{f}_{\mathbf{k},0,\uparrow},\hat{f}_{\mathbf{k},1,\uparrow},..,\hat{f}_{\mathbf{k},n-1,\uparrow})$ and $\hat{f}_{\mathbf{k},\downarrow}=(\hat{f}_{\mathbf{k},0,\downarrow},\hat{f}_{\mathbf{k},1,\downarrow},..,\hat{f}_{\mathbf{k},n-1,\downarrow})^T$. Here, $n=$ $2$, $8$ and $32$ for classes A, B and C, respectively. In this basis, the general mean-field Hamiltonian in $\mathbf{k}$-space can be cast in the following form:
\begin{equation}
    \hat{H}(\mathbf{k})=\hat{\psi}^\dagger_\mathbf{k}\hat{H}^{BdG}_{\mathbf{k}}\hat{\psi}_\mathbf{k}
    \label{eq:bdg_ham}
\end{equation}
with
\begin{eqnarray}
\hat{H}^{BdG}_\mathbf{k} =
\begin{bmatrix}
\hat{H}^{U(1)}_\mathbf{k} & \hat{H}^{\mathds{Z}_2}_\mathbf{k}  \\
(\hat{H}^{\mathds{Z}_2}_\mathbf{k})^\dagger & -\hat{H}^{U(1)}_{-\mathbf{k}}.
\end{bmatrix}
\label{eq:quadratic_ham_k}
\end{eqnarray}
Here, $\hat{H}^{U(1)}_\mathbf{k}$ and $\hat{H}^{\mathds{Z}_2}_\mathbf{k}$ due to the contribution coming from the hopping and pairing terms, respectively. The superscript $\mathds{Z}_2$ is used to denote that if $IGG\in \mathds{Z}_2$, $\hat{H}^{\mathds{Z}_2}_\mathbf{k}\neq0$.

First, let us consider the case of the hopping only Hamiltonian, as in the case for $SU(2)$ and $U(1)$ \textit{Ans\"atze} (in a suitable gauge), i.e., $\hat{H}^{\mathds{Z}_2}_\mathbf{k}=0$. In this case, the up and down sectors ($\hat{f}_{\mathbf{k},\uparrow}$ and $\hat{f}_{\mathbf{k},\downarrow}$) are decoupled from each other and $\hat{H}^{BdG}_\mathbf{k}$ split into two equal blocks. We need to consider only one sector and the Hamiltonian takes the following form:
\begin{equation}
    \hat{H}(\mathbf{k})= \sum_{\sigma=\uparrow,\downarrow}\hat{f}^\dagger_{\mathbf{k},\sigma}\hat{H}^{U(1)}\hat{f}_{\mathbf{k},\sigma}.
    \label{eq:u1_ham}
\end{equation}
In such cases, due to spinon number conservation, one does not need to consider chemical potential explicitly. The (mean-field) one particle per site constraint can be fulfilled by setting the Fermi level such that the lower half of the energy eigenstates are filled.

But for the $\mathds{Z}_2$ \textit{Ans\"atze}, we need to consider the general basis given by Eq.~\eqref{eq:bdg_basis}. The eigenvalues come in positive and negative pairs ($\pm\epsilon_{\mathbf{k},\mu}$) and in the diagonalized basis [$(\hat{\xi}_{\mathbf{k},\mu,\uparrow},\hat{\xi}^\dagger_{-\mathbf{k},\mu,\downarrow})^T=U_{\xi f}(\hat{f}_{\mathbf{k},\mu,\uparrow},\hat{f}^\dagger_{-\mathbf{k},\mu,\downarrow})^T$ where $U_{\xi f}$ is a unitary transformation] the Hamiltonian can be rewritten as 
\begin{equation}
\left.\begin{aligned}
    \hat{H}(\mathbf{k})&= \sum^n_{\mu=1}\epsilon_{\mathbf{k},\mu}(\hat{\xi}^\dagger_{\mathbf{k},\mu,\uparrow}\hat{\xi}_{\mathbf{k},\mu,\uparrow}-\hat{\xi}_{\mathbf{k},\mu,\downarrow}\hat{\xi}^\dagger_{\mathbf{k},\mu,\downarrow})\\
    &=\sum^n_{\mu=1}\sum_{\sigma=\uparrow,\downarrow}\epsilon_{\mathbf{k},\mu}(\hat{\xi}^\dagger_{\mathbf{k},\mu,\sigma}\hat{\xi}_{\mathbf{k},\mu,\sigma}-1).\\
\end{aligned}\right.
    \label{eq:z2_ham}
\end{equation}
These can be interpreted as Bogoliubov quasiparticles having positive excitation energy. Thus, the Fermi level lies at zero energy.

\section{Robustness of the 4-fold degenerate nodal loops in the spectrum of $0$-flux $\mathbf{SU(2)}$ state}
\label{app:robust_nodal_loops}
As discussed in Sec.~\ref{sec:class-a-su2}, the band structure of the $SU(2)$ $0$-flux state ($SA3$) comprises of three 4-fold degenerate nodal loops when the amplitudes are considered up to 3NN. The corresponding \textit{Ansatz} is given by Eq.~\eqref{eq:sa3}. In this section, we consider bonds beyond 3NN, to verify the robustness of the nodal manifold. The corresponding PSG is given by
\begin{align}\label{eq:sa3_psg_0}
     G_{T_i} (r,\mu) =& g_{T_i}, \; i=1,2,3 \; , \notag \\
    G_{\mathcal{O}}(r,\mu) =& (-1)^{\mu} g_{\mathcal{O}}, \; g_{\mathcal{O}}\in SU(2),\; \mathcal{O}=\{S_4,\bar{C}_3,\mathcal{T}\}\; .
\end{align}
The global $SU(2)$ matrices can be fixed as $g_\mathcal{O}=\tau^0$. Thus
\begin{align}\label{eq:sa3_psg}
     G_{T_i} (r,\mu) =& \tau^0, \; i=1,2,3 \; , \notag \\
    G_{\mathcal{O}}(r,\mu) =& (-1)^{\mu} \tau^0, \;  \mathcal{O}=\{S_4,\bar{C}_3,\mathcal{T}\}\; .
\end{align}
Akin to the vanishing 2NN \textit{Ansatz} ($u_2=0$), the structure of $G_\mathcal{O}(r,\mu)$ immediately sets mean-field parameters on the other even nearest neighbor bonds (4NN, 6NN, 8NN,...) to zero in order to satisfy
\begin{equation}
    g^{\dagger}_{\mathcal{T}}(i,\mu) u_{i,\mu;j,\mu} g_{\mathcal{T}}(j,\mu) = - u_{i,\mu;j,\mu}\; .
\end{equation}
Thus the only terms that contribute to the Hamiltonian are those corresponding to bonds that connect $\mu = 0$ sites to $\mu = 1$ sites, i.e., odd nearest neighbors such as 5NN, 7NN, 9NN, etc. We define the corresponding reference bonds as $u_5$, $u_7$ and $u_9$, respectively. Using the PSG given by Eq.~\eqref{eq:sa3_psg}, we obtain the following \textit{Ans\"atze} for $SA3$ up to 10NN:
\begin{align}
 u_1 =& \dot{\iota} h_1\tau^0,\; u_3=\dot{\iota} h_3\tau^0 ,\;u_5=\dot{\iota} h_5\tau^0 ,\;u_7=\dot{\iota} h_7\tau^0 ,\;  \notag \\
u_9=& \dot{\iota} h_9\tau^0,\;u_2=u_4=u_6=u_8=u_{10}=0 \; .
\end{align}
The Hamiltonian can then be written as
\begin{align}
\hat{H}(\mathbf{k}) = \sum_{\sigma=\uparrow,\downarrow}
\begin{bmatrix}
    \hat{f}^\dagger_{\mathbf{k},0,\sigma}\\
    \hat{f}^\dagger_{\mathbf{k},1,\sigma} 
\end{bmatrix}
^T
\begin{bmatrix}
    0 & \dot{\iota} A_\mathbf{k} \\
    -\dot{\iota} A^*_\mathbf{k} & 0 
\end{bmatrix}
\begin{bmatrix}
    \hat{f}_{\mathbf{k},0,\sigma}\\
    \hat{f}_{\mathbf{k},1,\sigma} 
\end{bmatrix},
\end{align}
where, $A_\mathbf{k}$ = $h_1A_{\mathbf{k},1} + h_3A_{\mathbf{k},3} + h_5A_{\mathbf{k},5} + \dots$ and $A_{\mathbf{k},n}$ = $\sum_{j \in nNN} e^{\dot{\iota} \mathbf{k} \cdot \delta_j}, \{n = 1,3,5,7,\dots\}$ and $h_n$ are the real hopping parameters.
\begin{widetext}
\begin{align}
A_{\mathbf{k},1} =& 4 \left(\cos \left(\frac{k_x}{4}\right) \cos \left(\frac{k_y}{4}\right) \cos \left(\frac{k_z}{4}\right) - i \sin \left(\frac{k_x}{4}\right) \sin \left(\frac{k_y}{4}\right) \sin \left(\frac{k_z}{4}\right) \right),
\end{align}
\begin{align}
A_{\mathbf{k},3} =& 4 \cos \left(\frac{k_x}{4}\right) \cos \left(\frac{k_y}{4}\right) \cos \left(\frac{k_z}{4}\right) \left(-3 + 2 \cos \left(\frac{k_x}{2}\right) + 2 \cos \left(\frac{k_y}{2}\right) + 2 \cos \left(\frac{k_z}{2}\right)\right), \; \notag \\
&+ 4 \dot{\iota} \left(3 + 2 \cos \left(\frac{k_x}{2}\right) + 2 \cos \left(\frac{k_y}{2}\right) + 2 \cos \left(\frac{k_z}{2}\right)\right) \sin \left(\frac{k_x}{4}\right) \sin \left(\frac{k_y}{4}\right) \sin \left(\frac{k_z}{4}\right),
\end{align}
\begin{align}
A_{\mathbf{k},5} =& 4 \cos \left(\frac{3 k_x}{4}\right) \cos \left(\frac{3 k_y}{4}\right) \cos \left(\frac{k_z}{4}\right) + 
 8 \cos\left(\frac{k_x}{4}\right) \cos\left(\frac{k_y}{4}\right) \left(-1 + \cos \left(\frac{k_x}{2}\right) + \cos\left(\frac{k_y}{2}\right)\right) \cos\left(\frac{3 k_z}{4}\right) \; \notag \\
& - 4 \dot{\iota} \sin\left(\frac{3 k_x}{4}\right) \sin\left(\frac{3 k_y}{4}\right) \sin\left(\frac{k_z}{4}\right) + 
 8 \dot{\iota} \left(1 + \cos\left(\frac{k_x}{2}\right) + \cos\left(\frac{k_y}{2}\right)\right) \sin\left(\frac{k_x}{4}\right) \sin\left(\frac{k_y}{4}\right) \sin\left(\frac{3 k_z}{4}\right),
\end{align}

\begin{align}
 A_{\mathbf{k},7} =& 4 \cos\left(\frac{3 k_x}{4}\right) \cos\left(\frac{3 k_y}{4}\right) \cos\left(\frac{3 k_z}{4}\right) + 4 \cos\left(\frac{k_x}{4}\right) \cos\left(\frac{k_y}{4}\right) \; \notag \\
&\times \left(2 \left(1 - \cos\left(\frac{k_x}{2}\right) + \cos\left(k_x\right) - \cos\left(\frac{k_y}{2}\right) + \cos\left(k_y\right) \right) \cos\left(\frac{k_z}{4}\right) + \cos\left(\frac{5 k_z}{4}\right)\right) \; \notag \\
&+ 4 \dot{\iota} \sin\left(\frac{3 k_x}{4}\right) \sin\left(\frac{3 k_y}{4}\right) \sin\left(\frac{3 k_z}{4}\right) -  4 \dot{\iota} \sin\left(\frac{k_x}{4}\right) \sin\left(\frac{k_y}{4}\right) \; \notag \\
&\times \left(2 \left(1 + \cos\left(\frac{k_x}{2}\right) + \cos\left(k_x\right) + \cos\left(\frac{k_y}{2}\right) + \cos\left(k_y\right)\right) \sin\left(\frac{k_z}{4}\right) + \sin\left(\frac{5 k_z}{4}\right)\right),
\end{align}

\begin{align}
A_{\mathbf{k},9} =& 4 \left(\cos \left(\frac{3 k_x}{4}\right) \cos \left(\frac{5 k_y}{4}\right) \cos \left(\frac{k_z}{4}\right) - \dot{\iota} \sin \left(\frac{3 k_x}{4}\right) \sin \left(\frac{5 k_y}{4}\right) \sin \left(\frac{k_z}{4}\right) \right) \; \notag \\
& + 4 \left(\cos \left(\frac{k_x}{4}\right) \cos \left(\frac{5 k_y}{4}\right) \cos \left(\frac{3 k_z}{4}\right) - \dot{\iota} \sin \left(\frac{k_x}{4}\right) \sin \left(\frac{5 k_y}{4}\right) \sin \left(\frac{3 k_z}{4}\right) \right) \; \notag \\
& + 4 \left(\cos \left(\frac{5 k_x}{4}\right) \cos \left(\frac{3 k_y}{4}\right) \cos \left(\frac{k_z}{4}\right) - \dot{\iota} \sin \left(\frac{5 k_x}{4}\right) \sin \left(\frac{3 k_y}{4}\right) \sin \left(\frac{k_z}{4}\right) \right) \; \notag \\
& + 4 \left(\cos \left(\frac{k_x}{4}\right) \cos \left(\frac{3 k_y}{4}\right) \cos \left(\frac{5 k_z}{4}\right) - \dot{\iota} \sin \left(\frac{k_x}{4}\right) \sin \left(\frac{3 k_y}{4}\right) \sin \left(\frac{5 k_z}{4}\right) \right) \; \notag \\
& + 4 \left(\cos \left(\frac{5 k_x}{4}\right) \cos \left(\frac{k_y}{4}\right) \cos \left(\frac{3 k_z}{4}\right) - \dot{\iota} \sin \left(\frac{5 k_x}{4}\right) \sin \left(\frac{k_y}{4}\right) \sin \left(\frac{3 k_z}{4}\right) \right) \; \notag \\
& + 4 \left(\cos \left(\frac{3 k_x}{4}\right) \cos \left(\frac{k_y}{4}\right) \cos \left(\frac{5 k_z}{4}\right) - \dot{\iota} \sin \left(\frac{3 k_x}{4}\right) \sin \left(\frac{k_y}{4}\right) \sin \left(\frac{5 k_z}{4}\right) \right) \; .
\end{align}
\end{widetext}

The eigenvalues are 2-fold degenerate and given by $\epsilon_{\mathbf{k},\mu}=\pm |A_\mathbf{k}|$. Now, it can be verified that $|A_{\mathbf{k},n}|=0\; \forall~n$ along the following twelve lines given by
\begin{equation}
\left.\begin{aligned}
&k_\alpha=0,\;k_{\beta\neq\alpha}=\pm2\pi\;\text{where} \;\alpha,\beta=x,y,z\; .\\
\end{aligned}\right.
\label{eq:degenerate_lines}
\end{equation}
Therefore these twelve lines are the trajectory of the 4-fold degenerate gapless $\mathbf{k}$-points which give rise to the three nodal loops. This strongly supports the robustness of the nodal manifold.  It is important to mention that the same projective action of time-reversal symmetry is also present in the $SU(2)$ $\pi$-flux state ($SB1$) where no such nodal structure can be seen, thus the underlying symmetry responsible for the robustness of the nodal structure is not the projective realization of time-reversal. Thus we conclude that it is the projective realization of lattice space group symmetries (here, screw and rotoinversion) that are responsible for the appearance of nodal band topology. Also, note that in the cases of the \textit{Ans\"atze} labeled by $UA01$, $ZA000$ the lattice symmetries are acting linearly. The presence of the nodal manifold in these cases implies such band topology is protected by the lattice symmetries as well as a particular projective implementations of those symmetries given by Eq.~\eqref{eq:sa3_psg_0}. Actually the non projective operations of the lattices symmetries can also be connected symmetrically to the PSG of the zero flux $SU(2)$ state ($SA3$). Thus the non projective class is also a subgroup of the PSGs of $SA3$. Thus actually protection of the nodal band structure is originating from the larger projective class.

 Furthermore, we also report another nodal band topology in the $\pi$-flux state. However, unlike zero-flux state, its appearance is not robust, but rather an artifact of restriction to short ranged couplings. We verified this by incorporating fifth-nearest neighbors (5NN) and found an opening up of a gap. The details regarding the sign configuration of the \textit{Ans\"atze} on 5NN bonds are as follows.
 
The twelve 5NN bonds can be defined by connecting $F(0,0,0,0)$ with the twelve 5NN sites given by $1(0,0,-2,1)$, $2(0,1,-2,1)$, $3(1,0,-2,1)$, $4(-2,0,0,1)$, $5(-2,1,0,1)$, $6(0,-2,0,1)$, $7(0,1,0,1)$, $8(1,-2,0,1)$, $9(1,0,0,1)$, $10(-2,0,1,1)$, $11(0,-2,1,1)$ and $12(0,0,1,1)$, and denoting the bonds by $u_{F\alpha}$ with $\alpha=1,2,3,...,12$. We choose $u_{F1}=u_5$ to be the reference bond for 5NNs.
$u_{F\alpha} = \{1,-1,1,1,-1,-1,-1,1,1,-1,-1,-1\} u_5$, where $\alpha = 1,2,3,...,12$.
The pattern of spatial modulation of the \textit{Ans\"atze} are given below for the 5NN bonds,

\begin{equation}
	\label{eq:sign_str_5nn}
	\left.\begin{aligned}
		\langle(r_1,r_2,r_3,0)&(r_1,r_2,r_3-2,1)\rangle=u_{F1},\\
		\langle(r_1,r_2,r_3,0)&(r_1,r_2+1,r_3-2,1)\rangle=\eta_1^{r_3} u_{F2},\\
        \langle(r_1,r_2,r_3,0)&(r_1+1,r_2,r_3-2,1)\rangle= \eta_1^{r_2+r_3} u_{F3},\\
        \langle(r_1,r_2,r_3,0)&(r_1-2,r_2,r_3,1)\rangle={u_{F4}},\\
        \langle(r_1,r_2,r_3,0)&(r_1-2,r_2+1,r_3,1)\rangle=\eta_1^{r_3} u_{F5},\\
        \langle(r_1,r_2,r_3,0)&(r_1,r_2-2,r_3,1)\rangle=u_{F6},\\
        \langle(r_1,r_2,r_3,0)&(r_1,r_2+1,r_3,1)\rangle=\eta_1^{r_3} u_{F7},\\
        \langle(r_1,r_2,r_3,0)&(r_1+1,r_2-2,r_3,1)\rangle=\eta_1^{r_2+r_3} u_{F8},\\
        \langle(r_1,r_2,r_3,0)&(r_1+1,r_2,r_3,1)\rangle=\eta_1^{r_2+r_3} u_{F9},\\
        \langle(r_1,r_2,r_3,0)&(r_1-2,r_2,r_3+1,1)\rangle=u_{F10},\\
        \langle(r_1,r_2,r_3,0)&(r_1,r_2-2,r_3+1,1)\rangle=u_{F11},\\
        \langle(r_1,r_2,r_3,0)&(r_1,r_2,r_3+1,1)\rangle=u_{F12}.\\
	\end{aligned}\right.
\end{equation}

\section{Calculation of the Dynamical Spin Structure Factor}
\label{app:calculation_dsf}
The dynamical spin structure factor is defined as
\begin{equation}\label{eq:dsf}
	\left.\begin{aligned}
\mathcal{S}^{\lambda\lambda'}(\mathbf{q},\omega)=\int^{+\infty}_{-\infty}\frac{d\tau e^{\dot{\iota}\omega\tau}}{2\pi \mathcal{N}}&\sum_{i,j}\sum^N_{m,n=1}e^{\dot{\iota}\mathbf{q}\cdot(\mathbf{r}_{i,m}-\mathbf{r}_{j,n})}\\
&\times\langle \hat{S}^\lambda_{i,m}(\tau)\hat{S}^{\lambda'}_{j,n}(0)\rangle\\
	\end{aligned}\right.
\end{equation}
where $\lambda,\lambda'\in\{x,y,z\}$ and $(m,n)$ is the sublattice index. Given the spin-rotational invariance of the model permits us to restrict the study to longitudinal components only. Thus we shall consider $\mathcal{S}^{zz}(\mathbf{q},\omega)$ only. Now, with the substitution of $\hat{S}^z_{i,m}(\tau)=e^{\dot{\iota}\hat{H}\tau}\hat{S}^z_{i,m}e^{-\dot{\iota}\hat{H}\tau}$, Eq.~\eqref{eq:dsf} reads
\begin{equation}\label{eq:dsf_zz}
	\left.\begin{aligned}
\mathcal{S}^{zz}(\mathbf{q},\omega)&=\int^{+\infty}_{-\infty}\frac{d\tau e^{\dot{\iota}\omega\tau}}{2\pi \mathcal{N}}\\
&\times\sum_{i,j,m,n}e^{\dot{\iota}\mathbf{q}\cdot(\mathbf{r}_{i,m}-\mathbf{r}_{j,n})}\langle e^{\dot{\iota}\hat{H}\tau}\hat{S}^z_{i,m}e^{-\dot{\iota}\hat{H}\tau}\hat{S}^z_{j,n}\rangle.\\
	\end{aligned}\right.
\end{equation}
This can be recast in terms of fermion operators after substituting Eq.~\eqref{eq:abrikosov} as
\begin{equation}\label{eq:dsf_zz_f}
	\left.\begin{aligned}
\mathcal{S}^{zz}(\mathbf{q},\omega)&=\int^{+\infty}_{-\infty}\frac{d\tau e^{\dot{\iota}\omega\tau}}{2\pi \mathcal{N}}\sum_{i,j,m,n}e^{\dot{\iota}\mathbf{q}\cdot(\mathbf{r}_{i,m}-\mathbf{r}_{j,n})}(\delta_{\alpha,\beta}-\delta_{\alpha,\bar{\beta}})\\
&\times\sum_{\alpha,\beta}\langle e^{\dot{\iota}\hat{H}\tau}\hat{f}^\dagger_{i,m,\alpha}\hat{f}_{i,m,\alpha}e^{-\dot{\iota}\hat{H}\tau}\hat{f}^\dagger_{j,n,\beta}\hat{f}_{j,n,\beta}\rangle.\\
	\end{aligned}\right.
\end{equation}
After Fourier transformation, this takes the form
\begin{equation}\label{eq:dsf_z2_uu}
	\left.\begin{aligned}
\mathcal{S}^{zz}(\mathbf{q},\omega)&=\int^{+\infty}_{-\infty}\frac{d\tau e^{\dot{\iota}\omega\tau}}{2\pi \mathcal{N}}\sum_{\mathbf{k},\mathbf{k}',m,n}e^{\dot{\iota}\mathbf{q}\cdot \mathbf{r}_{mn}}(\delta_{\alpha,\beta}-\delta_{\alpha,\bar{\beta}})\\
&\times\sum_{\alpha,\beta}\langle e^{\dot{\iota}\hat{H}\tau}\hat{f}^\dagger_{\mathbf{k},m,\alpha}\hat{f}_{\mathbf{k+q},m,\alpha}e^{-\dot{\iota}\hat{H}\tau}\hat{f}^\dagger_{\mathbf{k'+q},n,\beta}\hat{f}_{\mathbf{k'},n,\beta}\rangle.\\
	\end{aligned}\right.
\end{equation}
where $\mathbf{r}_{mn}=\mathbf{r}_m-\mathbf{r}_n$. To show the derivation of $\mathcal{S}^{zz}(\mathbf{q},\omega)$ for $\mathds{Z}_2$ \textit{Ans\"atze}, we demonstrate one term corresponding to $\alpha=\beta=\uparrow$:
\begin{equation}\label{eq:dsf_uu}
	\left.\begin{aligned}
\mathcal{S}^{zz}_{\uparrow\uparrow}(\mathbf{q},\omega)&=\int^{+\infty}_{-\infty}\frac{d\tau  e^{\dot{\iota}\omega\tau}}{2\pi \mathcal{N}}\sum_{\mathbf{k},\mathbf{k'},m,n}e^{\dot{\iota}\mathbf{q}\cdot \mathbf{r}_{mn}}\\
&\times\langle e^{\dot{\iota}\hat{H}\tau}\hat{f}^\dagger_{\mathbf{k},m,\uparrow}\hat{f}_{\mathbf{k+q},m,\uparrow}e^{-\dot{\iota}\hat{H}\tau}\hat{f}^\dagger_{\mathbf{k'+q},n,\uparrow}\hat{f}_{\mathbf{k'},n,\uparrow}\rangle.\\
	\end{aligned}\right.
\end{equation}
Let us consider the unitary matrix $U_{\mathbf{k}}$ that diagonalizes the Bogoliubov-deGennes (BdG) Hamiltonian as discussed in Appendix~\ref{app:quadratic_ham}:   
\begin{equation}
U^\dagger_\mathbf{k}\hat{H}(\mathbf{k})U_{\mathbf{k}}=\text{diag}(\epsilon_{\mathbf{k},1,\uparrow},..,\epsilon_{\mathbf{k},N,\uparrow},\epsilon_{-\mathbf{k},1,\downarrow},..,\epsilon_{-\mathbf{k},N,\downarrow}).
\end{equation}
where $N$ is the number of sublattices. Under $U_{\mathbf{k}}$ the fermion operators transform as
\begin{equation}
	\left.\begin{aligned}
&\hat{f}_{\mathbf{k},m,\uparrow}=U_{\mathbf{k}}(m,p)\hat{\xi}_{\mathbf{k},p,\uparrow}+U_{\mathbf{k}}(m,p+N)\hat{\xi}^\dagger_{-\mathbf{k},p,\downarrow}\\
&\hat{f}^\dagger_{-\mathbf{k},m,\downarrow}=U_{\mathbf{k}}(m+N,p)\hat{\xi}_{\mathbf{k},p,\uparrow}+U_{\mathbf{k}}(m+N,p+N)\hat{\xi}^\dagger_{-\mathbf{k},p,\downarrow}.\\
	\end{aligned}\right.
\end{equation}
Due to the absence of Bogoliubov quasiparticles in the ground state vacuum $|0\rangle$, the scattering mechanism follows the creation of a pair of Bogoliubov quasiparticles at time $\tau=0$ and annihilation at time $\tau$. Now, the $e^{-\dot{\iota}\hat{H}\tau}\hat{f}^\dagger_{\mathbf{k'+q},n,\uparrow}\hat{f}_{\mathbf{k'},n,\uparrow}|0\rangle$ can be recast in terms of quasiparticle creation and annihilation operators as
\begin{equation}
	\left.\begin{aligned}
&e^{-\dot{\iota}\hat{H}\tau}(U^*_{\mathbf{k'+q}}(n,\nu)\hat{\xi}^\dagger_{\mathbf{k'+q},\nu,\uparrow}+U^*_{\mathbf{k'+q}}(n,\nu+N)\hat{\xi}_{\mathbf{-k'-q},\nu,\downarrow})\\
&\times (U_{\mathbf{k'}}(n,\nu')\hat{\xi}_{\mathbf{k'},\nu',\uparrow}+U_{\mathbf{k'}}(n,\nu'+N)\hat{\xi}^\dagger_{-\mathbf{k'},\nu',\downarrow})|0\rangle.\\
	\end{aligned}\right.
\end{equation}
The quasiparticle-free vacuum leads to the only nonzero part of $e^{-\dot{\iota}\hat{H}\tau}\hat{f}^\dagger_{\mathbf{k'+q},n,\uparrow}\hat{f}_{\mathbf{k'},n,\uparrow}|0\rangle$ given by 
\begin{equation}\label{eq:dsf_1}
	\left.\begin{aligned}
&e^{-\dot{\iota}(\epsilon_{\mathbf{k'+q},\nu,\uparrow}+\epsilon_{-\mathbf{k'},\nu',\downarrow})\tau}\\
&\times U^*_{\mathbf{k'+q}}(n,\nu)U_{\mathbf{k'}}(n,\nu'+N)\hat{\xi}^\dagger_{\mathbf{k'+q},\nu,\uparrow}\hat{\xi}^\dagger_{-\mathbf{k'},\nu',\downarrow}|0\rangle.\\
	\end{aligned}\right.
\end{equation}
Now the annihilation of the pair $\hat{\xi}^\dagger_{\mathbf{k'+q},\nu,\uparrow}\hat{\xi}^\dagger_{-\mathbf{k'},\nu',\downarrow}$, $\langle0| e^{\dot{\iota}\hat{H}\tau}\hat{f}^\dagger_{\mathbf{k},m,\uparrow}\hat{f}_{\mathbf{k+q},m,\uparrow}$ should include two annihilation operators.
\begin{equation}\label{eq:dsf_2}
	\left.\begin{aligned}
&\langle0|e^{\dot{\iota}\hat{H}\tau}U^*_{\mathbf{k}}(m,\mu+N)U_{\mathbf{k+q}}(m,\mu')\hat{\xi}_{-\mathbf{k},\mu,\downarrow}\hat{\xi}_{\mathbf{k+q},\mu',\uparrow}.\\
	\end{aligned}\right.
\end{equation}
From Eq.~\eqref{eq:dsf_1} and Eq.~\eqref{eq:dsf_2}, the nonvanishing contribution of the matrix element $\langle e^{\dot{\iota}\hat{H}\tau}\hat{f}^\dagger_{\mathbf{k},m,\alpha}\hat{f}_{\mathbf{k+q},m,\alpha}e^{-\dot{\iota}\hat{H}\tau}\hat{f}^\dagger_{\mathbf{k'+q},n,\beta}\hat{f}_{\mathbf{k'},n,\beta}\rangle$ is given by
\begin{equation}\label{eq:dsf_3}
	\left.\begin{aligned}
&U^*_{\mathbf{k}}(m,\mu+N)U_{\mathbf{k+q}}(m,\mu')U^*_{\mathbf{k'+q}}(n,\nu)U_{\mathbf{k'}}(n,\nu'+N)\\ 
&e^{-\dot{\iota}(\epsilon_{\mathbf{k'+q},\nu,\uparrow}+\epsilon_{-\mathbf{k'},\nu',\downarrow})\tau}\times\delta(\mathbf{k}-\mathbf{k'})\delta_{\mu',\nu}\delta_{\nu',\mu}.\\
	\end{aligned}\right.  
\end{equation}
As we discussed in Appendix~\ref{app:quadratic_ham}, the energy eigenvalues follows $\epsilon_{\mathbf{k},\mu,\uparrow}=\epsilon_{-\mathbf{k},\mu,\downarrow}=\epsilon_{\mathbf{k},\mu}$. Using this fact and substituting the expression~\eqref{eq:dsf_3} in Eq.~\eqref{eq:dsf_uu} we obtain
\begin{equation}\label{eq:dsf_4}
	\left.\begin{aligned}
\mathcal{S}^{zz}_{\uparrow\uparrow}(\mathbf{q},\omega)&=\frac{1}{\mathcal{N}}\sum_{\mathbf{k},m,n}\delta(\omega-\epsilon_{\mathbf{k+q},\mu}-\epsilon_{\mathbf{k},\nu})e^{\dot{\iota}\mathbf{q}\cdot \mathbf{r}_{mn}}\\
&\times U^*_{\mathbf{k}}(m,\mu+N)U_{\mathbf{k+q}}(m,\nu)\\
&\times U^*_{\mathbf{k+q}}(n,\nu)U_{\mathbf{k}}(n,\mu+N).\\
	\end{aligned}\right.
\end{equation}
Similarly, one can derive the expressions for other terms as well. Thus, the final expression for $\mathcal{S}^{zz}(\mathbf{q},\omega)$ for a $\mathds{Z}_2$ \textit{Ans\"atze} reads as
\begin{widetext}
\begin{equation}\label{eq:dsf_z2}
    	\left.\begin{aligned} 
\mathcal{S}^{zz}_{\mathds{Z}_2}(\mathbf{q},\omega)&=\frac{2}{ \mathcal{N}}\sum_{\mathbf{k},m,n}\delta(\omega-\epsilon_{\mathbf{k+q},\mu}-\epsilon_{\mathbf{k},\nu})e^{\dot{\iota}\mathbf{q}\cdot \mathbf{r}_{mn}}\\
&\times \left[U^*_{\mathbf{k}}(m,\mu+N)U_{\mathbf{k+q}}(m,\nu)+U_{\mathbf{k+q}}(m+N,\nu)U^*_{\mathbf{k}}(m+N,\mu+N)\right]U^*_{\mathbf{k+q}}(n,\nu)U_{\mathbf{k}}(n,\mu+N).\\
	\end{aligned}\right.
\end{equation}
\end{widetext}
To derive the expression for the $U(1)$ \textit{Ans\"atze}, we recall the fact that BdG Hamiltonian is block diagonal with two equal blocks in the $\uparrow$ and $\downarrow$ sectors. Thus, the basis contains only annihilation operators and the unitary matrix $U_{\mathbf{k}}$ would be such that $U^\dagger_{\mathbf{k}}\hat{H}(\mathbf{k})U_{\mathbf{k}}=\text{diag}(\epsilon_{\mathbf{k},1},\epsilon_{\mathbf{k},2},..\epsilon_{\mathbf{k},N})$. Furthermore, the basis vectors will transform as $\hat{f}_{\mathbf{k},m,\alpha}=U_{\mathbf{k}}(m,\mu)\hat{\xi}_{\mathbf{k},\mu,\alpha}$. In terms of new operators Eq.~\eqref{eq:dsf_z2_uu} can be written as 
\begin{equation}\label{eq:dsf_u_1}
	\left.\begin{aligned}
&\mathcal{S}^{zz}_{U(1)}(\mathbf{q},\omega)=\int^{+\infty}_{-\infty}\frac{dte^{\dot{\iota}\omega\tau}}{2\pi N}\sum_{\mathbf{k},\mathbf{k'},m,n}e^{\dot{\iota}\mathbf{q}\cdot \mathbf{r}_{mn}}(\delta_{\alpha,\beta}-\delta_{\alpha,\bar{\beta}})\\
&\times U^*_{\mathbf{k}}(m,\mu)U_{\mathbf{k+q}}(m,\mu')U^*_{\mathbf{k'+q}}(n,\nu)U_{\mathbf{k'}}(n,\nu')\\
&\times\sum_{\alpha,\beta}\langle e^{\dot{\iota}\hat{H}\tau}\hat{\xi}^\dagger_{\mathbf{k},\mu,\alpha}\hat{\xi}_{\mathbf{k+q},\mu',\alpha}e^{-\dot{\iota}\hat{H}\tau}\hat{\xi}^\dagger_{\mathbf{k'+q},\nu,\beta}\hat{\xi}_{\mathbf{k'},\nu',\beta}\rangle.\\
	\end{aligned}\right.
\end{equation}
In contrary to $\mathds{Z}_2$ \textit{Ans\"atze}, the lowest half energy eigenstates are filled for $U(1)$ states. This reflects in the scattering mechanism. At time $\tau=0$, a pair of excitations will be created by removing a particle with a state $(\mathbf{k'},\beta)$ from the filled bands and creating a particle with a state $(\mathbf{k'+q},\beta)$ at the empty bands and then annihilate the pair of excitations at time $\tau$. This requires that $\langle e^{\dot{\iota}\hat{H}\tau}\hat{\xi}^\dagger_{\mathbf{k},\mu,\alpha}\hat{\xi}_{\mathbf{k+q},\mu',\alpha}e^{-\dot{\iota}\hat{H}\tau}\hat{\xi}^\dagger_{\mathbf{k'+q},\nu,\beta}\hat{\xi}_{\mathbf{k'},\nu',\beta}\rangle$ gives,
\begin{equation}\label{eq:dsf_u_2}
	\left.\begin{aligned}
&U^*_{\mathbf{k}}(m,\mu)U_{\mathbf{k+q}}(m,\mu')U^*_{\mathbf{k'+q}}(n,\nu)U_{\mathbf{k'}}(n,\nu')\\ 
&e^{-\dot{\iota}(\epsilon_{\mathbf{k'+q},\nu}-\epsilon_{\mathbf{k'},\nu'})\tau}\times\delta(\mathbf{k}-\mathbf{k'})\delta_{\mu',\nu}\delta_{\nu',\mu}\delta_{\alpha,\beta}.\\
	\end{aligned}\right.  
\end{equation}
With this substitution Eq.~\eqref{eq:dsf_u_1} can be written as,
\begin{equation}\label{eq:dsf_u_3}
	\left.\begin{aligned}
\mathcal{S}^{zz}_{U(1)}(\mathbf{q},\omega)&=\frac{2}{\mathcal{N}}\sum_{\mathbf{k},m,n}e^{\dot{\iota}\mathbf{q}\cdot \mathbf{r}_{mn}}\delta(\omega-\epsilon_{\mathbf{k+q},\nu}+\epsilon_{\mathbf{k},\mu})\\
&\times U^*_{\mathbf{k}}(m,\mu)U_{\mathbf{k+q}}(m,\nu)U^*_{\mathbf{k+q}}(n,\nu)U_{\mathbf{k}}(n,\mu)\\
&\times n_{\mathbf{k},\mu}(1-n_{\mathbf{k+q},\nu}),\\
	\end{aligned}\right.
\end{equation}
where $n_{\mathbf{k},\gamma}=\frac{1}{e^{\beta(\epsilon_{\mathbf{k},\gamma}-\epsilon_F)}+1}$ is the Fermi distribution function with Fermi energy $\epsilon_F$ which at $\beta=\infty$ can be written using step function as $n_{\mathbf{k},\mu}(1-n_{\mathbf{k+q},\nu})=\theta(\epsilon_F-\epsilon_{\mathbf{k},\mu})\theta(\epsilon_{\mathbf{k+q},\nu}-\epsilon_F)$. Thus the final expression of dynamical spin structure factor for $U(1)$ \textit{Ans\"atze} takes the form
\begin{equation}\label{eq:dsf_u1}
	\left.\begin{aligned}
\mathcal{S}^{zz}_{U(1)}(\mathbf{q},\omega)&=\frac{2}{\mathcal{N}}\sum_{\mathbf{k},m,n}e^{\dot{\iota}\mathbf{q}\cdot \mathbf{r}_{mn}}\delta(\omega-\epsilon_{\mathbf{k+q},\nu}+\epsilon_{\mathbf{k},\mu})\\
&\times U^*_{\mathbf{k}}(m,\mu)U_{\mathbf{k+q}}(m,\nu)U^*_{\mathbf{k+q}}(n,\nu)U_{\mathbf{k}}(n,\mu)\\
&\times\theta(\epsilon_{\mathbf{k+q},\nu}-\epsilon_F)\theta(\epsilon_F-\epsilon_{\mathbf{k},\mu}).\\
	\end{aligned}\right.
\end{equation}

\end{document}